%% file: disertace.tex
\documentclass[12pt,a4paper,oneside]{thesis}
\usepackage{amsfonts}%
\usepackage{amssymb}%
\usepackage{fancyhdr}
\usepackage{array}%
\usepackage{epsf}%
\usepackage[pdftex]{graphics} 
\usepackage[sl,small]{caption}
\usepackage{subfigure}
\usepackage{txfonts}
\usepackage{url}
\usepackage{rotating}
\usepackage{lscape}
\usepackage{natbib}
\usepackage{setspace}

\renewcommand{\baselinestretch}{1.0}
\newcommand{\vsp}[1]{\vspace{#1mm}}
\newcommand{\degree}[1][]{\ensuremath{{#1}^\circ}}
\newcommand{\diff}{\mathrm{d}}

\input cyracc.def

\begin{document}
\chapapp{Chapter}
\input{titlepage}
\input{declaration}
\pagestyle{fancy} \pagenumbering{roman} \lhead{} \fancyfoot[C]{\thepage}
\input{acknowledgements}
\input{abstract}
\tableofcontents \listoffigures \listoftables
\input{notation}
\pagestyle{fancy} \lhead{} \fancyhead[R]{\slshape \rightmark} \fancyfoot[C]{\thepage}
\renewcommand{\sectionmark}[1]{\markright{\thesection.\ #1}{}}
\input{introduction}
\input{survey}

\input{theory}

\input{apparatus}

\input{results}
\input{discussion}
\input{conclusion}
\input{references}
\input{appendix}
\end{document}

%% file: titlepage.tex
\begin{titlepage}
\begin{center}
Czech Technical University in Prague \\
Faculty of Electrical Engineering \\Department of Physics\\

 \vspace*{1.5in} {\LARGE The Study of a~Fibre $Z$--Pinch}

\par
\vspace*{1.5in} {\large Ing. Daniel Kl\'{\i}r}\\
Supervisor: Prof.~RNDr. Pavel Kube\v{s}, CSc.

\par
\vfill
A thesis submitted for the degree of Doctor of Philosophy\\
PhD Programme: Electrical Engineering and Informatics \\
Branch of Study: Plasma Physics\\
 \par \vspace*{0.5in} September 2005
\end{center} \end{titlepage}

%% file: declaration.tex
\chapter*{Declaration}
\thispagestyle{empty} \vfill \noindent I declare that this doctoral thesis has not been submitted as an exercise
for a degree at this or any other university and that it is entirely my own work. \vsp{10}

\noindent
\quad \hfill \qquad \\
Prague, September 2005 \hfill Daniel Kl\'{\i}r \qquad
\par
\pagebreak

%% file: acknowledgements.tex
\chapter*{Acknowledgements}
\thispagestyle{plain} \noindent First of all, I would like to make grateful acknowledgement to my supervisor
Prof.~RNDr. Pavel Kube\v{s}, CSc.~(Faculty of Electrical Engineering, Czech Technical University in Prague) for his
continued encouragement and kind help throughout my work on this thesis.

I also wish to express my gratitude to doc.~Ing.~Jozef Krav\'{a}rik, CSc.~(FEE, Czech Technical University in
Prague) and the research teams at the $Z$--pinch facility S--300 (Kurchatov Institute, Moscow) and plasma focus
device PF--1000 (Institute of Plasma Physics and Laser Microfusion, Warsaw) for their extensive assistance
during our experi\-ments.

I am also grateful to all who provided me with invaluable advice, inspiration, and encouragement. Prof. Dr. Hans
J. Kunze (Ruhr Universit\" at, Bochum) deserves special mention here.

I thank to the Ministry of Education of the Czech Republic (research programs INGO No.~1P2004LA235,
No.~1P05ME761 and LC528), the Grant Agency of the Czech Republic (grant No.~202--03--H162) and Czech Technical
University in Prague (research program No.~6840770016) for their financial support.

Finally, my sincere thanks go to my wife Michaela for her support.

%% file: abstract.tex
\chapter*{Abstract}
\thispagestyle{plain} \noindent This thesis presents the results of fibre $Z$--pinch experiments carried out on
the Z--150 device at the Czech Technical University in Prague. The generator that was used to drive the
experiments consisted of one capacitor of 3 $\muup$F capacitance. In the case of 20~kV charging voltage, the
current was peaking at 80 kA with a 850 ns quarter period. The $Z$--pinch was formed from carbon fibres of 15
$\muup$m diameter and 1 cm length. The discharge was observed by number of diagnostics which included a~voltage
probe, Rogowski coil, two filtered PIN diodes, a hard X--ray detector, gated XUV spectrograph, time-integrated
X--ray pinhole camera, gated VUV pinhole camera, and schlieren system. This comprehensive set of diagnostics
enabled us to describe the gross dynamics of the $Z$--pinch.

It was found out that after the breakdown a low density coronal plasma was formed while the fibre diameter
remained almost unchanged. This low density corona (ion density of about 10$^{16}$ cm$^{-3}$) was carrying
almost all the current of the order of 10 kA. When the current had built up, i.e.~after about 150 ns, the
implosion of the corona onto the central fibre occurred. The implosion velocity approached the value of
$2\times10^5$ m\,s$^{-1}$. When the imploded corona had reached the fibre, the dip in $\diff I/\diff t$, voltage
peak up to 10 kV, and XUV pulse of a $10\div30$~ns width were observed. XUV radiation was emitted from several
bright spots which corresponded to the interaction of  $m = 0$ instability necks with the dense core. The
electron temperature and density were approximately 80 eV and 10$^{19}$~cm$^{-3}$, respectively. Although the
presence of a fibre did not significantly suppress MHD instabilities, they were not disruptive.

After the fibre ablation, i.e.~after 500 ns, material evaporated from electrodes started to play a dominant
role. The observed plasma column seemed to be MHD unstable and when $m=0$ instabilities had developed, XUV, soft
X--ray and hard X--ray pulses were emitted from several hot spots, particularly near the anode. At that time the
voltage peak of up to 30 kV was detected.

In each phase of the discharge, the energetics of the $Z$--pinch and basic plasma parameters were estimated.

%% file: notation.tex
\chapter*{List of Symbols}
\fancyhead[R]{\slshape List of Symbols} \noindent
 Throughout the thesis we use the following symbols representing physical quantities and
constants. \vsp{10}

\begin{tabular}{l||l||l}
 Symbol &     Physical quantity or constant   &  Unit \\
 \hline \hline $a$ &diameter of stop in schlieren system&m\\
   $A$ & area& m$^{2}$\\
   $A$ & atomic mass number& \\
   $A$& constant in bremsstrahlung radiation $P_\mathrm{BS}$ in Eq.~\ref{Bremsstrahlung}&  kg$^{1/2}$\,m$^4$\,s$^{-2}$\\
   $A_\mathrm{u\rightarrow}$ & transition probability for emission from
     level ``u''& s$^{-1}$\\
   $A_\mathrm{{ul}}$ & transition probability for emission from
     level ``u'' to ``l'' & s$^{-1}$\\
   $b$ &radius of experimental chamber&m \\
   $\vec{B}$ & magnetic field & T \\
   $c$ & speed of light in vacuum& m\,s$^{-1}$\\
   $c_p$& isobaric specific heat& J\,K$^{-1}$ \\
$c_\mathrm{s}$ & ion sound speed& m\,s$^{-1}$\\
   $c_V$& isochoric specific heat& J\,K$^{-1}$ \\
   $C$& constant&  \\
   $C$& capacitance& F \\
   $C_0$& capacitance of capacitor bank& F \\
   $C_\mathrm{a}$& probability of collision of one ion A with electrons& s$^{-1}$\\
   $C_{\mathrm{u}\rightarrow}$& probability of collision depopulating upper level ``u''& s$^{-1}$\\
   $d$&groove spacing of grating&m\\
   $d_\mathrm{D}$& thickness of dead layer& m\\
   $d_\mathrm{F}$& thickness of filter& m\\
   $d_\mathrm{RES}$ & spatial resolution of pinhole camera&m\\
$d_\mathrm{S}$& thickness of sensitive layer& m\\
 \end{tabular}

\begin{tabular}{l||l||l}
 Symbol &     Physical quantity or constant   &  Unit \\
\hline \hline
$D$& distance between grating centre and detector plane& m\\
   $D_0$&fog density of film& \\
   $\mathcal{D}$&diameter of lens&m \\
$e$&electric charge&C\\
 $\vec{E}$&electric field&V\,m$^{-1}$\\
   $E_1$&energy of ground state&J\\
   $E_\mathrm{H}$&ionization energy of hydrogen&J\\
   $E_\mathrm{ion}$&ionization energy&J\\
   $E_\mathrm{l}$&energy of lower level&J\\
   $E_\mathrm{u}$&energy of upper level&J\\
  $f$&focal length of lens&m\\
   $f_\mathrm{e}(v)$ & electron velocity distribution& m$^{-1}$\,s\\
   $f_\mathrm{lu}$&absorption oscillator strength (l$\rightarrow$u)&\\
   $f_\mathrm{ul}$&emission oscillator strength (u$\rightarrow$l)&\\
   $F$ & calibration factor & V/A\,s\\
   $\vec{f}_p$ & pressure gradient, force density&N\,m$^{-3}$\\
   $\vec{f}_m$ & magnetic force density & N\,m$^{-3}$\\
   $g$ &gain coefficient& m$^{-1}$\\
   $g$ &gravitational acceleration& m\,s$^{-2}$\\
   $\bar{g}$ & average Gaunt--factor&\\
   $g_\mathrm{i}^z$&statistical weight of level ``i'' in ionic stage $z$&\\
   $g_\mathrm{l}$&statistical weight of lower level ``l''&\\
   $g_\mathrm{u}$&statistical weight of upper level ``u''&\\
   $G$& conductance & $\Omega^{-1}$\\
   $h$& Planckian constant& J\,K$^{-1}$\\
   $i_\mathrm{PIN}$& current generated in PIN diode& A \\
   $i_\mathrm{C}$& current induced in Rogowski coil& A \\
   $I$& relative intensity of spectral line& \\
   $I$& electric current& A\\
   $I(r)$ & current flowing inside radius $r$& A \\
   $I_0$ & total current in $Z$--pinch& A \\
   \end{tabular}

\begin{tabular}{l||l||l}
 Symbol &     Physical quantity or constant   &  Unit \\
\hline \hline
$I_\mathrm{PB}$& Pease--Braginskii current& A\\
 $\vec{j}$ & current density & A\,m$^{-2}$ \\
$k$ &axial wave number&m$^{-1}$\\
   $k$ & Boltzmann constant & J\,K$^{-1}$ \\
$l$ & length of pinch column or active media& m\\
 $l$ & distance between grating centre and spectral line & m\\
   $l$ & wave function of lower level ``l'' & \\
   $l_\Omega$&characteristic length along line of sight & m\\
   $L$& inductance& H \\
   $L_0$& inductance of a short circuit& H \\
   $L_\mathrm{C}$& inductance of Rogowski coil& H \\
   $L_\mathrm{P}$& plasma inductance& H \\
   $L_\mathrm{S}$& inductance of solenoid& H \\
   $L_\nu$& spectral radiance &W\,m$^{-2}$\,sr$^{-1}$\,Hz$^{-1}$\\
   $m$ &azimuthal wave number& rad$^{-1}$\\
   $m$ & diffraction order& \\
   $m_\mathrm{e}$ &electron mass& kg\\
   $M$ &magnification of image&\\
   $M_\mathrm{i}$ & ion mass & kg\\
   $n$, $n_\mathrm{i}$ & density of ions, plasma density & m$^{-3}$\\
   $n_\mathrm{A}$ & density of ions A& m$^{-3}$ \\
   $n_\mathrm{e}$ & electron density & m$^{-3}$ \\
   $n_\mathrm{TH}$ & thermal limit & \\
   $n_\mathrm{u}$ & number density of electrons in upper level & m$^{-3}$ \\
   $n^z$& density of ionization stage $z$ & m$^{-3}$ \\
   $N$ & refractive index &\\
   $N_\mathrm{a}, N_\mathrm{i}$ & atom line density, ion line density & m$^{-1}$ \\
$p$ & kinetic pressure & Pa \\
   $P_\mathrm{J}$& Ohmic power & W\\
   $P_\mathrm{R}$& radiated power& W\\
$Q$ &fusion gain& \\
   $Q^z(T_\mathrm{e})$& partition function of ionic stage $z$&\\

\end{tabular}

\begin{tabular}{l||l||l}
 Symbol &     Physical quantity or constant   &  Unit \\
\hline \hline
$r$ & radial coordinate & m \\
   $r_\mathrm{B}$&Bohr radius& m\\
$\vec{r}$ & displacement vector& m\\
   $\vec{r}^{\,0}$ & unit vector in radial direction & m \\
   $R$ & pinch radius & m \\
   $R$& resistance& $\Omega$ \\
   $R_\mathrm{a}$ & number of transitions ``a'' per unit time & \\
    &  and per unit volume&s$^{-1}$\,m$^{-3}$ \\
   $R^\mathrm{c}_{\mathrm{u}\rightarrow}$& rate of collisions depopulating upper level ``u'' & s$^{-1}$\,m$^{-3}$\\
   $R^\mathrm{c}_{\rightarrow \mathrm{u}}$ & rate of collisions filling upper level ``u''& s$^{-1}$\,m$^{-3}$\\
   $R^\mathrm{r}_{\rightarrow \mathrm{u}}$& rate of radiative processes filling upper level ``u''& s$^{-1}$\,m$^{-3}$\\
   $R_0$ & initial pinch radius & m \\
   $R_0$& initial resistance& $\Omega$ \\
   $R_\mathrm{C}$& resistance of Rogowski coil& $\Omega$ \\
   $R_\mathrm{G}$& radius of grating curvature& m \\
   $R_\mathrm{L}$& resistance of load& $\Omega$ \\
   $R_\mathrm{P}$& plasma resistance& $\Omega$ \\
   $R_\mathrm{R}$& radius of Rowland circle& m \\
   $\mathcal{R}$& resolving power of spectrograph &\\
   $\mathcal{R_\mathrm{max}}$& maximum attainable resolving power&\\
   ${\cal R}_i$ & responsivity& A\,W$^{-1}$ \\
   $s$ & distance between entrance slit and grating
   centre& m\\
   $S$ & ionization rate coefficient& m$^{3}$\,s$^{-1}$\\
$S$ & Lundquist number& \\
$S*$ & critical Lundquist number& \\
   $S_{\nu}(\vec{r})$ & source function& W\,m$^{-2}$\,sr$^{-1}$\,Hz$^{-1}$\\
   $\cal S_\lambda$ & film sensitivity & J\,cm$^{-2}$ \\
   $\mathbb{S}$ & total electron spin quantum number& \\
   $t$ & time & s\\
   $t_0$ & beginning of implosion & s\\
   $t_{10\%-90\%}$ & rise time & s\\
   \end{tabular}

\begin{tabular}{l||l||l}
 Symbol &     Physical quantity or constant   &  Unit \\
\hline \hline
$T$ & temperature& K, eV\\
   $T_\mathrm{B}$ & Bennett temperature & K, eV\\
   $T_{1/4}$ & quarter period of oscillation circuit & s\\
$T_\mathrm{e}$ & electron temperature & K, eV\\
  $T_\mathrm{i}$ & ion temperature & K, eV\\
  $u$ & wave function of upper level ``u''& \\
   $U$ &  voltage & V\\
   $U_0$ & charging voltage & V\\
   $U_{osc}$ & voltage at oscilloscope&V\\
   $v$ & velocity& m\,s$^{-1}$\\
   $v_\mathrm{A}$ & Alfv\'{e}n velocity& m\,s$^{-1}$\\
   $v_{\mathrm{De}}$ & electron drift velocity & m\,s$^{-1}$ \\
$v_{\rm{exp}}$ & expansion velocity&m\,s$^{-1}$\\
   $v_\mathrm{imp}$ & implosion velocity&m\,s$^{-1}$\\
   $v_{T\mathrm{i}}$ & thermal ion velocity & m\,s$^{-1}$ \\
   $V$ & volume & m$^{3}$ \\
   $w$&energy need for creation of one electron--hole pair in Si& J \\
$W$&width of grating in dispersion plane& m \\
$W_\mathrm{imp}$&energy delivered to the axis during the implosion& J \\
   $x$& distance between spectral line $\lambda_0$ & \\
      &and the zeroth order&m\\
   $x_{0}$& distance between the zeroth order& \\
       & and perpendicular from grating centre upon detector&m\\
   $x_{\lambda}$&distance between spectral line $\lambda$&\\
    & and perpendicular from grating centre upon detector&m\\
   $X$& collisional excitation/deexcitation rate coefficient&m$^{3}$\,s$^{-1}$\\
   $X_{\rightarrow\mathrm{u}}$& excitation rate coefficient of upper level ``u''&m$^{3}$\,s$^{-1}$\\
   $X_{\mathrm{l}\rightarrow\mathrm{u}}$& excitation rate coefficient of upper level from lower level&m$^{3}$\,s$^{-1}$\\
   $X_{\mathrm{u}\rightarrow}$& deexcitation rate coefficient of upper level ``u''&m$^{3}$\,s$^{-1}$\\
   $X_{\mathrm{u}\rightarrow \mathrm{l}}$& deexcitation rate coefficient of upper level to lower level&m$^{3}$\,s$^{-1}$\\
    $y$ & location of imaged point on entrance slit& m\\

\end{tabular}

\begin{tabular}{l||l||l}
 Symbol &     Physical quantity or constant   &  Unit \\
\hline \hline
  $z$ & axial coordinate & m \\
   $z$ & effective nuclear charge, ionic charge & \\
   $\overline{z}$ & average ion charge & \\
   $\vec{z}^{\,0}$ & unit vector in axil direction & m \\
   $Z$& atomic number&  \\
   $Z_0$& characteristic impedance& $\Omega$ \\
   $\alpha$&recombination coefficient &m$^{3}$\,s$^{-1}$\\
   $\alpha$& angle between grating plane and~spectrograph axis&m\\
   $\beta(r)$ & 2$\muup p(r)/B(r)$&  \\
   $\gamma$ & ratio of specific heats&\\
   $\gamma_{\lambda}$ & film contrast&\\
   $\Gamma$ & exponential factor& s$^{-1}$\\
   $\Gamma_z$ & flux density of ions of charge $z$& m$^{-2}$\,s$^{-1}$\\
   $\delta \lambda$ & spectral resolution & m, nm\\
   $\Delta E_\mathrm{ul}$ &energy difference between level ``u'' and ``l'' &J\\
   $\epsilon_\nu$& emissivity / volume emission coefficient& W\,m$^{-3}$\,sr$^{-1}$\,Hz$^{-1}$\\
   $\varepsilon_0$ &vacuum permitivity&F\,m$^{-1}$\\
   $\eta$ & viscosity& kg\,s$^{-1}$\,m$^{-1}$\\
   $\eta$ & quantum efficiency& \\
   $\kappa(\nu,\vec{r})$ & effective absorption coefficient & m$^{-1}$ \\
   $\lambda$ & period of perturbation & m$^{-1}$ \\
    $\lambda$ & wavelength & m, nm \\
   $\lambda_0$ & focused wavelength& m, nm \\
   $\lambda_0$ & wavelength of line centre& m, nm \\
   $\lambda_\mathrm{B}$ & blaze wavelength of grating & m, nm \\
   $\Lambda$ & $\ln \Lambda$ is Coulomb logarithm &  \\
   $\mu$ & permeability & H\,m$^{-1}$\\
   $\mu_\mathrm{D}$ & absorption coefficients of dead layer& m$^{-1}$\\
   $\mu_\mathrm{F}$ & absorption coefficients of filter& m$^{-1}$\\
   $\mu_\mathrm{S}$ & absorption coefficients of sensitive layer& m$^{-1}$\\
\end{tabular}

\begin{tabular}{l||l||l}
 Symbol &     Physical quantity or constant   &  Unit \\
\hline \hline
$\nu$ & photon frequency & Hz\\
   $\nu_0$ & photon frequency of line centre& Hz\\
   $\pi$ & Ludolphian number& \\
   $\rho$ & mass density& kg\,m$^{-3}$\\
   $\rho_\mathrm{i}$ & average ion Larmor radius& m\\
   $\overline{\rho}$ & average mass density& kg\,m$^{-3}$\\
 $\sigma$& specific conductivity & S\,m$^{-1}$ \\
   $ \sigma_{\rm{Spitzer}}$ &  Spitzer conductivity \citep{Spitzer62} & S\,m$^{-1}$\\
   $\sigma (v)$ & microscopic cross--section of process& m$^{-2}$\\
   $\sigma_0$&coefficient in Spitzer conductivity $\sigma_0(kT_\mathrm{e})^{3/2}/(\bar{z}\ln\Lambda)$& S\,m$^{-1}$\,J$^{-3/2}$\\
   $\tau$ & confinement time &s\\
   $\tau$ & time when instabilities first appears &s\\
   $\tau(\nu,\vec{r})$&optical depth&\\
   $\tau_\mathrm{A}$& Alfv\'{e}n transit time& s\\
   $\tau_\mathrm{i}$& ion--ion collision time& s\\
   $\tau_\mathrm{u}$ & life time of upper level ``u''& s\\
   $\phi_{0}$&grazing angle of incidence&m\\
   $\phi$&grazing angle of diffraction&m\\
   $\Phi$& incident power of radiation& W \\
    $\theta$ & azimuthal angle, coordinate & rad \\
   $\vec{\theta}^{\,0}$ & unit vector in azimuthal direction & m\\
   $\vartheta$ & deflection angle of laser beam & rad \\
   $\vec{\xi}$ & Lagrangian displacement& m\\
   $\omega$ & angular frequency & rad\,s$^{-1}$\\
 $\omega_\mathrm{ci}$ & ion cyclotron frequency &rad\,s$^{-1}$\\
   $\Omega$ & solid angle& srad\\
   $\varnothing$ & diameter of pinhole& m, $\muup$m\\
\end{tabular}

\pagebreak

%% file: introduction.tex
\chapter*{Introduction}

\addcontentsline{toc}{chapter}{Introduction}
\fancyhead[R]{\slshape Introduction}

\pagenumbering{arabic} \noindent The major concern of this thesis is a~$Z$--pinch which belongs to one of the
most fascinating objects in plasma physics due to its natural occurrence and variety of applications.
A~\emph{pinch} is defined as plasma configuration in which an electric current generates magnetic field that
tends to confine the plasma (cf.~\citealp{Spielman01}). The prefix ``$Z$'' was added in the 1950s to denote the
confinement driven by an axial (\emph{z}) current.

\section*{Scope of Thesis: Fibre $Z$--pinch Experiments in Vacuum}
\noindent First, it should be stressed that this thesis does not focus on $Z$--pinches in general but is limited
to the so--called \emph{fibre $Z$--pinch}. $Z$--pinches could be initiated by electrical power sources from
various discharge loads (gas, wire, low--density foam, cylindrical array of wires, cylindrical foil, etc.),
however, we are studying mainly the pinch formed from a~fibre with few micrometers in diameter. In our work we
distinguish between a~fibre and a wire. The expression ``fibre'' here means a wire with a very low conductivity,
such as cryogenic deuterium, deuterated polyethylene, carbon, etc. The reason for making such distinction is
that the dynamics of a~fibre $Z$--pinch substantially differs from the behaviour of a metal wire, such as
aluminium, copper, gold.

Furthermore, $Z$--pinch discharge can occur in vacuum, gas, water, etc. So far our research has been restricted
to discharges \emph{in vacuum}.

As far as the experimental study is concerned, the fibre $Z$--pinch was carried out on a slow current generator
$Z$--150 at the Czech Technical University in Prague. Nevertheless, we also performed experiments\footnote{To
draw a comparison between these two experiments, the results from S--300 device are included in
appendix~\ref{appendixS300}.} on more powerful S--300 generator at Kurchatov Institute, Moscow.

\section*{Aim of Thesis}
\noindent We aim at presenting the most important results of our $Z$--pinch research at the Department of
Physics, Faculty of Electrical Engineering, CTU in Prague. The primary objective of our research is the
\emph{description of our fibre $Z$--pinch dynamics} because to our knowledge no systematic effort has been made
to study a~$Z$--pinch initiated from a~carbon fibre in vacuum using a low voltage capacitive discharge. We
believe that our findings help us to get deeper insight into processes taking place in $Z$--pinches.

\section*{Layout of Thesis}
\noindent The thesis is organised as follows:

Chapter 1 presents a brief survey of the past and present $Z$--pinch research. The emphasis is put on
applications of a~$Z$--pinch in controlled thermonuclear fusion and as a powerful source of X--ray radiation.
The last part of this chapter is devoted to fibre $Z$--pinches which were treated in more detail.

Chapter 2 provides basic theoretical background for equilibrium and dynamic $Z$--pinches and thus creates the
ground to which following chapters inevitably refer.

Chapter 3 describes in detail the current generator and diagnostics used in our experiment at the Czech
Technical University.

Chapter 4 brings forward the most important results we have obtained. The results are shown chronologically,
i.e.~according to the phase 
in which they were recorded.
Experimental data interpretation is given after each phase of our $Z$--pinch.

Chapter 5 contains the overall discussion within the framework of other experiments too.

Finally, the main results are summarised, and conclusions together with research prospects are presented.

\section*{Author's Share in Experiment}
\noindent The experiment itself has been carried out mainly by doc.~Jozef Krav\'{a}rik, CSc. The author's place was
in data acquisition, data processing (XUV spectroscopy, X--ray optics, theory of detection) and elementary
interpretation of results ($Z$--pinch physics). Chapters 3\,--\,5 together with conclusions are examples of his
own work.

%% file: survey.tex
\newpage

\chapter{Brief Survey of $Z$--Pinch Research and Application}

\input{history} \input{present} \input{fibre}

%% file: history.tex
\fancyhead[R]{\slshape \rightmark}

\section[Historical Perspective of
$Z$--Pinches] {Historical Perspective of $Z$--Pinches} \noindent The following section was based on
\citet{Sethian97,Liberman99,Ryutov00,Haines00,Spielman01,Sweeney02}.

\subsubsection{Early beginnings of $Z$--pinches}
\begin{figure}[!h]
\begin{center}
\includegraphics*{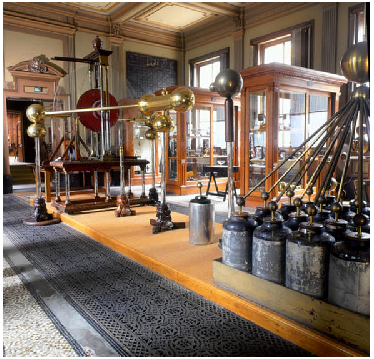} 
\end{center}
\caption[Van Marum's electrostatic generator in Teylers
Museum]{Van Marum's electrostatic generator in Teylers Museum \citep{marumwww}.} \label{Marum}
\end{figure}

\noindent The first experimental study of ``$Z$--pinch'' could be assigned to the Dutch chemist and physicist
Martinus van Marum. In the late of the 17th century, he designed and used a~large electrostatic generator (see
Fig.~\ref{Marum}) that was built by an English instrument--maker John Cuthbertson in Amsterdam.

The machine consisted of two parallel rotating glass discs (1.62 m in diameter) and four--man power. Its 100
capacitors (``Leyden Jars'') were 4 mm thick Bohemian glass bottles of about 35 cm diameter and 60 cm hight. The
capacitance of a~single jar was around 5 nF and hence the total capacitance was 0.5 $\muup$F. The permissible
voltage might have been over 60 kV, in which case the bank's energy was at least 1 kJ\footnote{With this
machine, van Marum was able to create 60 cm long sparks in the air. In this experiment, the voltage was over 330
kV and the energy was 30 kJ~\citep{Spielman01}.}. At the time when van Marum carried out experiments with
exploding wires, the expected discharge time was about 0.5 $\muup$s and a peak current of 60 kA.

The term ``pinch'' and also the first systematic research of $Z$--pinches, however, began in the
20$^{\mathrm{th}}$ century. In 1905, Pollock and Barraclough observed the distortion of a lighting conductor
caused by the $\vec{j} \times \vec{B}$ force. In 1934, W.\,H. Bennett derived the well--known \emph{equilibrium}
of magnetically self--focusing streams of charged particles with a temperature $T_\mathrm{B}$ (see subsection
\ref{odkazBennett} and \citealp{Bennett34}). About three years later, L. Tonks resumed the Bennett theory and
introduced the term ``pinch'' to describe the self--constricted plasma.

\subsubsection{$Z$--pinches for fusion}
\label{history} \noindent The growing interest in $Z$--pinches started in the 1950s in conjunction with
controlled thermonuclear fusion research. The D--T $Z$--pinch, in which an electric current both heats and
confines a plasma, was suggested to be a possible fusion system (see \citealp{Haines96}). This idea was
supported by a~high number of neutrons ($10^8$ per one pulse) observed in the first experiments with deuterium
gas (20 kV capacitor bank and 1.5 $\muup$s implosion time). Later, Anderson with his colleagues showed that
neutrons were not produced by a Maxwellian plasma. He proposed that the deuterons were accelerated by axial
electric fields created by the growth of $m=0$ instabilities~\citep{Anderson58}. This was consistent with
Kurchatov's explanation~\citep{Kurchatov57}, Kruskal's and Schwarzschild's theoretical work~\citep{Kruskal54},
and earlier experiments carried by \citealp{Carruthers57}.

Despite this fact, further experiments were carried out on the Columbus-II machine at Los Alamos and the
SuperFast Pinch (300~kA, 20 kA/ns) at Space Technology Laboratories. The idea was to use a fast driver to reach
fusion conditions before instabilities could grow. Unfortunately, this approach did not achieve the results
either and a plasma still exhibited MHD instabilities.

The problem of MHD stability was investigated to a~large extent also in the 1950s. The basic stability criteria
were established \citep{Kadomtsev63}\footnote{Especially the energy principle helped in quantifying plasma
instabilities.}. These derived criteria showed that it is impossible to eliminate instabilities in cylindrical
configurations. For instance, toroidal systems of magnetic confinement that were also free from electrode
phenomena seemed to be more stable. All these facts together with the conclusion that neutrons were not of
thermal origin led to the abandonment of \emph{linear} $Z$--pinch as a fusion power source. As a~result, more
complex schemes of magnetically confined plasma devices (such as the Tokamak) were suggested and researched in
an attempt to reduce MHD instabilities.

\subsubsection{$Z$--pinches as sources of radiation}
\noindent In the 1960s, $Z$--pinches were used as efficient UV, XUV and soft X--ray sources. The first
$Z$--pinch X--ray sources were exploding wires of $10\div100$ $\muup$m diameters~\citep{Chace}. The typical
parameter of discharges was the charging voltage of tens of kilovolts. The capacitor banks were discharged into
a load via spark gaps with microsecond discharge time. The $Z$--pinch plasma reached usually the temperature of
$10\div100$ eV. The peak electron density was up to 10$^{21}$ cm$^{-3}$. Because the absorption mean free path
was less than the diameter of a plasma column, the plasma radiation was close to the one of the blackbody.

Greater interest in $Z$--pinches was renewed in association with the development of nanosecond pulsed--power
technologies in the 1970s. These drivers were able to deliver MV voltages, MA currents and TW electrical powers
to a load. The new pulsed--power generators were used to drive single wires. The limitation of a~single wire was
its high initial impedance of $\approx1$ $\Omega$ that limited the conversion of the stored electrical energy to
radiation. It was found that the implosion of a plasma onto its axis is a more effective way to generate
radiation than resistive and compressional heating of an equilibrium plasma
column~\citep{Linhart61,Turchi73,Stallings76}. For that reason, a~single wire was replaced by e.g.~a cylindrical
array of wires, a thin cylindrical foil, an annular gas puff, a low--density foam cylinder, etc. In these
configurations, the stored electrical energy is converted into the kinetic energy of magnetically confined,
imploding plasma. At the stagnation, the kinetic energy is thermalised and the significant part of energy is
radiated in sub-keV and keV radiation. This way, $Z$--pinches have become powerful radiation sources with X--ray
powers at levels previously attainable only on the largest laser facility. Some current applications of
$Z$--pinches as radiation sources will be given in subsection~\ref{zareni}.

As regards the physics of a cylindrical load, it substantially differs from that of a~single wire or fibre. In
the case of a~single wire with a diameter of the order of ten micrometers, the time needed for establishing
radial equilibrium is usually much shorter than the duration of a current pulse. That is why the pinch is
assumed to be in the Bennett equilibrium. Ohmic and compressional heating are the most important. Such pinch is
classified as \emph{equilibrium}.

In the case of a cylindrical load with the initial diameter of a few centimetres, the implosion time is
comparable with the driving current pulse. When the implosion stops, the plasma starts to expand and never
reaches the Bennett--type equilibrium. The kinetic energy of implosion is the most dominant process of energy
deposition into a plasma. Such pinches are called \emph{dynamic}.

\subsubsection{Rebirth of $Z$--pinch fusion research}
\noindent The progress in pulsed-power technology led to new attempts in using $Z$--pinches in fusion research.
The ``classical'' effort pursued at Imperial College, Naval Research Laboratory, and Los Alamos National
Laboratory was to develop a stable linear $Z$--pinch.

One of the first experiments was carried out with \emph{compressional} $Z$--pinch at Imperial College, London in
the late 1970s. The $Z$--pinch was created in a quartz tube of 1 cm diameter containing a deuterium filling. The
discharge started on the insulating wall and the electric current (100 kA peak, 100 ns rise time) created the
imploding plasma shell which was preceded by a shock--wave. Though a~high number of neutrons was observed
(10$^{12}$ neutrons per shot), a low ion temperature indicated that neutrons were not produced by a Maxwellian
plasma. Similar results were obtained at a 3 MA current on the Angara~5-1 device in Troitsk, Soviet
Union~\citep{Batyunin90}.

Another attempt was not only to use a fast driver to reach fusion conditions before instabilities could grow but
also to employ the effect of radiative collapse (cf.~subsection~\ref{collapse} on page~\pageref{collapse}). In
order to achieve magnetic fusion, $Z$--pinches should operate at the Pease--Braginski current (1.4 MA for
hydrogen). Haines and Hammel derived the current waveform that is necessary to follow so as to keep the plasma
column at a constant diameter. The temperature of 10 keV requires a current channels of $\approx100$ $\muup$m
diameter~\citep{Hammel83}. In order to have $100$ $\muup$m current channel, it seemed convenient to start with a
small initial diameter of a plasma column. Therefore, the pinch was initiated by a focused laser or electron
beam that ionised a channel on the axis --- this is how a \emph{gas--embedded} pinch is formed.\pagebreak

One of experiments with gas--embedded pinches was pursued at Los Alamos (High Density $Z$--Pinch experiments).
These gas--embedded experiments demonstrated the possibility of producing $m=0$ stable
$Z$--pinches~\citep{Hammel83}. However, these pinches could not reach multi--hundred eV temperatures because
they rapidly accreted particles from the surrounding gas. In order to overcome this problem, the research group
in Naval Research Laboratory formed the pinch inside small quartz capillary of 100 $\muup$m inner diameter. A
capillary reduced the accretion and the occurrence of MHD instabilities but its wall also caused rapid cooling
of a plasma. It was a bad news for fusion purposes but on the other hand it seemed promisingly for a
recombination mechanism for X--ray lasers~\citep{Rocca88}.

Different suggestion how to overcome the accretion was to initiate $Z$--pinch from very small diameter fibres of
cryogenic solid deuterium in vacuum~\citep{Scudder85}. As it is the subject of our thesis, the research of
\emph{fibre $Z$--pinches} will be described in its own section~\ref{fibres}.

Nowadays, a simple linear $Z$--pinch is not considered to be a feasible source for fusion
energy~\citep{Sethian97}. Current applications of $Z$--pinches for fusion will be described in the following
section.

%% file: present.tex
\newpage

\section{Present State and Applications of $Z$--Pinches}
\noindent There are numerous applications of $Z$--pinches at present but among the most important ones belong
the pinches used (i) in controlled thermonuclear fusion research and (ii) as efficient sources of radiation.
Since the use of a~$Z$--pinch in fusion research is closely connected with its application as a radiation
source, we begin with the description of the latter one first.

\subsection{Source of radiation}
\label{zareni} \noindent Currently, the most important application of (dynamic) $Z$--pinches is their use as
X--ray sources. Attractive properties of $Z$--pinches are mainly the simplicity of design, high power density,
high X--ray energy output, and very high efficiency of transformation of stored electrical energy to radiation.

\subsubsection{Source of blackbody radiation}
\noindent Sandia's $Z$ machine is now the most powerful X--ray source that is capable to generate nearly 2 MJ of
thermal X--rays with a peak power of 280 TW in one pulse~\citep{Matzen97,Deeney98,Matzen99}. More than 15\% of
the stored electrical energy of 11.5 MJ is converted into X--rays. Such a~high peak power and radiated energy
were achieved as a~result of the implosion of very high--wire--number arrays \citep{Sanford96, Deeney97,
Spielman98} and nested wire--arrays~\citep{Deeney98}. As the implosion of high--$Z$ material happened in the
closed cavity, the radiation became close to the one of the blackbody.

The primary interest of the research is to use $Z$--pinch sources for inertial confinement fusion (see
paragraph~\ref{icf}). However, there are other reasons for studying these X--ray sources which are also unique
for other high--energy-density physics experiments:
\begin{itemize}
\item[--] the study of radiation hydrodynamics~\citep{Olson97}, radiation transport, opacity
measurements~\citep{Bailey02a},\vspace{-0.2cm}

\item[--] the study of material properties under extreme conditions~\citep{Olson97}, shock physics, measurement
of the equation of state under multi-mega-bar pressure,\vspace{-0.2cm}

\item[--] the laboratory study of weapons processes, survivability of weapons systems, etc.
\end{itemize}

\begin{landscape}
\begin{table}\centering
\begin{onehalfspacing}
\begin{tabular}{|l| l| c| c| c| c| c| c|}
\hline \hline Name & Location & Current & Rise time & Energy & Power & Impedance & Ref.\\
 &  & [MA] & [ns] & [kJ] & [TW] & [$\Omega$]&\\
\hline
Z & SNL, USA & 20 & 105  & 11400 & 50 & 0.12 & \citep{Spielman97}\\
\hline
Saturn & SNL, USA & 10.5  & 40  & 4000 & 32 & 0.11 & \citep{Spielman89}\\
\hline
Angara 5-1 & Triniti & 5  & 90  & 1400 & 9 & 0.25 & \citep{Grabovsky01} \\
& Troitsk, Russia &   &   &  &  & & \\
\hline
S--300 & Kurchatov Institute & 4  & 70  & 920  & 8 & 0.15 & \citep{Chernenko96}\\
 & Moscow, Russia &  &   &   &  &  & \\
\hline
Double & Physics International & 4  & 100  &   & 7 & 0.33 & \citep{Sincerny85}\\
Eagle & San Leandro, USA &   &   &   &  & & \\
\hline
Ambiorix & CEA   & 2.4  & 50  & 300  & 6 & 0.5 &\\
 & Bruyeres Le Chatel, France &   &   &   &  &  &\\
\hline
MIG & HCEI   & 1.9  & 80  &   &  & 0.65 &\citep{Sorokin02}\\
 & Tomsk, Russia &   &   &   &  &  &\\
 \hline
MAGPIE & Imperial College & 1.4 & 150  &  336 & 1 & 1.25& \citep{Mitchell96}\\
 & London, Great Britain &  &   &  &  &  & \\
\hline
NTF ``Zebra'' & University of Nevada& 1.2 & 100  & 200 & 1 & 1.9 & \citep{Nevada05}\\
 & Reno, USA&  &   &  &  &  &\\
\hline \hline
\end{tabular}
\end{onehalfspacing} \caption[Electrical characteristics of the largest sub-microsecond $Z$--pinches]{Electrical
characteristics of the largest sub-microsecond $Z$--pinches.} \label{Devices1}
\end{table}
\end{landscape}
 At this point it might be useful to give the list of the
largest pulsed power facilities used in $Z$--pinch mode (see Tab.~\ref{Devices1}). The list of devices may not
be complete since there are more pulsed power generators that can be used in $Z$--pinch mode. However, they are
mainly used as bremsstrahlung diodes or as accelerators of charged particles.

For next generation of $Z$--pinch facilities it is important to know scaling laws. Therefore, recent experiments
 on Sandia's $Z$ machine (and be it said that not only there) were focused on the scaling of X--ray emission with a current~\citep{Stygar04}.
They found that the total radiated X--ray energy varied as $I^{1.73\pm0.18}$ and the peak power varied as
$I^{1.24\pm0.18}$. Nevertheless, recent experiments close to $I^2$ law are more perspective for future
generators.

\subsubsection{Source of kiloelectronvolt radiation}
\noindent Besides being excellent Planckian--like X--ray sources, $Z$--pinches are also sources of non--LTE,
K--shell or L--shell radiation. Ne, Kr or Ar gas puffs and Al, Ni or Ti wire--arrays have proven to be powerful
sources of kiloelectronvolt emission (see a nice review written by \citealp{Pereira88}). As an outstanding
example we could mention Sandia's experiments with an argon double-shell $Z$--pinch. \citet{Sze01} reported
K--shell X--ray output of 274 kJ in a 15 TW peak power and 12 ns FWHM pulse.

\subsubsection{Long--implosion--time $Z$--pinches}
\noindent As was mentioned above, the interest in $Z$--pinches was renewed with the development of pulsed power
generators that are low impedance ($\lesssim 1 \Omega$) water--line systems producing mega--ampere currents with
100--ns rise time. These fast high--current generators are very large and require high voltages. In order to
reduce the complexity and cost of generators, lower voltage, long--implosion--time (0.1 $\muup$s $\div$ 2
$\muup$s) plasma radiation sources (PRS) have been studied. A comprehensive review was given
by~\citet{Deeney01}.

\begin{table}[!h]
\begin{onehalfspacing}
\centering
\begin{tabular}{|l| l| l| l| l|}
\hline \hline Name & Location & Current & Rise time & Energy\\
\hline
Atlas & LANL, USA & $40\div50$ MA & $5\div10$ $\muup$s & 36 MJ \\
\hline
Shiva & Air Force Research Laboratory, USA & $12\div30$ MA & $8\div10$ $\muup$s & 9.4 MJ \\
\hline
Pegasus& LANL, USA & $6\div12$ MA & $4\div8$ $\muup$s & 4.3 MJ \\
\hline
\hline
\end{tabular}
\end{onehalfspacing}
 \caption[Basic parameters of the largest microsecond $Z$--pinches]{Basic parameters of the largest microsecond $Z$--pinches.}
\label{Devices2}
\end{table}

Experiments on Shiva and Pegasus drivers\footnote{Their characteristics are displayed in
Tab.~\ref{Devices2}.}$^,$\footnote{The largest devices, such as Atlas, Shiva, etc.~were not primarily built  for
radiation experiments but for conducting hydrodynamic experiments supporting weapons--physics experiments.}
showed that $Z$--pinches with very long implosion times ($\approx1$ $\muup$s) were efficient, high--yield X--ray
sources but that the peak power was not so high as in the case of 100--ns--class implosions.

The significant increase in the radiated power was achieved when plasma flow swit\-ches and flux compression
were used to reduce the implosion times from microseconds to the hundreds of nanoseconds. Currently, experiments
with plasma radiating sources are being pursued on the devices listed below in Tab.~\ref{Devices3}.

\begin{table}[!h]
\begin{onehalfspacing}
\centering
\begin{tabular}{|l| l| l| l| l|}
\hline \hline Name & Location & Current & Rise time & Note\\
\hline
Decade Quad & Titan& 8 MA & 300 ns & Flux compr.\\
& San Leandro, USA &  &  & 15 MA  in constr.\\
\hline ECF2 &  Centre d'Etudes & 5 MA & 200 ns & Flux compr.\\
& Gramat, France & & 1200 ns & 10 MA in constr. \\
\hline
ACE--4& Maxwell Laboratories& 2.5 MA& 100 ns& POS\\
&  San Diego, USA & 4 MA &  2000 ns &  \\
\hline
GIT--12& HCEI & 2.5 MA & 70 ns &  POS\\
& Tomsk, Russia &  4.6 MA &  1600 ns &  \\
\hline \hline
\end{tabular}
\end{onehalfspacing}
 \caption[Basic parameters of some $Z$--pinches with ``pulse compression'']{Basic parameters of several $Z$--pinches with ``pulse compression''.}
\label{Devices3}
\end{table}

A longer ($\approx200$ ns) implosion time is researched not only as a low--cost plasma radiating source but also
as a possible approach\footnote{The rise time of a current is given by $\diff I/\diff t\thickapprox U/L_0$.
Therefore, if we want to achieve higher currents with an unchanged voltage $U$ and inductance $L_0$, it is
necessary to prolong rise time.} to the next generation of drivers with the current above 20 MA. It was one of
the reasons why a long--pulse mode is also investigated on Double Eagle, Saturn and $Z$
drivers~\citep{Deeney99b, Spielman98}. Although it is clear that long--implosion--time $Z$--pinches still need
to be improved, the energy output and radiated power are comparable to those achieved with less than 100--ns
implosion--times.

\subsubsection{X--ray laser}
\noindent During the 1980s, some researchers held belief that powerful $Z$--pinches (such as Double Eagle,
BlackJack 5, Gamble II, and Saturn) could be used for X--ray laser pumping. There were two main reasons
supporting this idea. Firstly, the $Z$--pinch geometry with plasma lengths in centimetres and diameters in
millimetres is naturally favourable for X--ray lasers~\citep{Elton90}. Secondly, $Z$--pinches can effectively
deliver high energy density (up to 10 MJ/cm$^{-3}$) into a plasma.

As far as a pinch plasma is concerned, the population inversion was demonstrated by resonant
photopumping~\citep{Porter92}, recombination pumping~\citep{Steden90, Kunze94, Shin96, Ellwi00, Ellwi01,
Metzner96, Koshelev94, Wagner96, Eberl98}, and pumping by charge--exchange recombination~\citep{Koshelev94,
Koshelev97}. However, the noticeable gain length product $gl$ and, by extension, the significant X--ray laser
output were achieved only by collisional excitation in a capillary discharge~\citep{Rocca94}.

\subsubsection{EUV lithography}
\noindent Last but not least application that will be mentioned is EUV lithography~\citep{Bailey82, McGeoch98}.
Several companies (such as XTREME technologies, CYMER Inc., Philips, Intel, Cutting Edge Optronics) make an
effort to develop the most suitable EUV source at 13.5 nm. Gas discharge--produced plasma and laser--produced
plasma belong among two prospective candidates. Both major technologies compete for high volume manufacturing
and have their pros and cons~\citep{Stamm04}.

\newpage
\input{fusion}

\newpage
\subsection{Other applications}
\noindent Except the above mentioned applications of a $Z$--pinch, there are yet other possibilities of using
this device. The following list briefly summarises few of them:

\begin{itemize}
\item[--] Lab--astrophysics~\citep{Ryutov02}. The generation of highly supersonic plasma jets with MACH number
of 50 can serve as an example \citep{Lebedev02}.\vspace{-0.2cm}
    \item[--] Controlled generation of high magnetic fields ($\approx 1000$ T) by magnetic field compression in an imploding cylindrical shell~\citep{Felber85}.\vspace{-0.2cm}
    \item[--] Generation of charged and neutral particles for industrial,
    scientific and medical applications.\vspace{-0.2cm}
    \item[--] Optical guidance of terrawatt laser pulses in a capillary discharge~\citep{Hosokai00}.\vspace{-0.2cm}
    \item[--] Spectroscopy of highly charged ions~\citep{Pereira88}.
\end{itemize}

%% file: fusion.tex
\subsection{Controlled thermonuclear fusion}
\noindent For more than 50 years, scientists have been trying to harness the fusion energy for peaceful
purposes. The fusion reaction, the fundamental energy--producing process in stars, occurs when two nuclei of
lighter elements are fused together to form a heavier element. Among possible reactions, the highest rate has
the fusion of deuterium and tritium to helium

\vspace{0.2cm}
\begin{tabular}{l l l l l l l}
\hspace{1.5cm} D\ &+& $$T\ \, &$\longrightarrow$& $_2^4$He\ (3.5\
MeV)\ &+& n\ (14.1\ MeV)
\end{tabular}
\vspace{0.2cm}

\noindent The free energy of 17.6 MeV is given to the $\alpha$--particle and the remaining neutron. This fast
escaping neutron can produce tritium from lithium (the material that is intended to form the blanket surrounding
a plasma)

\vspace{0.2cm}
\begin{tabular}{l l l l l l l}
\hspace{1.5cm} n\ &+& $_3^6$Li &$\longrightarrow$& $_2^4$He\ (2.1\
MeV)\ &+&T\ (2.7\ MeV)
\end{tabular}
\vspace{0.2cm}

As deuterium can be easily extracted from water and lithium can be
found in the earth's crust, this kind of energy could be cheap and
accessible worldwide.

Unfortunately, fusion reactions take place only when the fusing nuclei are brought close together. In order to
overcome the Coulomb barrier of charged particles, the nuclei need to have the sufficient kinetic energy --- the
case which can happen e.g.~during collisions of ions in a~high temperature plasma. However, to create a plasma
with very high temperature is not the only condition that we need in order to achieve fusion. It is also
necessary to sustain the plasma at certain density $n$ for sufficiently long time $\tau$ so as a~large number of
fusion reactions could occur. The British physicist J.\,D. Lawson derived a simple criterion under which a
plasma will generate $Q$--times more energy than is required to create and sustain it: the product of the
particle density $n$ and the confinement time $\tau$ should be higher than a certain value, in the case of a
D--T mixture
\begin{equation}
 n\tau\ >\ Q\cdot10^{14}\ \mathrm{cm}^{-3}\, \mathrm{s}
\end{equation}
Currently, the most promising concepts seem to be magnetic confinement fusion (MCF) with a long confinement time
($\tau\gtrsim$~0.1~s) and inertial confinement fusion (ICF) with a~high density plasma ($n\gtrsim 10^{26}$
cm$^{-3}$). The development of both concepts has passed the stage of \emph{Proof--of--Principle}. The basic
science is understood and the physics of the concept near fusion--relevant regimes is being explored
(i.e.~\emph{Performance Extension}). \pagebreak

 As far as $Z$--pinches are concerned, there are two
substantially different ways of their application in the fusion research. The historically first way is the
\emph{direct} heating and confining of a D--T plasma via $Z$--pinch effect. The second approach is the use of
a~$Z$--pinch as a driver for ICF. We shall describe the latter first.

\subsubsection{$Z$--pinch as driver for ICF}
\label{icf} \noindent $Z$--pinch, more precisely Sandia's $Z$ machine, is the world's most powerful laboratory
\mbox{X--ray} source. Such a~large volume, near--Planckian X--ray source provides a well--cha\-racteri\-sed
indirect driver for experiments relevant to the \emph{Inertial Confinement Fusion} programme\footnote{It is
possible to study ablator physics and radiation symmetrization experiments~\citep{Matzen99b}.}. The high X--ray
production efficiency ranked $Z$--pinch among three major drivers for \emph{Inertial Fusion Energy}
(IFE)\footnote{Inertial fusion science and applications have come to be referred to as ``inertial fusion
energy'' or IFE, whereas ``inertial confinement fusion'' or ICF denotes high energy density phenomena produced
by either multiple high--energy laser beams or energetic pulsed power systems~\citep{IFEaICF}.}. Besides
$Z$--pinches, lasers (KrF or DPSSL) and heavy ion drivers (induction linacs) are considered.

In one of the most perspective concepts, the $Z$--pinch radiation is produced during the collision of an
imploding liner (high--$Z$ wire--array) with an inner shell (a low density foam ``convertor''). The inward
travelling shock--wave generates radiation that is trapped by the outer imploding high--$Z$ shell. Therefore,
the shell acts as a hohlraum wall and becomes a~high temperature blackbody radiator. Later, the generated
blackbody radiation drives a spherically symmetric DT capsule in very similar way as in ICF indirectly driven by
lasers. Since the shell is moving, we call it ``dynamic hohlraum'' or ``flying radiation
case''~\citep{Smirnov91}.

\begin{figure}[h]
 \begin{minipage}[t]{0.5\linewidth}
   \centering
   \includegraphics{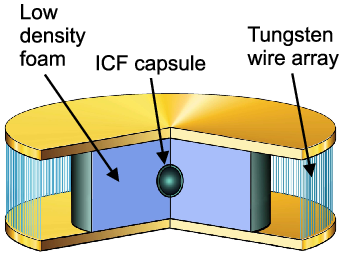}
   \caption[Dynamic hohlraum]{Dynamic hohlraum.}
   \label{icf0}
 \end{minipage}%
 \begin{minipage}[t]{0.5\linewidth}
   \centering
    \includegraphics{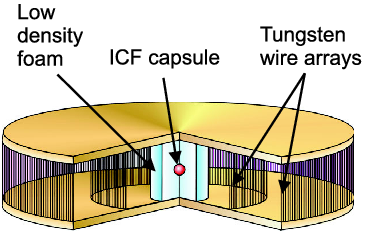}
   \caption[Dynamic hohlraum created from two nested wire--arrays]{Dynamic hohlraum created from two nested wire--arrays.}
    \label{icf3}
 \end{minipage}
  \end{figure}
\pagebreak

 Recently, dynamic hohlraum ICF experiments on $Z$ accelerator have been created from two annular
tungsten wire--arrays (a 240--wire 40--mm diameter outer array and a~120--wire 20--mm diameter inner array, see
Fig.~\ref{icf3}, \citealp{Bailey02b, Slutz03, Bailey04, Ruiz04}). This way, a deuterium--filled  capsule
absorbed $\approx24$ kJ of X--rays from a $\approx220$ eV dynamic hohlraum. The thermonuclear neutron yield from
the D--D reaction was up to $8\times10^{10}$ which is the highest D--D neutron yield ever reached with X--ray
drivers. Argon spectra confirmed a hot fuel with electron temperature and density of 1 keV and
$(1\div4)\times10^{23}$ cm$^{-3}$, respectively. The temperature of ions inferred from the neutron energy
distribution was $4.8\pm1.5$ keV. All these values show that the dynamic hohlraum can produce high temperature
radiation and deliver a~large amount of energy into the ICF capsule. However, in order to achieve higher
convergence ratio, it is necessary to improve radiation symmetry control. Therefore, other complementary
indirect drive concepts for high--yield IFE are being explored.

\begin{figure}[h]
   \centering
    \includegraphics{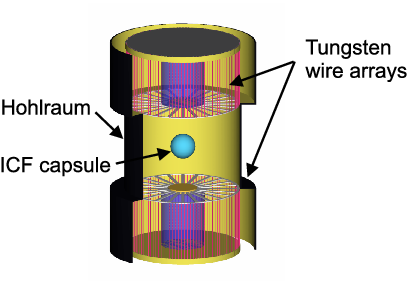}
   \caption[Static hohlraum with two $Z$--pinch radiation sources at the ends]{Static hohlraum with two $Z$--pinch radiation sources at the ends.}
    \label{icf2}
\end{figure}

The seriously investigated concept of a double--pinch hohlraum is
shown in Fig.~\ref{icf2}. This concept used two dynamic hohlraums
to drive a static--walled hohlraum. Experiments on $Z$ machine
demonstrated a 70 eV hohlraum temperature, 3\% drive symmetry and
$14\div20$ convergence ratio~\citep{Cuneo02}.

The results of the past ten years were so attractive that the $Z$--pinch IFE Road Map and the concept of power
plant were developed~\citep{Rochau03}. The design of high--yield dynamic hohlraum includes a~single 12 MJ, 54 MA
$Z$--pinch driver that produces a 350 eV radiation temperature, 2.4 MJ absorbed capsule energy and 550 MJ
thermonuclear yield~\citep{Lash00}. Yet, before the high--yield 60 MA driver ``ZX'' can be constructed, the 26
MA refurbished $Z$ machine ``ZR'' must first provide scaling information on a capsule~\citep{Weinbrecht03}.

The present effort is thus concentrated on demonstrating single--shot, high--yield fusion capsules. However, the
rep--rated ($\approx0.1$ Hz) $Z$--pinch is needed for IFE~\citep{Slutz03b}. As a possible rep--rated pulsed
power technology, linear transformer drivers (LTD) could be applied. The necessary separation between the fusion
target and the power plant should be provided by a recyclable transmission line \citep{Slutz03c}. It is planned
that 1 MA driver Z--PoP (Proof--of--Principle, 1 MV voltage, 100 ns rise time, 0.1 Hz repetition) will be
constructed in near future.

In spite of a rapid progress in $Z$--pinch driven ICF, many questions have remained to be answered. We mention
at least some of them here:
\begin{itemize}
\label{ICFquestions} \item[--] How to achieve spherically symmetric irradiation of the capsule when a~$Z$--pinch
is cylindrical by nature?\vspace{-0.2cm}
 \item[--] How to control the shape and intensity of the X--ray pulse?\vspace{-0.2cm}
  \item[--] How to prevent the early preheating of the capsule?\vspace{-0.2cm}
 \item[--] How to mitigate gross hydrodynamic instabilities, e.g.~Rayleigh--Taylor
 instability?
\end{itemize}

It is apparent that more experiments on these issues are still needed. Unfortunately, the number of shots on
powerful devices is limited. Smaller--scale experiments are therefore used to elucidate $Z$--pinch physics.

Several smaller--scale experiments focus on the effect of a~single wire because it helps to understand the
initiation stage of wire--array $Z$--pinches (e.g.~initial energy deposition, expansion rates, seeding of
instabilities that later influence Rayleigh--Taylor instability, etc.). These experiments are pursued e.g.~at
Sandia National Laboratories, Albuquerque~\citep{Sarkisov04}, Imperial College, London~\citep{Lebedev00,
Lebedev01}, University of Michigan, Ann Arbor~\citep{Johnston03}, Cornell University, Ithaca~\citep{Kalantar93,
Pikuz99a, Pikuz99b, Sinars00a, Sinars00b, Hammer01, Duselis04}, University of Nevada, Reno~\citep{Nevada05}.

The study of the initial stage of a discharge allowed to improve MHD capability including a ``cold--start''.
Recently, 3D resistive MHD simulation of wire--array $Z$--pinches have been developed (Eulerian code ``Gorgon'',
~\citealp{Chittenden04}, and Arbitrary--Lagrangian--Eulerian code ALEGRA--HEDP,~\citealp{Garasi04}).

\newpage

\subsubsection{Direct heating and confining of a fusion plasma via $Z$--pinch effect}
\label{direct} \noindent The other mentioned approach of the $Z$--pinch use in the fusion research is the direct
heating and confining of a fusion plasma via pinch effect. The accomplishment of this goal has been somewhat
problematic from the beginning of $Z$--pinch research (see subsection~\ref{history}). There were serious doubts
about issues of crucial importance, such as how to suppress magnetohydrodynamic instabilities and the
interaction of the plasma with a wall? Since the early attempts, a~large number of ideas have been proposed to
overcome these difficulties:
\begin{itemize}
    \item[--]A \emph{quasiequilibrium $Z$--pinch} with direct heating of a DT mixture
    (e.g.~cryogenic DT fibre, see subsection~\ref{fibres}).\vspace{-0.2cm}

    \item[--]\emph{Staged pinch}. In this case, the azimuthal magnetic field
    (generated by the current flowing in a DT fibre) is compressed by
    an imploding liner. When the imploding liner approaches the
    fibre, the inductance of the fibre decreases, the current in the
    fibre rapidly increases, and the fibre is heated and compressed~\citep{Rahman95}.\vspace{-0.2cm}

    \item[--]\emph{$Z$--pinch stabilised by a strong axial flow}. What happens here is that a plasma is accelerated
    between two coaxial electrodes and after that the $Z$--pinch effect
    occurs. This concept is very similar to a plasma focus.\vspace{-0.2cm}

    \item[--]\emph{Adiabatic compression} and heating of a DT plasma during
    cylindrically or spherically symmetric implosion of a liner~\citep{Degnan95}.\vspace{-0.2cm}

    \item[--]\emph{Magnetised Target Fusion}. This concept is based on adiabatic compression of a~\emph{magnetised} plasma.
    In comparison with the previous concept, the heat losses from
    the fusion plasma to walls are suppressed by weak
    magnetic field with $\beta\gtrsim1$. As examples of
    the MTF concept we can mention (i) the compression of field reverse
    configuration~\citep{Ryutov01} and (ii) the MAGO\footnote{MAGnitnoye Obzhatiye = Magnetised Target Fusion.} project of Los Alamos National Laboratory and
    All--Russian Scientific Research Institute of Experimental
    Physics~\citep{Lindemuth96}.\vspace{-0.2cm}

    \item[--]\emph{High yield systems}. They are characterised by the propagating of
    a nuclear burn wave that is ignited in some part of a cylindrical column,
          e.g.~in the neck of an $m=0$ instability~\citep{Linhart94, Chittenden05}.\vspace{-0.2cm}
\end{itemize}

All these concepts have not been explored to such an extent as ICF and MFC were. Thus one may ask what the
motive for such alternative approaches is.

One of the main reasons is that the region between conventional magnetic fusion ($n_\mathrm{e}\lesssim10^{14}$
cm$^{-3}$) and inertial fusion ($n_\mathrm{e}\gtrsim10^{26}$ cm$^{-3}$) offers vast possibilities for smaller
and cheaper fusion experiments~\citep{Siemon99}. We often speak about \emph{Magneto--Inertial Fusion} that
represents the all--inclusive set of inertially confined approaches to fusion that involve magnetic field in an
essential way. Magnetic field is then used to suppress the heat flow from the inertially confined plasma and
therefore requirements on fusion drivers could be reduced.

%% file: fibre.tex
\newpage
\section{Fibre $Z$--pinch}
\label{fibres}

\subsection{Cryogenic deuterium fibres}
\noindent The high density $Z$--pinches formed from dielectric fibres were investigated in the 1980s and 1990s
in connection with the research of controlled thermonuclear fusion and radiative collapse (see subsection
\ref{collapse}). The notion was to heat and ionise the fibre from frozen deuterium and to confine the high
density and high temperature plasma column within a~small diameter.

The first experiments were performed at Naval Research Laboratory~\citep{Sethian87} and Los Alamos National
Laboratory~\citep{Scudder85, Hammel87}. $Z$--pinch plasma in these experiments seemed to be stable for many
radial Alfv\'{e}n transit times (cf.~Eq.~\ref{timeA} on page \pageref{timeA}), whereas the neutron yield approached
$10^{10}$. This promising result led to the modification of the $Z$--pinch theory when the arbitrary current
distribution~\citep{Coppins89}, resistivity~\citep{Cochran90a,Culverwell90}, viscosity~\citep{Cochran90b}, the
Hall term and electron pressure gradient in Ohm's law~\citep{Coppins84}, pressure anisotropy~\citep{Coppins92},
etc. were included in fluid models. Further, finite ion Larmor radius effects were calculated via Vlasov--fluid
model~\citep{Scheffel97}. Nevertheless, none of these improvements could explain the experimentally observed
stability (see a nice review of the $Z$--pinch stability written by \citealp{Coppins97}). With the use of
a~``cold--start'' MHD simulation,
 \citet{Lindemuth89} calculated that the enhanced stability was due
to the presence of the solid core of a~fibre which lasted for a~very long time. The subsequent calculation
showed that rather than~enhancing the stability, the core partially masked the evidence of
instabilities~\citep{Lindemuth90, Sheehey92}. Similarly discouraging results were obtained from further and
better diagnosed experiments on new generators at NRL~\citep{Sethian90b} and Los Alamos~\citep{Scudder90} as
well as on KALIF at Kernforschungszentrum Karlsruhe~\citep{Kies91} and on MAGPIE at Imperial College in
London~\citep{Lebedev98}.

\subsection{Carbon and deuterated polyethylene fibres}
\noindent The fibres from carbon and deuterated polymer were also employed in the $Z$--pinch experiments because
their discharge behaviour was roughly the same as~the frozen deuterium ones~\citep{Lebedev98,Riley96}; at the
same time they were easily available and could be handled much easier. The main difference is an atomic number
($Z=1$ for hydrogen or deuterium, and $Z=6$ for carbon). Furthermore, carbon or deuterated polymer fibres
radiate more than pure deuterium fibres. The carbon and polymer fibres also require more energy for producing
fully stripped ions.

\citet{Stephanakis72} pursued the first experiments with deuterated polymer fibre on the Gamble II facility in
the 1970s. The number of detected neutrons exceeded the value of $10^{10}$. From that time on, several tens of
experiments with carbon and deuterated polymer fibres were performed but the neutron yield has never been
higher. The list of most important experiments is displayed in~Tab.~\ref{DevicesFibre1}, \ref{DevicesFibre2} and
\ref{DevicesFibre3}\footnote{The list does not include experiments with glass and polyamide fibres because only
fragmentary results were published. The only exception is Figura's experiment on the IMP generator at Imperial
College in London~\citep{Figura91}.}. The results of these experiments -- those that are relevant to our work --
are given in the discussion part in chapter~\ref{discussion}.

\subsection{Contemporary state of art}
\noindent To our knowledge, the last experiments with deuterium fibres were performed in the late 1990s and the
idea of nuclear fusion and radiative collapse in fibre $Z$--pinches was abandoned. During that period of time,
the theory and MHD modelling were improved to such a~degree that it was possible to explain gross dynamics of
the discharge as well as a~few ``fine'' phenomena (e.g.~\citealp{Chittenden97}) of fibre $Z$--pinches. However,
several key--points (such as the mechanism of neutron production, the generation of electron and ion beams,
etc.) remained still unresolved.

\subsection{Fibre $Z$--pinch drivers}
\noindent All the above cited experiments utilised high--voltage pulsed--power systems instead of a~simple
capacitor bank. According to~\citet{Mitchell96}, there were two principal reasons:
\begin{itemize}
\item[--]High voltage transients caused by changes of plasma inductance and resistance could damage low
impedance and low voltage capacitors. \vspace{-0.2cm}\item[--]The current rise must be fast enough to prevent
the early expansion of a~plasma column.
\end{itemize}

\subsection{Our experiment}
\noindent Our research is based on the results from a~carbon fibre $Z$--pinch driven by a~slow
(microsecond--long) capacitive discharge. The purpose of this small scale experiment is a~detailed description
of fibre $Z$--pinch dynamics\footnote{Our experiment also serves for preparing diagnostics which is used on more
powerful current generators (e.g.~S--300 generator at Kurchatov Institute in Moscow and plasma focus PF--1000 at
IPPLM institute in Warsaw.).}. On one hand, the use of a~capacitor as a~current generator has the aforementioned
disadvantages, but on the other hand, it could provide new experimental results which can~be compared with those
obtained at more powerful devices. We hope that this comparison will enable us better understanding of general
characteristics of fibre $Z$--pinches. Also, we would like to take the advantage of a~small--scale experiment
which could be easily modified and, in some cases, better diagnosed, while it remains interesting from the
physical point of view.

\begin{landscape}
\begin{table}
\centering
\begin{onehalfspacing}  \small{
\begin{tabular}{|l| l| c| c| c| c| c|c|c|c|l|}
\hline \hline Name & Location & Current & Rise time & Energy & Voltage &  Fibre & Diameter & Length & Neutrons & Ref./Note\\
 &  & [kA] & [ns] & [kJ] & [kV] & & [$\muup$m] & [cm] & & \\
\hline
Gamble II & NRL, USA & 1200  & 50  &   & 600 &  CH, CD & $10\div190$& &$10^{10}$&\citet{Stephanakis72}\\
\hline
Gamble II & NRL, USA & 700  & 50  &  50 & 1000 &  CH, CD & $8\div150$& &$10^{10}$&\citet{Young77}\\
\hline
Poseidon & NRL, USA & 640  & 130  &   & 500 &   D & $60\div125$& 5 &$8\times10^{9}$&\citet{Sethian87}\\
\hline
ZFX & NRL, USA & 450  & 640  & 60  & 210 &   D & 125 &5 & $4\times10^{8}$& \citet{Sethian89}\\
\hline
ZFX & NRL, USA & 1000  & 670  &   &  &   D & 125  & & $2\times10^{9}$& \citet{Sethian90a}\\
 &  &   &  &   &  &    CD & 40 & & & \\
 &  &   &   &   &  &   C & 7 & & & \\
\hline
ZFX & NRL, USA & 920  & 840  &   &  &   D & 125 & & & \citet{Sethian90b}\\
\hline
HDZP--I & LANL, USA & 350  & 150  &  12 & 600   & D & 40  & 5 & $10^{7}\div10^{8}$& \citet{Scudder85}\\
 \hline
HDZP--II & LANL, USA & 650  & 100  &  100 &    & D & 20  & & $3\times10^{9}$ & \citet{Shlachter90}\\
Zebra  &  &   &   &   &    &  &   & &  & \\
  \hline
HDZP--II  & LANL, USA & 800  &  &  &    & D &   & & $5\times10^{9}$ & \citet{Scudder90}\\
Zebra  &  &   &   &   &    &  &   & &  & \\
  \hline
HDZP--II & LANL, USA & 700  & 100  & 210  & 3200 &   CD & $15\div150$  & 5& & \citet{Scudder94}\\
Zebra  &  &   &   &   &    &  &   & &  & \\
\hline
HDZP--II & LANL, USA & 800  & 100  & 100  & 2200 &   CD & $3\div60$  & 5& $2\times10^{9}$ & \citet{Riley96}\\
Zebra  &  &   &   &   &    &  &   & &  & \\
\hline \hline
\end{tabular}}
\end{onehalfspacing} \caption[Electrical characteristics of C, CH, CD and D fibre $Z$--pinches I]{Electrical
characteristics of carbon, polymer, and deuterium fibre $Z$--pinches I.} \label{DevicesFibre1}
\end{table}
\end{landscape}

\begin{landscape}
\begin{table}
\centering
\begin{onehalfspacing}  \small{
\begin{tabular}{|l| l| c| c| c| c| c|c|c|c|l|}
\hline \hline Name & Location & Current & Rise time & Energy & Voltage &  Fibre & Diameter & Length & Neutrons & Ref./Note\\
 &  & [kA] & [ns] & [kJ] & [kV] & & [$\muup$m] & [cm] & & \\
\hline
IMP & Imperial College & 100  & 55  & 5  & 400 &   C & 7, 33  & 2& & \citet{Beg97}\\
 & London, GB &   &   &   &   &  & & & &\\
 \hline
IMP & Imperial College & 160  & 80  & 11  & 600 &   C & 360  & 1.4& & \citet{Davies97}\\
 & London, GB &   &   &   &    &  & & & &\\
 \hline
IMP & Imperial College & 200  & 60  & 20  & 800 &   C & 7  & 2& & \citet{Lorenz98, Lorenz99}\\
 & London, GB &   &   &   &    &  & & & &\\
\hline
MAGPIE & Imperial College & 1400  & 150  & 200  & 1900 & CD & $50\div200$  & 2.5& $10^{9}$& \citet{Mitchell98}\\
& London, GB &   &   &   &    &  & & & &\\
\hline
MAGPIE & Imperial College & 1000  & 150  & 120  & 1400 &   C & 7, 33 & 2.5& & \citet{Rossel98}\\
& London, GB &   &   &   &  &    & 300 & & &\\
\hline
MAGPIE & Imperial College &  1000 & 200 &  & &   D &  100 & 2.3 & $5\times10^{9}$& \citet{Haines97}\\
& London, GB &  &   &   &  &  C  & 7, 33 & & &\citet{Lebedev98}\\
\hline
Kalif & KFZ, Karlsruhe & 800  & 60  &   & 2000 &   D & $25\div50$  & $3\div5$& $10^{10}$ & \citet{Decker89}\\
& Germany &   &   &   &  &  &   & & &\\
\hline
Kalif & KFZ, Karlsruhe & 1000  & 90  &   & 2000 &   C, CD & $7\div300$ & $3\div6$& $10^{10}$ & \citet{Kies91}\\
& Germany &   &   &   &  & hair &  & & &\\
\hline
Kalif & KFZ, Karlsruhe & 1000  & 90  &   &  &   C & 7  & &  & \citet{Kies91}\\
& Germany &   &   &   &  &  &   & & &\\
\hline \hline
\end{tabular}}
\end{onehalfspacing} \caption[Electrical characteristics of C, CH, CD and D fibre $Z$--pinches II]{Electrical
characteristics of carbon, polymer, and deuterium fibre $Z$--pinches II.} \label{DevicesFibre2}
\end{table}
\end{landscape}

\begin{landscape}
\begin{table}
\centering
\begin{onehalfspacing} \small{
\begin{tabular}{|l| l| c| c| c| c| c|c|c|c|l|}
\hline \hline Name & Location & Current & Rise time & Energy & Voltage &  Fibre & Diameter & Length & Neutrons & Ref./Note\\
 &  & [kA] & [ns] & [kJ] & [kV] & & [$\muup$m] & [cm] & & \\
\hline
SPEED 2 & Duesseldorf Univ. & 3000  & 400  &   & 300 &  CD, D &   & & $10^{9}$ & \citet{Decker89}\\
& Germany &   &   &   &  &  &  & & &\\
\hline
S--300  & Kurchatov Inst. & 2000  & 190  &   &  & CD & 100  & 1 &  $10^{8}\div10^9$& see Appendix \ref{appendixS300} \\
& Moscow, Russia &   &   &   &  &    & & & &\\
\hline
LIMAY-I & Nagoya Univ.& 150  & 100  &  5 & 1000 &   C & 7  & 2 &  & \citet{Ishii89}\\
& Japan &   &   &   &    &  & & & &\\
\hline
LIMAY-I & Nagoya Univ.& 300  & 80  &  5 & 1000 &   C & 7  & 2 &  & \citet{Ishii93}\\
& Japan &   &   &   &    &  & & & &\\
\hline
& Inst. of Techn.& 1 & 100  &  & 30 &   C & 7  & 2 &  & \citet{Ishii89}\\
& Tokyo, Japan &   &   &   &    &  & & & &atmosph., water, vacuum\\
\hline
  & Bochum Univ. & 30  & 1500  &  1 & 20 &   C & 270  & 2.5 &  & \citet{Haun01}\\
& Germany &   &   &   &    &  & & & &\citet{Haun02}\\
&  &   &   &   &    &  & & & &atmosph. discharge\\
\hline
$Z$--150  & CTU, Prague & 100  & 800  &  1 & $20\div30$ &  C & 6, 15 & 1 &  & \citet{Klir04b}\\
& Czech Republic &   &      &  &  &  & 80 & &
&\citet{Klir02}\\
\hline \hline
\end{tabular}}
\end{onehalfspacing} \caption[Electrical characteristics of C, CH, CD and D fibre $Z$--pinches III]{Electrical
characteristics of carbon, polymer, and deuterium fibre $Z$--pinches III.} \label{DevicesFibre3}
\end{table}
\end{landscape}

%% file: theory.tex
\newpage

\chapter{Theoretical Background}
\noindent This chapter provides theoretical background upon which the thesis is based. Because the dominant
physical processes of equilibrium and dynamic pinches are substantially different, it might raise a question of
what kind of pinch we are actually describing here. The title clearly suggests that our work is devoted mainly
to a~fibre $Z$--pinch. One can therefore expect an equilibrium pinch. However, as it will be shown later, many
phenomena of dynamic pinches were also observed in our experiments. That's why this theoretical background
comprises both equilibrium pinches and dynamic pinches.

\input{bennett}
\input{dynamic}

%% file: bennett.tex
\section{Equilibrium $Z$--pinches}

\subsection[Bennett equilibrium]{Bennett equilibrium~\citep{Liberman99}}
\label{odkazBennett} \noindent
 Steady--state equilibrium of a~$Z$--pinch can
be derived from Maxwell's equations and the ideal magnetohydrodynamic equations in which we set the velocity and
all time derivatives to zero. We obtain so--called magnetostatic equations for perfectly conducting plasma
\begin{eqnarray}
  \label{rovnovaha1}
  \nabla p - \vec{j}\times \vec{B} &=&0\\
 \label{rovnovaha2}
 \nabla \times \vec{B}&=&\mu\vec{j} \\
 \label{rovnovaha3}
 \nabla \cdot \vec{B}&=&0
\end{eqnarray}
By excluding the current density $\vec{j}$ from equations \ref{rovnovaha1} and \ref{rovnovaha2}, we get
\begin{equation}
\nabla p + \nabla
\frac{\vec{B}^{2}}{2\mu}-\frac{1}{\mu}(\vec{B}\nabla)\vec{B}=0
\label{rovnovaha4}
\end{equation}
In cylindrical geometry, we can rewrite it as follows:
\begin{eqnarray}
(\vec{B}\nabla)\vec{B} &=& \vec{r}^{\,0}(B_{r}\frac{\partial
B_{r}}{\partial r}+\frac{B_{\theta}}{r}\frac{\partial
B_{r}}{\partial \theta}+B_{z}\frac{\partial B_{r}}{\partial
z}-\frac{1}{r}B_{\theta}B_{\theta})  \nonumber \\
 &+& \vec{\theta}^{\,0}(B_{r}\frac{\partial B_{\theta}}{\partial
r}+\frac{B_{\theta}}{r}\frac{\partial B_{\theta}}{\partial
\theta}+B_{z}\frac{\partial B_{\theta}}{\partial
z}+\frac{1}{r}B_{\theta}B_{r}) \nonumber \\
&+& \vec{z}^{\,0}(B_{r}\frac{\partial B_{z}}{\partial
r}+\frac{B_{\theta}}{r}\frac{\partial B_{z}}{\partial
\theta}+B_{z}\frac{\partial B_{z}}{\partial z})
\end{eqnarray}
The $Z$--pinch is a cylindrically symmetric plasma column and therefore all quantities are function only of the
radial coordinate $r$, i.e.~$\partial/\partial z=0$ and $\partial/\partial \theta=0$. Thus, the basic radial
pressure balance relation for a~$Z$--pinch is
\begin{eqnarray}
\label{rovnovaha5}
    \frac{\partial}{\partial r}\Big(p+\frac{B^{2}}{2\mu}-\frac{B_{r}^{2}}{2\mu}\Big)
    + \frac{1}{\mu} \frac{B^{2}_{\theta}}{r} & = & 0 \\
 \label{rovnovaha6}
     \frac{\partial}{\partial r}\Big(p+\frac{B_{\theta}^{2}}{2\mu}+\frac{B_{z}^{2}}{2\mu}\Big)
    + \frac{1}{\mu} \frac{B^{2}_{\theta}}{r} & = & 0
\end{eqnarray}
Next, we assume the current flowing along the $z$ axis. Since we do not consider external magnetic field
$B_{z}$, Amper's law \ref{rovnovaha2} gives only one non--zero component of magnetic field, namely azimuthal one
$B_{\theta}$. Then equation \ref{rovnovaha6} reduces to
\begin{eqnarray}
\label{rovnovaha7} \frac{\partial}{\partial
r}\Big(p+\frac{B_{\theta}^{2}}{2\mu}\Big)
 + \frac{1}{\mu} \frac{B^{2}_{\theta}}{r} & = & 0 \\
 \label{rovnovaha8}
\frac{\partial p}{\partial r} + \frac{1}{2 \mu
r^{2}}\frac{\partial}{\partial r}(B^{2}_{\theta}r^{2})&=&0
\end{eqnarray}
The first term in equation \ref{rovnovaha7} represents the gradient of the magnetic and kinetic pressure. The
second term is the tension force generated by the curvature of the azimuthal magnetic field. This ``curvature
force'' acts towards field's line centre of curvature. Since we have only one equation \ref{rovnovaha8} for two
independent functions $p(r)$ and $B_{\theta}(r)$, one of them can be taken arbitrary. From experimentalist's
point of view, it is convenient to choose arbitrarily the magnetic field $B_{\theta}(r)$ and hence the current
$I(r)$. If we introduce Amper's law
\begin{equation}
\label{pole}
B_{\theta}(r)=\frac{\mu I(r)}{2\pi r}\\
\end{equation}
into equation \ref{rovnovaha8}, we find that
\begin{equation}
\label{rovnovaha8b}
 \frac{\partial p}{\partial r} + \frac{\mu}{8
\pi^{2} r^{2}}\frac{\partial I^{2}(r)}{\partial r}=0
\end{equation}
By integrating this equation \emph{per--partes} and using the boundary condition at the pinch radius
\mbox{$p(R)=0$}, we obtain
\begin{equation}
\label{rovnovaha9} \mu I^{2}_{0}=16 \pi^{2} \int_{0}^{R}r p(r)
\diff r
\end{equation}
At this point, it is appropriate to define the ion line density as
\begin{equation}
N_\mathrm{i}=2 \pi \int_{0}^{R} n_\mathrm{i}(r)r\diff r
\end{equation}
and the Bennett temperature\footnote{It is the weighted temperature. The weighting goes over plasma density
\begin{displaymath}
\nonumber
 T_\mathrm{B}=\frac{2 \pi}{kN_\mathrm{i} (1+\overline{z})}\int_{0}^{R} p(r)r \diff r=
 \frac{1}{k (1+\overline{z})}\frac{\int_{0}^{R} p(r)r
\diff r}{\int_{0}^{R} n_\mathrm{i}(r)r \diff r}=\frac{\int_{0}^{R}
[T_\mathrm{i}(r)n_\mathrm{i}(r)+T_\mathrm{e}(r)n_\mathrm{e}(r)]r \diff r}{\int_{0}^{R} (1+\overline{z})
n_\mathrm{i}(r)r \diff r}
\end{displaymath}
 For an isothermal plasma, where
$T_\mathrm{e}=T_\mathrm{i}$, the weighted temperature equals the average temperature
$T_\mathrm{B}=\overline{T_\mathrm{e}}=\overline{T_\mathrm{i}}$.}, as
\begin{equation}
T_\mathrm{B}=\frac{2 \pi}{kN_\mathrm{i} (1+\overline{z})}\int_{0}^{R} p(r)r \diff r
\end{equation}
Substituting $N_\mathrm{i}$ and $T_\mathrm{B}$ in \ref{rovnovaha9}, we easily find the well--known Bennett
condition in the following form
\begin{equation}
\label{Bennet}
 \frac{\mu I^{2}_{0}}{8\pi}=kT_\mathrm{B}N_\mathrm{i}(1+\overline{z})
\end{equation}
This equation represents an integral equilibrium valid for any profiles of $p(r)$ and $I(r)$. For practical use,
the Bennett relation can be rewritten as
\begin{equation}
T_\mathrm{B} \doteq 3.12 \frac{I_{0}^2}{(1+\overline{z})N_\mathrm{i}} \qquad [\textrm{keV; MA, $10^{18}$ cm$
^{-1}$}]
\end{equation}
It follows that the lower line density $N_\mathrm{i}$ is, the higher plasma temperature $T_\mathrm{B}$ can be
achieved. However, if the line density $N_\mathrm{i}$ is too low, microinstabilities driven by the current can
be triggered (see paragraph~\ref{turbulences}).

\subsection{Pease--Braginskii current and radiative collapse}
\label{collapse} \noindent The balance between the kinetic and magnetic pressure is not the only condition that
is necessary for a steady--state equilibrium. The $Z$--pinch in radial equilibrium should satisfy also the
energy balance. If we neglect axial heat losses absorbed by the electrodes, the condition of the balance between
ohmic heating $P_\mathrm{J}$ and radiative cooling $P_\mathrm{R}$ must be fulfilled. To evaluate radiated power
and Joule heating power, we will consider optically thin plasma, uniform temperature, uniform current density,
and bremsstrahlung radiation only
\begin{eqnarray}
\label{Bremsstrahlung}
P_\mathrm{R}&=&P_\mathrm{BS}=A\overline{z}^{3}(kT_\mathrm{e})^{1/2}l\int_{0}^{R}\int_{0}^{2\pi}
  n_\mathrm{i}^{2}(r)r\diff\theta\diff r\\
P_\mathrm{J}&=&\frac{1}{G}I^{2}_{0}=\frac{l\overline{z}\ln\Lambda}{\pi R^{2} \sigma_{0}
(kT_\mathrm{e})^{3/2}}I^{2}_{0}
\end{eqnarray}
By using equation~\ref{Bennet} and a parabolic density profile $n_\mathrm{i}(r)=n_\mathrm{max}(1-r^2/R^2)$, we
obtain
\begin{equation}
\frac{P_\mathrm{J}}{P_\mathrm{BS}}=\frac{I^{2}_{PB}}{I^{2}_{0}}
\end{equation}
where
\begin{equation}
I_\mathrm{PB}=\frac{\pi}{\mu}\sqrt{\frac{48\ln \Lambda}{\sigma_{0}A}}\frac{1+\overline{z}}{\overline{z}}
\end{equation}
is known as the Pease--Braginskii current. Its value for a deuterium plasma\footnote{Coulomb logarithm equals
$\ln\Lambda\approx10$.} is $I_\mathrm{PB}\approx1.4$~MA.

The pinch radius does not change only if the current is equal to the Pease--Braginskii one. At currents above
this value, radiation losses exceed Joule heating, which results, according to the first law of thermodynamics,
in the negative value of the external work $p\diff V$. Hence, the pinch radius decreases. In other words, the
pinch undergoes a spontaneous contraction and the radiative collapse occurs. The collapse will continue until
the opacity and degeneracy of electron gas begin to play their role.

In this connection it is important to note that the value of the Pease--Braginskii current could be
substantially changed by anomalous resistivity, leakage currents, axial magnetic field, resonant line and
recombination radiation, opacity, axial and radial heat losses, particle losses, etc. From the experimental
point of view, we could mention the influence of the decreasing pinch radius and increasing plasma inductance
which can lead to the drop of the electric current delivered from an external generator. However, what we would
like to point out is that whatever the critical current is, the energy balance determines whether the pinch
expands or collapses.

\subsection{Ideal MHD stability of equilibrium $Z$--pinch}
\noindent The stability of $Z$--pinches is a primary object of interest because the plasma has to be stable for
a sufficiently long confinement time in order to burn thermonuclear reactions. During the early stage of fusion
research it was shown that linear $Z$--pinches are highly MHD unstable. During the last decades, linear and
non--linear analysis of magnetohydrodynamic and kinetic instabilities was undertaken and stability criteria were
deduced. With reference to this thesis, we will restrict ourselves to basic definitions and results of linear
theory.

\vspace{-0.3cm}
\subsubsection{Linear perturbation theory}
\noindent As far as the linear perturbation theory is concerned, the starting point is a set of linearised MHD
equations. We assume the plasma to be in the steady--state equilibrium. After that we perturb slightly this
equilibrium. We take the advantage of linear theory and use the Fourier analysis to express the Lagrangian
displacement
\begin{equation}
\vec{\xi}(\vec{r},t)=\vec{\xi}_{0}(\vec{r})\mathrm{e}^{i(m\theta +kz)+\Gamma t} \label{nomenclature}
\end{equation}
where Re($\Gamma$) is the growth rate, $m$ and $k$ are azimuthal and axial wave numbers.

The nomenclature of instabilities could be derived from the previous equation~\ref{nomenclature}: MHD
instabilities are often classified according to the azimuthal wave number $m$. The sausage (axisymmetric)
instability corresponds to $m=0$, whereas the kink (helical) instability to $m=1$.

Since all perturbed quantities are expressed in terms of displacement vector $\vec{\xi}(\vec{r},t)$, the set of
differential MHD equations can be converted to algebraical ones (with respect to time). Consequently, the
solution of algebraical equations could provide eigenvalues $\Gamma$ and thus the answer about stability. If
Re($\Gamma$) is negative, we speak about the exponentially stable system. Unfortunately, in the case of the
Bennett equilibrium and the ideal MHD model, the growth rate of an $m=0$ instability is positive of the order of
\begin{equation}
\mathrm{Re}(\Gamma)\sim \frac{v_\mathrm{A}}{R}
\end{equation}
where $R$ is the pinch radius and $v_A$ is the Alfv\'{e}n velocity
\begin{equation}
 v_\mathrm{A}\approx \frac{B_\mathrm{max}}{\sqrt{\mu \overline{\rho}}}
 \label{Alfven0}
\end{equation}
Substituting
\begin{eqnarray}
B_\mathrm{max}&=&\frac{\mu I_{0}}{2\pi R} \\
\overline{\rho}&=&\frac{N_\mathrm{i}M_\mathrm{i}}{\pi R^{2}} \label{prumernahustota}
\end{eqnarray}
into equation \ref{Alfven0}, we obtain the Alfv\'{e}n velocity
\begin{equation}
v_\mathrm{A}\approx\sqrt{\frac{\mu I^2_{0}}{4 \pi N_\mathrm{i} M_\mathrm{i}}} \label{Alfven}
\end{equation}

To compare theoretical predictions with experimental observations, it is convenient to define the number of
\emph{ideal MHD times} that have elapsed when instability first appeared at time $\tau$
\begin{equation}
\label{timeA} \int_{0}^{\tau} \Gamma(t)\diff t=\int_{0}^{\tau}
\frac{v_\mathrm{A}(t)}{R(t)}\diff
t=\sqrt{\frac{\mu}{4\pi^{2}}}\int_{0}^{\tau}
\frac{I_{0}(t)}{\sqrt{\overline{\rho}(t)}R^{2}(t)}\diff t
\end{equation}

\subsubsection{Physical mechanisms of the sausage an kink instabilities}
\begin{figure}[h!]
 \begin{minipage}[t]{0.5\linewidth}
   \centering
   \includegraphics{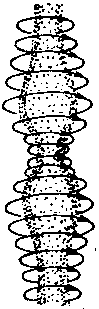}
   \caption[Sausage instability, $m=0$]{Sausage instability, $m=0$.}
   \label{1Nestabilita}
 \end{minipage}%
 \begin{minipage}[t]{0.5\linewidth}
   \centering
    \includegraphics{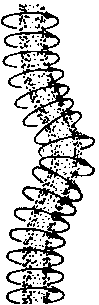}
   \caption[Kink instability, $m=1$]{Kink instability, $m=1$.}
    \label{2Nestabilita}
 \end{minipage}
  \end{figure}

\noindent A simple physical interpretation of the sausage and kink instability is illustrated in
Fig.~\ref{1Nestabilita} and~\ref{2Nestabilita}. If we imagine a plasma column in MHD equilibrium and with the
axisymmetric perturbation (cf.~Fig.~\ref{1Nestabilita},~\citealp{Horak81}) then the toroidal loops of magnetic
field will be shrunk in the neck. The magnetic field in the neck increases and cannot be balanced by the kinetic
pressure. The rise of the Lorentz force $\vec{j}\times\vec{B}$ causes the growth of a perturbation. The
characteristic velocity of the perturbation growth is that of a fast magnetosonic wave $v_\mathrm{A}$.
\pagebreak

 The bending of a plasma column (cf.~Fig.~\ref{2Nestabilita}) demonstrates the same effect. The field
strength is intensified along the inner face of the bend and reduced along the outer face. Thereby the
instability will be amplified.

\subsubsection{Stability criteria}
\noindent The necessary and sufficient condition for the $m=0$ and $m\geq1$ stability of $Z$--pinch was given by
\citet{Kadomtsev63}. He showed that the $Z$--pinch is stable against  $m=0$ instabilities if the pressure
profile fulfils the following inequality for all $r$
\begin{equation}
-\frac{\diff \ln p(r)}{\diff \ln r}=-\frac{r}{p(r)}\frac{\diff
p(r)}{\diff r}<\frac{4\gamma}{2+\gamma\beta(r)}
\end{equation}
where $\gamma=c_p/c_V$ is the ratio of specific heats and $\beta(r)=2\mu p(r)/B^2_\theta(r)$ is the ratio of the
kinetic and magnetic pressure. Since the pressure of a~fibre pinch\footnote{In the case of gas--embedded
$Z$--pinches, the Kadomtsev's criterion can be satisfied.} falls to zero at the vacuum--plasma boundary, the
pinch is always $m=0$ unstable.

As far as $m\geq 1$ modes are concerned, the current density
profile has to satisfy the condition~\citep{Liberman99}
\begin{equation}
\frac{r^2}{B_\theta}\frac{\diff}{\diff
r}\Bigg(\frac{B_\theta}{r}\Bigg)<\frac{1}{2}(m^2-4)
\end{equation}
The analysis of this expression shows that every $Z$--pinch with the finite (non--singular) current density at
the axis is $m=1$ unstable.

\subsubsection{Applicability of ideal MHD and non-ideal regimes of $Z$--pinches}
\noindent The ideal MHD stability theory predicts the growth of instabilities on the Alfv\'{e}n time scale. However,
ideal MHD neglects resistivity, viscosity, thermal conduction, the finite ion Larmor radius, and other phenomena
which may modify the instability growth time. In order to be valid, ideal MHD requires that the plasma fulfils
the following conditions:
\begin{itemize}
\item[--]First, the plasma must have a low resistivity which is usually quantified by the Lundquist number
$S=\mu\sigma v_\mathrm{A}R$ (the ratio between the Lorentz force and resistive magnetic diffusion force).
\citet{Culverwell90} found the critical Lundquist number $S^*$ below which the enhanced stability is predicted.
This critical Lundquist number depends on $kR$ ($k$ is the axial wave number and $R$ is the pinch radius) and,
typically, one can take this number of $10\div100$ (cf.~\citealp{Coppins97b}).\vspace{-0.2cm}
 \item[--]Second, the plasma must be highly collisional. In other words, the ion--ion
collision time should be lower than the characteristic MHD time, i.e.~$\Gamma\tau_\mathrm{i}<1$.
\vspace{-0.2cm}\item[--]Third, the ion Larmor radius must be small in comparison to the pinch radius,
i.e.~$\rho_\mathrm{i}/R\ll1$. Otherwise it is necessary to describe a collisional plasma by the Hall fluid model
where the Hall term and the electron pressure gradient are included in Ohm's law. When a plasma is
collisionless, the Vlasov fluid model is more appropriate. In the regime where the ion Larmor radius is large
$\rho_\mathrm{i}/R>0.1$ and ions are magnetised $\omega_\mathrm{ci}\tau_\mathrm{i}>1$, there is some theoretical
and experimental evidence of reduction of instability growth rates.
\end{itemize}
Assuming the pinch in the Bennett equilibrium, these three dimensionless parameters ($S$,
$\Gamma\tau_\mathrm{i}$, $\rho_\mathrm{i}/R$) are functions of only two parameters, namely the ion line density
$N_\mathrm{i}$ and the product of fourth power of the current and the pinch radius $I^4R$. If we plot $I^4R$
against $N_i$, we obtain the diagram of the stability regimes~\citep{Haines91}. In this diagram, the ideal MHD
region occupies a fairly small space of the experimentally relevant regions.

\subsubsection{Microinstabilities and turbulences}
\label{turbulences} \noindent Microinstabilities\footnote{Lower hybrid and ion acoustic instabilities are
usually most relevant in high--density $Z$--pinches~\citep{Ryutov00}.} are excited in a plasma when the drift
velocity of electrons $v_\mathrm{De}=j_z/en_{e}$ exceeds the critical velocity which is typically of the order
of the ion sound speed $c_{s}=\sqrt{k(T_\mathrm{i}+\bar{z}T_\mathrm{e})/M_\mathrm{i}}$. As a~result, turbulences
of electrostatic and electromagnetic field cause drifting electrons to be scattered. Since electrons are now
scattered both from individual ions and from turbulent electromagnetic field, the anomalous contribution to
resistivity has to be considered. This anomalous resistivity can occur not only in the neck of an $m=0$
instability, but also in a low density plasma boundary. In this region, the drift velocity $v_\mathrm{De}$ can
grow to a~higher value because the electron density approaches zero $n_\mathrm{e}\rightarrow0$ and the current
density $j_z$ is finite.

%% file: dynamic.tex
\newpage

\section{Dynamic $Z$--pinches}
\noindent Most of the $Z$--pinches, including our fibre $Z$--pinch, are in radial equilibrium for a very short
period, if at all. Therefore, the fully consistent approach requires to consider $Z$--pinch dynamics. The fluid
dynamics can be described by the Euler equation of motion
\begin{equation}
\rho \frac{\diff^2 \vec{r}}{\diff t^2}=\vec{f}_p+\vec{f}_m=-\nabla
p+\vec{j}\times \vec{B}
\end{equation}
where the pressure gradient $\vec{f}_p=-\nabla p$ and the magnetic force density $\vec{f}_m=\vec{j}\times
\vec{B}$ are included, whereas the momentum\footnote{This term cannot be neglected in the case of snowplough
implosion where the mass is swept up by the collapsing shell.} of accreting material $-\rho \dot{r}^2$, viscous
force $\eta\Delta \vec{v}$ and other terms are neglected. To express the radial component of the previous
equation, we can use the same derivation as we did for equation \ref{rovnovaha8b}
\begin{equation}
\rho(r) \frac{\diff^2 r}{\diff t^2}=-\frac{\partial p(r)}{\partial
r} - \frac{\mu}{8 \pi^{2} r^{2}}\frac{\partial I^{2}(r)}{\partial
r}
\end{equation}

\subsection{Magnetic compression}
\noindent One of the usual cases of a non--equilibrium state is the magnetic compression when the pressure
gradient is small compared with the Lorentz force. It means
\begin{equation}
\rho (r) \frac{\diff^2 r}{\diff t^2}=-\frac{\mu}{8 \pi^{2}
r^{2}}\frac{\partial I^{2}(r)}{\partial r}
\end{equation}
Next, we assume the uniform current density and constant current\footnote{The implosion time is often much
shorter than the current pulse width.}
\begin{equation}
I\Big(r(t)\Big)=I_0\frac{r(t)^2}{R(t)^2}
\end{equation}
and therefore
\begin{equation}
\rho(r) \frac{\diff^2 r(t)}{\diff t^2}=-\frac{\mu I_0^{2}r(t)}{2 \pi^{2}R^4(t)} \label{difr}
\end{equation}

\noindent It is evident that one has to know the density profile in order to solve this equation. To simplify
the problem and to find an approximative solution, we can replace some quantities in Eq.~\ref{difr}
\begin{eqnarray}
\rho(r)&\longrightarrow&\bar{\rho}\\
r&\longrightarrow&R(t)
\end{eqnarray}
Then we get
\begin{equation}
\bar{\rho} \frac{\diff^2 R(t)}{\diff t^2}\approx-\frac{\mu
I_0^{2}}{2 \pi^{2}R^3(t)}
\end{equation}
By using equation~\ref{prumernahustota} and~\ref{Alfven}, we can derive the equation
\begin{equation}
 R\frac{\diff^2 R}{\diff t^2}\approx-\frac{\mu I_0^{2}}{2
\pi N_\mathrm{i} M_\mathrm{i}}=-2v^2_\mathrm{A}
\end{equation}
This equation can be easily integrated by multiplication with the velocity $\diff R(t)/\diff t$, and by using
the initial conditions $R|_{t=0}=R_0$ and $\dot{R}|_{t=0}=0$
\begin{eqnarray}
 R\dot{R}\ddot{R}&\approx&-2v^2_\mathrm{A}\dot{R}\\
  \frac{1}{2}\frac{\diff}{\diff t}\Bigg(\frac{\diff R(t)}{\diff t} \Bigg)^2&\approx&-2v^2_\mathrm{A}\frac{\diff}{\diff t}\ln R(t)\\
\Bigg(\frac{\diff R(t)}{\diff
t}\Bigg)^2&\approx&4v^2_\mathrm{A}\ln\frac{R_0}{R(t)}\\
\Bigg|\frac{\diff R(t)}{\diff t}\Bigg|&\approx&2v_\mathrm{A}\sqrt{\ln\frac{R_0}{R(t)}}
\end{eqnarray}
It follows that the implosion velocity is increasing with the decreasing pinch radius $R(t)$. Because the
dependence is logarithmic and in addition to that in the square root, the implosion velocity $v_\mathrm{imp}$ is
of the order of the Alfv\'{e}n velocity
\begin{equation}
\label{alfven}
 v_\mathrm{imp}\approx2v_\mathrm{A}=\sqrt{\frac{\mu I^2_{0}}{\pi N_\mathrm{i} M_\mathrm{i}}}
\end{equation}

\subsection{Stability of dynamic pinches}
\noindent The dynamic and equilibrium pinches differ in the extent to which instabilities matter in each case.
The important difference arises from the fact that implosion/explosion takes a finite time, whereas a
steady--state plasma is supposed to last for ever. For instance, instabilities with growth rates $\Gamma \ll
v_A/R$ are much more important in equilibrium pinches than in dynamic ones~\citep{Ryutov00}. Regarding the
Rayleigh--Taylor instability, it plays a crucial role in all dynamic pinches, whereas in equilibrium pinches it
does not develop.

\subsubsection{Rayleigh--Taylor instability}
\noindent The RT instability is universal and occurs in all accelerating/decelerating systems where a~high
density fluid is ``supported" by a lower density one. The growth rate of perturbations for a liquid supported by
gravity is~\citep{Liberman99}
\begin{equation}
\label{growthRT}
 \Gamma=\sqrt{gk}
\end{equation}
where $k=2\pi/\lambda$ is the wave number of instability and $g$ is the (gravitational) acceleration. In dynamic
pinches, the gravity is represented by the magnetic field during the implosion and by the kinetic pressure
during the explosion. It follows from equation~\ref{growthRT} that the growth rate $\Gamma$ is unlimited with
increasing wave numbers $k$. However, for high wave numbers the dissipative effects (such as finite viscosity,
thermal conductivity and resistivity) have to be considered. As a~result, the maximum growth rate is finite.

\label{snowplow} Finally, we would like to mention the stabilising effect of the snowplough implosion. Although
the mass accretion during the implosion could be destabilising, the reduction of the acceleration and the effect
of the foregoing shock wave finally lead to the effective mitigation of RT instabilities.



%% file: apparatus.tex
\newpage

\chapter{Apparatus and Diagnostics}

\input{z150}

%% file: z150.tex
\begin{figure}[!h]
\centerline{\includegraphics{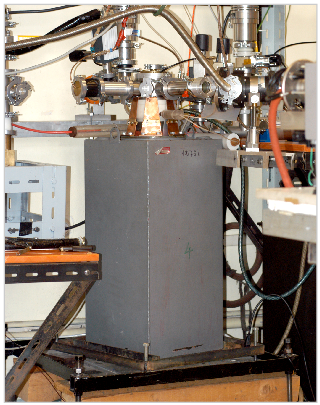}} \caption[$Z$--pinch device $Z$--150, Czech Technical
University, Prague]{$Z$--pinch device $Z$--150, Czech Technical University in Prague.} \label{aparatura}
\end{figure}

\noindent This chapter deals with the experimental set--up of a $Z$--150 $Z$--pinch\footnote{The name $Z$--150
was derived from the expected current of 150 kA when 30 kV voltage is applied.} (see Fig.~\ref{aparatura}),
which is located at the Czech Technical University in Prague, Faculty of Electrical Engineering. This small
device was employed for our fibre $Z$--pinch experiments, the results of which are presented in this thesis.
During the past four years, we carried out more than 500 shots in a few series under different experimental
conditions. The following sections describe the current generator, experimental chamber, $Z$--pinch load, and
diagnostics. First of all, however, we introduce the block diagram and electrical scheme of the $Z$--150.

\section{$Z$--pinch Device $Z$--150}
\begin{figure}[!h]
\centerline{\includegraphics{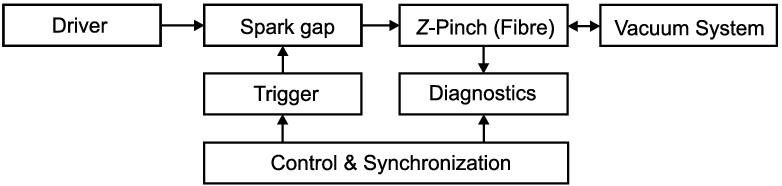}} \caption[Block diagram of the $Z$--150 device]{Block
diagram of the $Z$--150 device.} \label{blockscheme}
\end{figure}

\begin{figure}[!h]
\centerline{\includegraphics[width=141mm]{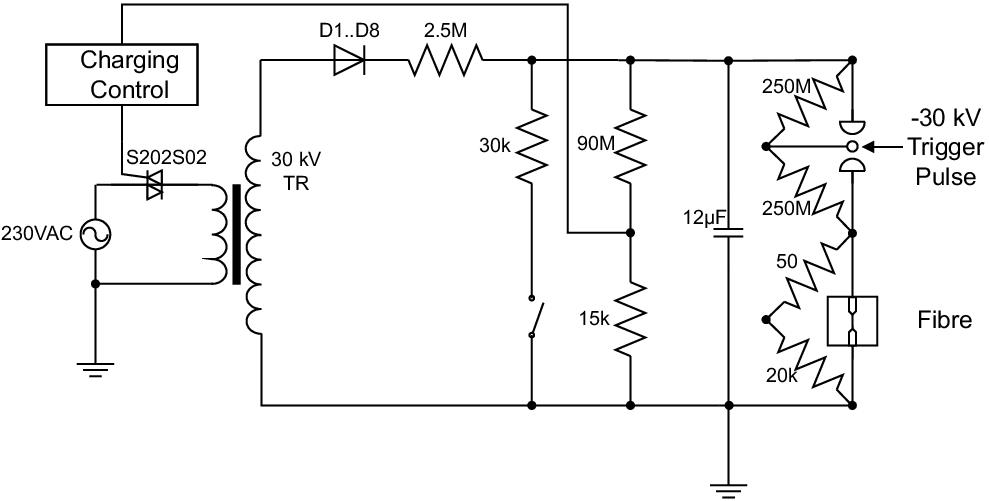}} \caption[Electric scheme of the $Z$--150
device]{Electric scheme of the $Z$--150 device.} \label{aparaturaschema}
\end{figure}

\noindent The block diagram in Fig.~\ref{blockscheme} shows that the $Z$--150 device can be divided into the
subsystems presented in Tab.~\ref{Subsystems}. The electric scheme of the $Z$--150 is displayed in
Fig.~\ref{aparaturaschema}.
\begin{table}[h!]
\centering
\begin{onehalfspacing}
\begin{tabular}{|l| l|}
\hline \hline
     Driver &capacitor, high voltage transformer, rectifier, control unit\\
          \hline
     Spark gap &operates in nitrogen\\
     \hline
     Trigger & thyratron based, voltage $2\times15$ kV \\
     \hline
     Fibre $Z$--pinch & experimental chamber, electrode system, diagnostic
     ports\\
     \hline
     Vacuum system & rotary pump, turbo pumps, pressure gauges, valves\\
     \hline
     Diagnostics &Rogowski coil, voltage monitor, filtered PIN diodes, \\
     &scintillator--photomultiplier hard X--ray detector, \\
     &XUV spectrometer, X--ray and VUV pinhole cameras, \\
     &laser diagnostics, digital storage oscilloscopes \\
     \hline
     Control \&&master pulse generator, delay units\\
       synchronization&\\
      \hline \hline
\end{tabular}
\end{onehalfspacing}
 \caption[Sub-systems of the $Z$--150 device]{Sub-systems of the $Z$--150 device.}
\label{Subsystems}
\end{table}

\newpage
\section{Current Generator}
\label{generator} 
\begin{figure}[!h]
\centerline{\includegraphics[width=49mm]{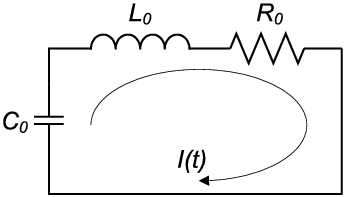}} \caption[``Short" circuit]{``Short" circuit.}
\label{RLCshort}
\end{figure}

\begin{figure}[!h]
\centerline{\includegraphics{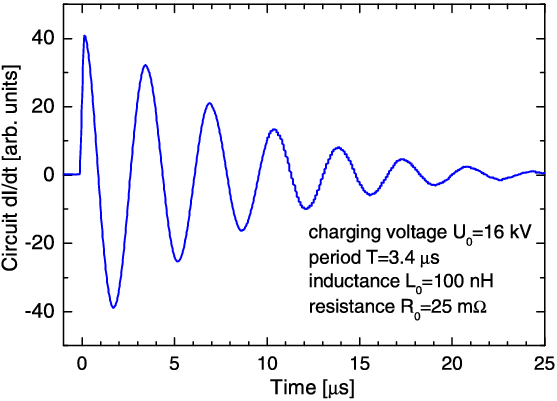}} \caption[Damped oscillation of a~``short" circuit]{Damped
oscillation of a~``short" circuit.} \label{short}
\end{figure}

\noindent The generator used to drive recent experiments\footnote{The first experiments were
performed with four capacitors of 12 $\muup$F capacitance. After multiple damages of an insulator we decided to
use only one capacitor. Most of the results were very similar.} consisted of one capacitor with the capacitance
of $C_0=3~\muup$F. The basic parameters of the generator were estimated from the damped oscillation of
a~``short" circuit, i.e.~with short--circuited electrodes (cf.~Fig.~\ref{RLCshort} and~\ref{short}).

Since it was a simple $R_0-L_0-C_0$ circuit in which the capacitor $C_0$ was discharged to a~$L_0-R_0$ load, we
could expect the time dependence of $\diff I/\diff t$ to be
\begin{equation}
\frac{\diff I}{\diff t}\propto
\mathrm{e}^{-\frac{R_0}{2L_0}t}\cos\omega t
\end{equation}
where the angular frequency $\omega$ is equal to
\begin{equation}
\omega=\sqrt{\frac{1}{L_0C_0}-\frac{R_0^2}{4L_0^2}}\approx\sqrt{\frac{1}{L_0C_0}}
\end{equation}
For the measured period of $T=2\pi\sqrt{L_0C_0}\approx3.4$~$\muup$s and damping
$\mathrm{e}^{-(R_0/2L_0)T/2}\approx0.8$ (cf.~Fig.~\ref{short}), we calculated the inductance of
$L_0\approx100$~nH and the resistance of $R_0\approx25$~m$\Omega$.
\begin{figure}[!h]
\centerline{\includegraphics[width=100mm]{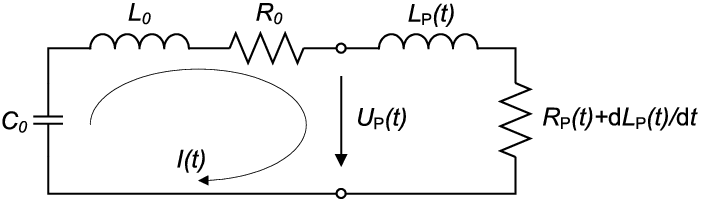}} \caption[Equivalent electric circuit of the discharge
with a~fibre]{Equivalent electric circuit of the discharge with a~fibre.} \label{RLC}
\end{figure}

When a~15~$\muup$m carbon fibre was used as a load (see Fig.~\ref{RLC}), the plasma resistance $R_\mathrm{P}$
and inductance $L_\mathrm{P}$ deeply affected the current waveform, especially $\diff I/\diff t$. In the case of
a 20~kV charging voltage, the current was peaking at $\approx 80$~kA with a rise time of $650 \div 800$~ns. More
details about our calibration of an electric current will be given in the subsection \ref{Icalibration}.

The basic parameters of the $Z$--150 device are summarised in
Tab.~\ref{Parameters1}.
\begin{table}[!h]
\centering
\begin{onehalfspacing}
\begin{tabular}{|l| l|}
\hline \hline
Capacitance &$C_0=3$ $\muup$F \\
\hline
Inductance &$L_0\approx100$~nH \\
\hline
Characteristic impedance &$Z_0=\sqrt{L_0/C_0}\approx 180$~m$\Omega$ \\
\hline
Resistance &$R_0\approx25$~m$\Omega$ \\
\hline
Charging voltage &$U_0=16\div30$ kV\\
\hline
Peak current in short--circuit&$I_{\mathrm{max}}\approx90$~kA for $U_0=18$ kV\\
\hline
Quarter period for short--circuit&$T_{1/4}\approx 850$~ns \\
\hline
Peak current for 15~$\muup$m carbon fibre&$I_{\mathrm{max}}\approx80$~kA for $U_0=20$ kV\\
\hline
Rise time for 15~$\muup$m carbon fibre&$t_{10\%-90\%}=650\div 800$ ns \\
\hline \hline
\end{tabular}
\end{onehalfspacing}
 \caption[Basic parameters of the $Z$--150 device]{Basic parameters of the $Z$--150 device.}
\label{Parameters1}
\end{table}

\newpage
\section{Experimental Chamber and Vacuum System}
\begin{figure}[h!]
   \centering
   \includegraphics{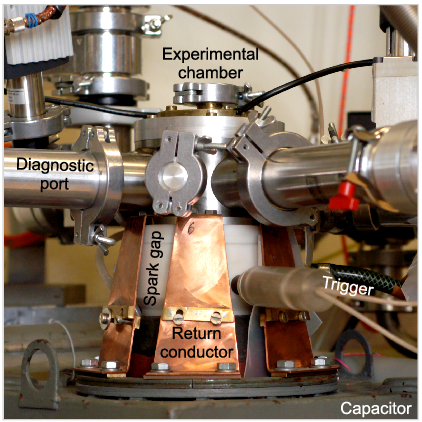}
    \caption[Side-on view of the experimental chamber and spark gap]{Side-on view of the experi\-ment\-al chamber and spark gap.}
   \label{chamber2}
 \end{figure}

\begin{figure}[h!]
\centering
   \includegraphics{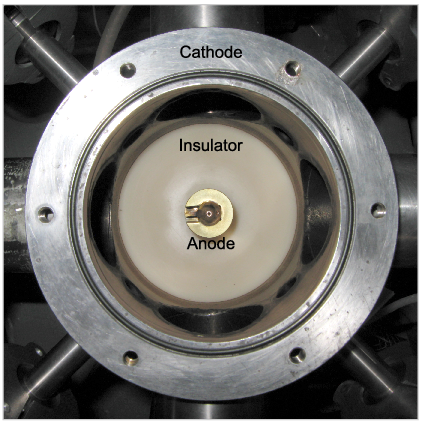}
    \caption[End-on view of the experimental chamber]{End-on view of the experimental chamber (without the upper cover).}
 \label{chamber}
 \end{figure}

\noindent The experimental chamber of a 6.5 cm height and 8 cm diameter was made of stainless steel (see
Fig.~\ref{chamber2} and \ref{chamber}). Its diameter was small in order to keep an inductance low. Altogether
eight diagnostic ports were available. The size of four diagnostic ports was matched with the vacuum fitting
NW40 and the size of the other four ports fitted with NW16.

\begin{figure}[h!]
   \centering
   \includegraphics{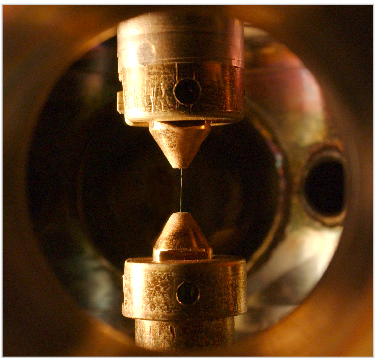} \caption[Brass conical electrodes with a~carbon fibre]{Brass
    conical electrodes with a~carbon fibre (15 $\muup$m diameter and 9 mm length).} \label{electrodes}
  \end{figure}

\newpage
The electrode system and fixing of carbon fibres varied during our experiment. As we wanted to distinguish
general features of the experiment from specific ones, we performed a~few experimental series with electrodes
from different materials (copper, steel, brass, and bronze) and of different shapes (flat, conical,
``bisectional"). The example of brass conical electrodes can be seen in Fig.~\ref{electrodes}.

\begin{figure}[h!]
\centering
   \includegraphics{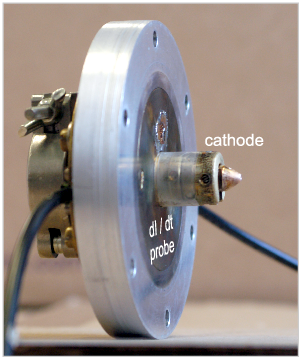}
    \caption[Upper cover of the experimental chamber]{Upper cover of the experimental chamber.}
 \label{chamber3}
 \end{figure}

Carbon fibres were mounted between the live anode and the cathode that were isolated from each other by an
alkaline polyamide spacer. The $\diff I/\diff t$ probe was placed near the cathode, in the upper cover of the
experimental chamber (see Fig.~\ref{chamber3}).

The chamber was evacuated to a pressure of $10^{-2}$ Pa. The vacuum was reached with the aid of three Leybold
Turbovac50 turbo pumps  
 which were backed by one rotary pump.

\newpage
\section{$Z$--pinch Load}
\noindent Our $Z$--pinch was initiated from \emph{carbon fibres} of various lengths and diameters. The carbon
fibres of 15~$\muup$m diameter and $7\div9$ mm length were most frequently used but also fibres with 6 and 80
$\muup$m diameter have been tried\footnote{We preferred 15~$\muup$m fibres for three reasons: (i) they could be
entirely ablated, (ii) the peak electron temperature exceeded 60~eV, and (iii) the discharge reproducibility was
the best among tested fibres.}. Some of important properties of carbon are presented in Tab.~\ref{properties1}
and \ref{properties2}.
\begin{table}[!h]
\centering
\begin{onehalfspacing}
\begin{tabular}{|l| l|}
\hline \hline
Density & $1.9\div2.3$ g\,cm$^{-3}$ \\
\hline
Concentration &$1.1\times10^{29}$ atoms\,m$^{-3}$\\
\hline
Specific resistance & 60 $\muup\Omega $\,m \\
\hline
Melting point&3825 \degree C (sublimate)\\
\hline
Specific heat&709 J\,kg$^{-1}$\,K$^{-1}$\\
\hline \hline
\end{tabular}
\end{onehalfspacing}
 \caption[Significant physical properties of carbon at 20\degree C and 100 kPa]{Significant physical properties of carbon at 20\degree C and a pressure of
100~kPa~\citep{Lide99}.} \label{properties1}
\end{table}

It is well known that $Z$--pinch dynamics is substantially determined by initial physical properties of
a~$Z$--pinch load. The characteristic initial physical property of carbon fibres is their high resistivity. The
measured resistance of a~15~$\muup$m carbon fibre was approximately 7~k$\Omega\,$cm$^{-1}$. As far as the
inductance of a 1 cm long fibre is concerned, the calculated value of~10 nH is small compared with the short
circuit inductance of 100~nH.
\begin{table}[h!]
\centering
\begin{minipage}[b]{0.5\linewidth}
\begin{onehalfspacing}
   \begin{tabular}{|l| l|}
\hline \hline
Atomization energy& 7.4 eV \\
\hline
1$^{\mathrm{st}}$ ionization energy& 11.3 eV\\
\hline
2$^{\mathrm{nd}}$ ionization energy& 24.4 eV\\
\hline
3$^{\mathrm{rd}}$ ionization energy& 47.9 eV\\
\hline
4$^{\mathrm{th}}$ ionization energy& 64.5 eV\\
\hline
5$^{\mathrm{th}}$ ionization energy& 392.1 eV\\
\hline
6$^{\mathrm{th}}$ ionization energy& 490.0 eV\\
\hline \hline
\end{tabular}
\par\vspace{0pt}
\end{onehalfspacing}
 \end{minipage}%
 \begin{minipage}[b]{0.5\linewidth}
  \begin{onehalfspacing}
  \begin{tabular}{|l| l|}
\hline \hline
K--${\alpha_1}$ transition energy& 277.0 eV\\
\hline
K--edge energy& 284.2 eV\\
\hline
He--$\alpha$ transition energy& 304.3 eV ($^3$P$_1$)\\
\hline
He--$\alpha$ transition energy&  307.8 eV ($^1$P$_1$)\\
\hline
Ly--$\alpha$ transition energy& 367.5 eV\\
 \hline \hline
 \end{tabular}
\par\vspace{0pt}
\end{onehalfspacing}
 \end{minipage}
 \caption[Significant energies of carbon atoms]{Significant energies of a~carbon atom~\citep{Carbonwww, Attwood99}.} \label{properties2}
\end{table}

\section{Diagnostics}
\begin{figure}[h!]
\centering
\includegraphics{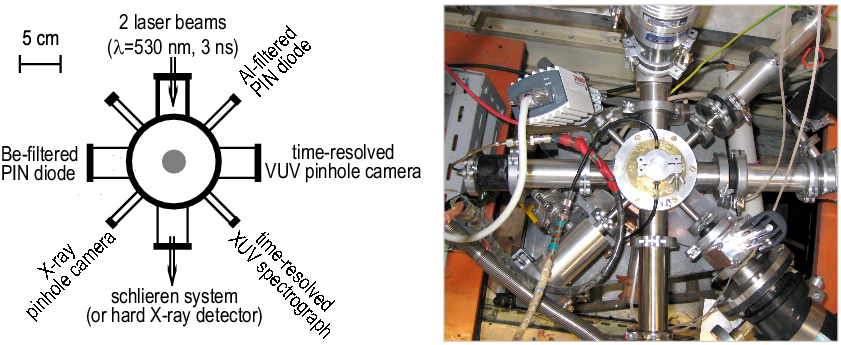}
\caption[Schematic diagram and picture of diagnostic set--up, end--on view]{Schematic diagram and picture of
diagnostic set--up, end--on view.} \label{diagnosticsschema}
\end{figure}

\noindent This subsection brings forward the description of various diagnostic tools that were employed to
observe plasma dynamics. The diagnostic set--up that we have used most recently is displayed in
Fig.~\ref{diagnosticsschema}.

The figure shows that the radiation emitted from the plasma was recorded with two filtered PIN diodes, a hard
X--ray detector (PMT and scintillator BCF-12), a gated XUV spectrograph, a time--integrated X--ray pinhole
camera, and a gated VUV pinhole camera. The laser probing enabled the visualization of the electron density
gradient by the schlieren method. The electrical characteristics of the $Z$--pinch discharge were monitored by a
voltage probe and Rogowski coil.

\subsection{Voltage probe}
\begin{table}[h]
\centering
\begin{onehalfspacing}
\begin{tabular}{|l|l|}
\hline \hline
  Attenuation & $1000\times$ ($\pm 5\%$) \\
  \hline
  Bandwidth & 75 MHz \\
  \hline
  Rise time & $\leq4.5$ ns \\
  \hline
  Delay time & approx.~10 ns \\
      \hline
  Loading  & 100 M$\Omega$, 3 pF  \\
  \hline
  Max. voltage & 20 kV DC, 40 kV pulse\\
  \hline
  Compensation & $12\div60$ pF\\
  \hline
  \hline
\end{tabular}
\end{onehalfspacing}
 \caption[Basic parameters of the P6015 Tektronix HV probe]{Basic parameters of the P6015 Tektronix HV probe~\citep{Tektronix}.}
\label{voltage}
\end{table}
\noindent The P6015 Tektronix high--voltage probe (attenuation 1000$\times$, loading 100 M$\Omega$ and 3 pF) was
used to monitor the voltage between the electrodes. This probe was developed for measurement of pulses up to 40
kV. The 75 MHz bandwidth and 4 ns rise--time enabled to capture fast changes. The most important characteristics
are displayed in Tab.~\ref{voltage}.

\subsection{Rogowski coil}
\noindent The Rogowski coil -- in other words a toroidal coil with multiple turns, was used to measure a pulsed
current in the discharge, or to be more precise the circuit $\diff I/\diff t$. The current $I$ flowing through
the upper (grounded) electrode also flew through the Rogowski coil. The alternating current $\diff I/\diff t$
and hence alternating magnetic flux then induced the voltage $L_\mathrm{C}\diff i_\mathrm{C}/\diff t$ inside the
current probe.
\begin{figure}[h!]
\centering
\includegraphics[width=95mm]{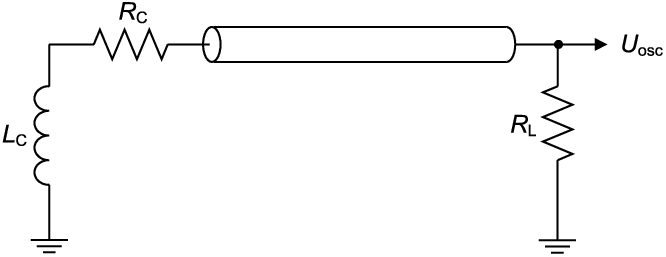}
 \caption[Equivalent circuit of $\diff I/\diff t$ probe]{Equivalent circuit of $\diff I/\diff t$ probe.}
\label{Rogowski}
\end{figure}

As the coil was connected via 50~$\Omega$ coaxial cable to the resistor $R_\mathrm{L}=50$~$\Omega$, we can write
the simplified circuit equation as
\begin{eqnarray}
\frac{\diff I}{\diff t}\propto L_\mathrm{C}\frac{\diff
i_\mathrm{C}}{\diff t}+(R_\mathrm{L}+R_\mathrm{C})i_\mathrm{C}
\end{eqnarray}
where $R_\mathrm{C}$  and $L_\mathrm{C}$ are the resistance and inductance of the Rogowski coil, respectively.
In our case, the inductance of the Rogowski coil $L_\mathrm{C}$ was about 400~nH. Hence, the term
$L_\mathrm{C}\diff i_\mathrm{C}/\diff t$ was negligible compared with $R_\mathrm{L}i_\mathrm{C}$ for frequencies
smaller than $R_\mathrm{L}/(2\pi L_\mathrm{C})\approx 20$~MHz. Under this frequency condition and for
$R_\mathrm{C}\ll R_\mathrm{L}$, the output voltage at the oscilloscope was proportional to the derivative of the
current
\begin{equation}
U_\mathrm{osc}\propto R_\mathrm{L}i_\mathrm{C}\propto\frac{R_\mathrm{L}}{R_\mathrm{L}+R_\mathrm{C}}\frac{\diff
I}{\diff t}\propto\frac{\diff I}{\diff t}
 \label{Uosc}
\end{equation}
\pagebreak

 During the last experiments, the Rogowski coil was replaced by the solenoid with the inductance of
$L_\mathrm{S}=30$~nH. It means that we measured $\diff I/\diff t$ again, but for frequencies up to
$R_\mathrm{L}/(2\pi L_\mathrm{S})\approx 250$~MHz.

\subsubsection{Calibration}
\label{Icalibration} \noindent The Rogowski coil was calibrated by discharging a capacitor into a short circuit
(cf. Fig.~\ref{RLCshort} and subsection~\ref{generator}). In this simple case, the absolute value of a current
could be calculated and thus the current probe could be calibrated against this known value. The current in a
short circuit was evaluated as follows:

The equation describing a damped oscillation is
\begin{equation}
U_0-\frac{\int_0^t I(t') \diff t'}{C_0}=R_0 I(t)+ L_0\frac{\diff
I(t)}{\diff t}
\end{equation}
Its solution for an under--damped circuit with
\begin{equation}
\frac{R_0}{2L_0}<\frac{1}{\sqrt{L_0C_0}}
\end{equation}
equals to
\begin{equation}
\label{calibration2} I=\frac{U_0}{\omega L_0}\mathrm{e}^{-(R_0/2L_0)t}\sin\omega t\;\;,\;\;
\omega=\sqrt{\frac{1}{L_0C_0}-\frac{R_0^2}{4L_0^2}}
\end{equation}
Since the charging voltage $U_0$ and capacitance $C_0$ were known, and since the parameters $\omega$, $L_0$, and
$R_0$ could be evaluated from the damped sinusoid (see Fig.~\ref{short}), the current in a short circuit was
simply obtained\footnote{For instance, in the case of 18 kV charging voltage, we calculated the peak current of
90 kA.}.

However, our Rogowski coil did not measure $I$ but $\diff I/\diff
t$
\begin{eqnarray}
\frac{\diff I}{\diff t}=F U_\mathrm{osc}
 \label{Uosc2}
\end{eqnarray}
Therefore, the calibration factor $F$ was gained after the numerical integration of the voltage measured by
oscilloscope
\begin{equation}
I(t)\approx F\int_0^t U_\mathrm{osc} \diff t
\end{equation}
Using this method we determined the calibration factor $F\approx2.8\times10^8$~A/V\,s. This value was applied
for the estimation of the current in discharges initiated from a~fibre.

\subsection{Filtered PIN diodes}
\begin{figure}[h!]
\centering
\includegraphics[width=108mm]{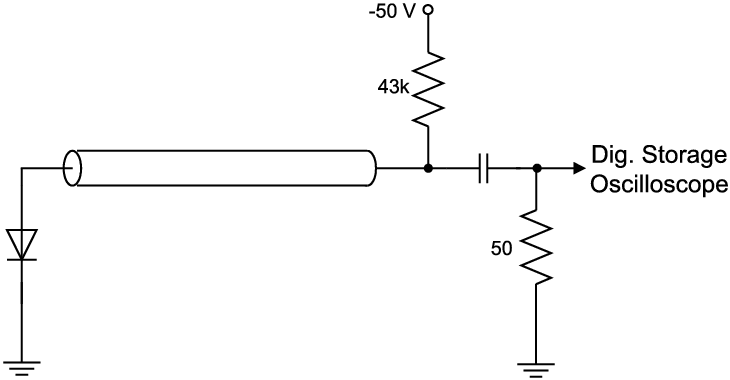}
 \caption[Reverse bias connection of PIN diodes]{Reverse bias connection of PIN diodes.}
\label{PINcircuit}
\end{figure}

\noindent The temporal evolution of XUV and X--ray radiation was detected by two filtered \mbox{PIN--Si} diodes.
These PIN diodes were produced by RNDr. Ivo Benc, CSc. in ``V\'{U}VET'' institution. They were operating as current
generators in the reverse bias mode. The equivalent circuit is displayed in Fig.~\ref{PINcircuit}.

These two PIN diodes were differentially filtered by a~15~$\muup$m thick Be filter and 0.8~$\muup$m thick Al
foil. The spectral responsivity ${\cal R}_i$ of filtered PIN diodes was calculated as follows:

 In order to create electron--hole pairs, photons have to be transmitted by a filter and
a~dead layer. At the same time, photons have to be absorbed in the sensitive (intrinsic) layer of a PIN diode.
The energy needed for the creation of one electron--hole pair in silicon is $w\approx3.62$~eV at room
temperature. The quantum efficiency\footnote{The ratio of number of incident photons to resulting photoelectrons
in the output current.} $\eta$ can be thus estimated
\begin{equation}
\label{ucinnost} \eta=\frac{h\nu}{w}\mathrm{e}^{-(\mu_\mathrm{F}
d_\mathrm{F}+\mu_\mathrm{D}
d_\mathrm{D})}(1-\mathrm{e}^{-\mu_\mathrm{S} d_\mathrm{S}})
\end{equation}
where $\mu_\mathrm{F}$, $\mu_\mathrm{D}$ and $\mu_\mathrm{S}$ represent absorption coefficients of a filter,
dead layer and sensitive layer, respectively. Quantities $d_\mathrm{F}$, $d_\mathrm{D}$ and $d_\mathrm{S}$ are
thicknesses of relevant layers.

\begin{figure}[h!]
\centering
\includegraphics{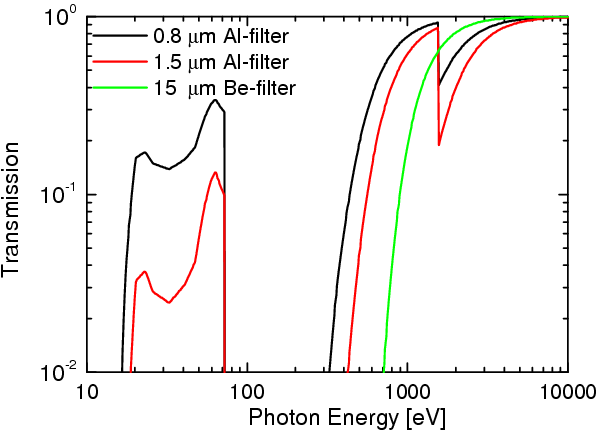}
\caption[Transmission of an Al and Be filter]{Transmission of an Al and Be filter.} \label{filters}
\end{figure}

\begin{figure}[h!]
\centering
\includegraphics{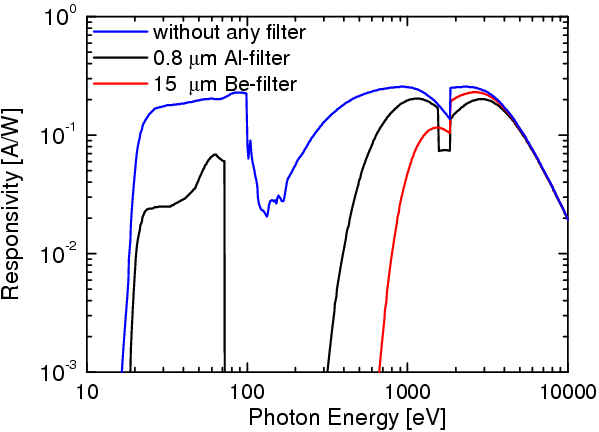}
\caption[Spectral responsivities of PIN diodes]{Spectral responsivities of PIN diodes.} \label{responsivities}
\end{figure}

Once the quantum efficiency is known, the spectral responsivity ${\cal R}_i$ can be derived from the ratio of
the photocurrent $i_\mathrm{PIN}$ and incident power $\Phi$ as
\begin{equation}
\label{responsivita} {\cal
R}_i=\frac{i_\mathrm{{PIN}}}{\Phi}=\frac{\eta e}{h\nu}
\end{equation}
Substituting $\eta$ from Eq.~\ref{ucinnost} into Eq.~\ref{responsivita} we obtain
\begin{equation}
{\cal R}_i=\frac{e}{w}\mathrm{e}^{-(\mu_\mathrm{F}
d_\mathrm{F}+\mu_\mathrm{D}
d_\mathrm{D})}(1-\mathrm{e}^{-\mu_\mathrm{S} d_\mathrm{S}})
\end{equation}
To calculate responsivities of our PIN diodes, we considered the thickness of a Si sensitive layer to be
$d_\mathrm{S}=10$~$\muup$m and the thickness of a dead layer to be $d_\mathrm{D}=0.1$ $\muup$m. The spectral
dependencies of transmission for filters, a sensitive layer and dead layer were calculated with Henke's
data~\citep{Henke93} provided by a Berkeley lab's website~\citep{CXRO}. The transmissions of two Al foils and a
15~$\muup$m thick Be filter are displayed in Fig.~\ref{filters}. The spectral responsivities of PIN diodes
 with these filters are shown in
Fig.~\ref{responsivities}.

The sensitive area of our PIN diodes was 1 mm$^2$ and hence we expected the time--response of approximately
$t_{10\%-90\%}\approx1$ ns. As far as the location is concerned, PIN diodes were placed approximately $20\div25$
cm away from the plasma. Most often we employed uncollimated PIN diodes. However, collimators were used once or
twice to avoid the radiation from electrodes. Finally we make a note that PIN diodes are sensitive to high
energy particles, particularly the influence of electrons should be considered.

\subsection{Hard X--ray detector}
\noindent The combination of a scintillator and photomultiplier was used to detect the time evolution of hard
X--ray emission. To put it simply, the principle of the detection can be described as follows: an X--ray photon
falls upon a scintillation material and produces higher number of visible photons which can be detected with a
photomultiplier. In our case, the scintillating fibre Bicron BCF--12 (1 mm diameter and 4 cm length) was
followed by the HAMAMATSU H1949--51 photomultiplier tube.

The Bicron fibre consists of a polystyrene--based core and a PMMA cladding (see table~\ref{scintillator}).
Scintillation efficiency for the blue emitting BCF--12 is approximately 2.4\%.

PMT's parameters are summarised in Tab.~\ref{PMT}. The HV power supply (HVPS--4) for the PMT tube was developed
in the Institute of Plasma Physics and Laser Microfusion (IPPLM) in Warsaw. The applied voltage determined the
transit time of electrons. In the case of often used $1.8\div2.0$ kV, the delay of the photomultiplier tube was
$35\div30$ ns.

\begin{figure}[p]
\centering
\includegraphics{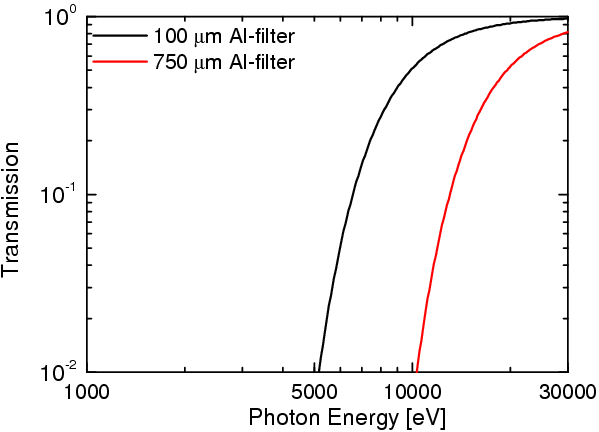}
\caption[Transmission of Al filters used together with a hard X--ray detector]{Transmission of Al filters used
together with a hard X--ray detector.} \label{ALfilter}
\end{figure}

In order to shield background radiation, the scintillation fibre was surrounded by a~plastic cone of 6 cm
diameter. As far as the energy of detected photons is concerned, it was predetermined by a filter placed in
front of the scintillator. The transmission of used filters (100 $\muup$m or 750 $\muup$m thick Al filter) is
displayed in Fig.~\ref{ALfilter}. This figure clearly shows that a 100 $\muup$m and 750 $\muup$m Al filter
transmits photons with energy higher than 5~keV and 10 keV, respectively.
\begin{table}[p]
 \begin{minipage}[t]{0.49\textwidth}
\centering
 \begin{onehalfspacing}
\begin{tabular}{|l| l|}
\hline \hline
Fibre & BCF--12\\
\hline
Emission colour & Blue\\
     \hline
Emission peak & 435 nm\\
     \hline
Decay time & 3.2 ns\\
     \hline
1/e Length & 2.7 m\\
     \hline
No.~of photons & 8000\\
per MeV & \\
     \hline
Core material & Polystyrene\\
     \hline
Core refractive index & 1.60\\
     \hline
Density & 1.05\\
     \hline
Cladding material & Acrylic\\
     \hline
Cladding\,refrac.\,index & 1.49\\
\hline
Numerical aperture & 0.58\\
     \hline
No.~of H atoms& $4.82\times10^{28}$\,m$^{-3}$\\
     \hline
No.~of C atoms& $4.85\times10^{28}$\,m$^{-3}$\\
     \hline
     \hline
\end{tabular}
\end{onehalfspacing}
\end{minipage}
\hfill
\begin{minipage}[t]{0.47\textwidth}
 \begin{onehalfspacing}
\centering
\begin{tabular}{|l| l|}
\hline \hline
Part number & H1949-1\\
\hline
Gain & $2\times10^7$\\
     \hline
Rise time & 1.3 ns\\
     \hline
Transit time & 28 ns\\
     \hline
Dark current & 50 nA\\
     \hline
Diameter & 51 mm\\
     \hline
Effective diameter & 46 mm\\
     \hline
Window\,material& Borosilicate\,glass\\
     \hline
Cathode type & Bialkali\\
     \hline
Number of stages & 12\\
     \hline
Peak wavelength & 420 nm\\
     \hline
Peak responsivity & 85 mA/W\\
     \hline
Min. wavelength & 300 nm\\
     \hline
Max. wavelength & 650 nm\\
     \hline
Output type& Current output\\
     \hline
     \hline
\end{tabular}
\end{onehalfspacing}
\end{minipage}
\begin{minipage}[t]{0.49\textwidth}
\caption[Basic parameters of the Bicron BCF--12 scintillating fibre]{Basic parameters of the Bicron BCF--12
scintillating fibre~\citep{Bicron}.} \label{scintillator}
\end{minipage}
\hfill
\begin{minipage}[t]{0.47\textwidth}
 \caption[Basic parameters of the HAMAMATSU PMT
tube]{Basic parameters of the HA\-MAMATSU PMT tube \citep{Hamamatsu}.} \label{PMT}
\end{minipage}
\end{table}

\input{spectrograph}

\subsection{X--ray pinhole camera}
\begin{figure}[!h]
\centering
\includegraphics{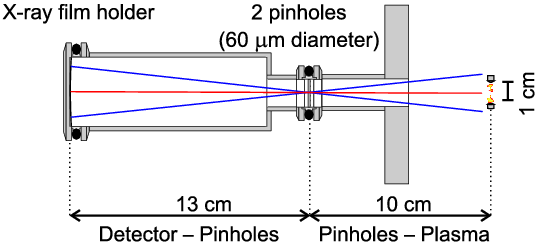}
\caption[X--ray pinhole camera for time--integrated imaging]{X--ray pinhole camera for time--integrated imaging.
Magnification 1.3, spatial resolution 105 $\muup$m.} \label{XPINHOLEsetup}
\end{figure}

\noindent The time--integrated pinhole camera was used to obtain X--ray images of an emitting plasma.
Fig.~\ref{XPINHOLEsetup} shows the cross-section of the pinhole camera construction and its position relative to
the emitting plasma.

The camera consists of the diagnostic channel, pinhole mount, and
film holder. The spatial resolution of the camera is given by
\begin{equation}
d_\mathrm{RES}=\varnothing\Bigg(1+\frac{1}{M}\Bigg)
\end{equation}
where $\varnothing$ is the diameter of the pinhole and $M$ is the magnification of the image. For
$\varnothing=60$~$\muup$m and $M=1.3$, we obtained the spatial resolution of $105$~$\muup$m.

In order to gain more information about the emitting plasma we used two pinholes --- in other words a two
channel pinhole camera. These two channels were differentially filtered with a~15~$\muup$m thick Be filter and
a~1.5$~\muup$m thick Al foil. The same filters were chosen to filter our PIN diodes (see Fig.~\ref{filters} on
page~\pageref{filters}). The X--ray images were recorded on the Kodak Industrex CX film.
\begin{figure}[h!]
\centering
\includegraphics{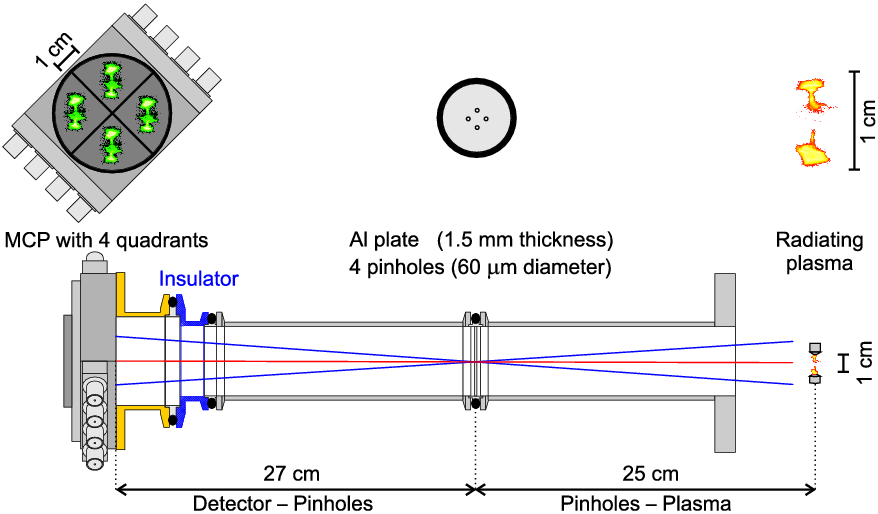}
\caption[VUV pinhole camera for time--resolved imaging]{VUV pinhole camera for time--resolved imaging.
Magnification 1.1, spatial resolution 200~$\muup$m.} \label{PINHOLEsetup}
\end{figure}

\subsection{Gated VUV pinhole camera}
\noindent The time--resolved information about radiation from the plasma and electrodes was obtained with the
gated pinhole camera. Its set--up can be seen in Fig.~\ref{PINHOLEsetup}. The temporal resolution of the gated
pinhole camera was carried out with a 4~frame MCP detector LPS--MCP/4F--D52. The exposure time was 2~ns. The
parameters of this microchannel plate were the same as for the microchannel plate LPS--MCP/4S--D56 (see
Table~\ref{MCP} on page~\pageref{MCP}). 4 quadrants instead of 4 stripes is the only diference. To record MCP
images, the Nikon D1 digital camera was used. The spatial resolution (given by the pinhole diameter,
magnification, and resolution of a CCD detector) approached 200 $\muup$m. The gated pinhole camera was mostly
used without any filter for the detection of VUV radiation ($\lambda<200$~nm, see Fig.~\ref{MCPefficiency} on
page~\pageref{MCPefficiency}). Sometimes we used an~aluminium filter for the detection of XUV emission.

\subsection{Laser probing}
\noindent The passive diagnostics described above gave some information about an electric current, voltage and
radiation in a discharge. To image the \emph{density} of a plasma, we used \emph{active} probing by a frequency
doubled Nd:YAG laser (530~nm wavelength, 3~ns pulse width\footnote{The pulse duration was reduced to 3 ns using
a Pockels cell.}, 1 mJ energy). In most cases, the optical probing in the schlieren set--up was applied. This
technique is sensitive to the refractive index and hence to the electron density gradient. The refractive index
$N(r,\theta,z)$ is dependent on the electron density $n_{\mathrm{e}}(r,\theta,z)$ by the equation
\begin{eqnarray}
N(r,\theta,z)&=&\sqrt{1-\frac{e^2n_{\mathrm{e}}(r,\theta,z)}{ m_\mathrm{e}\varepsilon_0\omega^2}}\\
N(r,\theta,z)&\approx&1-\frac{e^2n_{\mathrm{e}}(r,\theta,z)}{2
m_\mathrm{e}\varepsilon_0\omega^2}\qquad \mathrm{for}\quad
N(r,\theta,z)\approx1 \label{aproximace}
\end{eqnarray}
where $\omega$ is the angular frequency of the laser beam, $e$ is the electric charge and $m_\mathrm{e}$ is the
mass of an electron.

Variations in the electron density deflect the light passing through a plasma. After the passage through a
plasma, the light deviates by an angle $\vartheta$ given by
\begin{equation}
\vartheta=\int  \nabla_{\bot} N(r,\theta,z) \diff l
\end{equation}
where $\nabla_{\bot} N(r,\theta,z)$ is the gradient of the refractive index transverse to the propagation of
a~laser beam and $\int \diff l$ is the path length along the line of sight.

\begin{landscape}
\begin{figure}
    \centering
    \rotatebox{-90}{\includegraphics{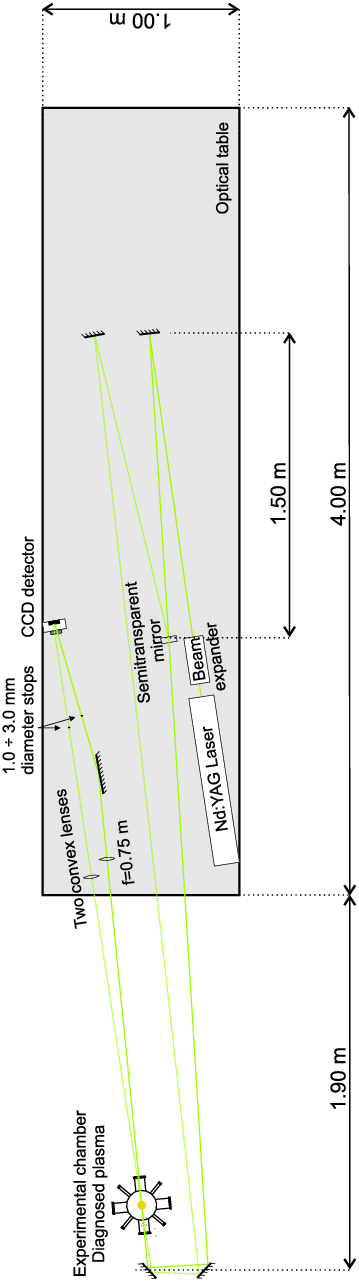}}
    \caption[Optical layout of the schlieren system, scale 1:30]{Optical layout of the schlieren system, scale 1:30.}
\label{Schliersetup}
\end{figure}
\end{landscape}

The layout of the schlieren system used most recently is displayed in Fig.~\ref{Schliersetup}. The diameters of
lenses and stops determined the plasma density that was imaged. In our experiment we used two convex lenses
(diameter $\mathcal{D}=8$ cm and focal length $f=0.75$~m) and stops (diameter $a=1.0\div3.0$ mm) in their focus.
It means that we imaged the regions with deflections
\begin{eqnarray}
\vartheta_\mathrm{MIN}=\frac{a}{2f}\leq&\vartheta&\leq\frac{\mathcal{D}}{2f}=\vartheta_\mathrm{max}\\
(7\div20) \times 10^{-4}\leq&\vartheta&\leq5 \times 10^{-2}
\end{eqnarray}
Using the approximative equation~\ref{aproximace} and the angular frequency of the Nd:YAG laser
$\omega=3.6\times10^{15}$~rad\,s$^{-1}$, we get for the electron density
\begin{equation}
(5\div15) \times 10^{24}\, \textrm{m}^{-3}\leq\int \nabla_{\bot} n_\mathrm{e}(r,\theta,z) \diff l \leq 4\times
10^{26}\, \textrm{m}^{-3} \label{schliersensitivity}
\end{equation}
To record the development of a plasma, the laser beam was split into two beams and one of them was delayed by 10
ns with respect to the other. The plasma was imaged on the charge coupled device (digital camera Nikon D100,
active sensor area $23.4\times15.6$ mm, 6.0 million pixels $3000\times2000$) with a magnification of
$0.6\div0.7$. The visible light emitted from the plasma was reduced by polarising and grey filters. The spatial
resolution of the schlieren system was less than 30 $\muup$m.

\section{Control and Synchronising Units}
\begin{figure}[h]
    \centering
    \includegraphics{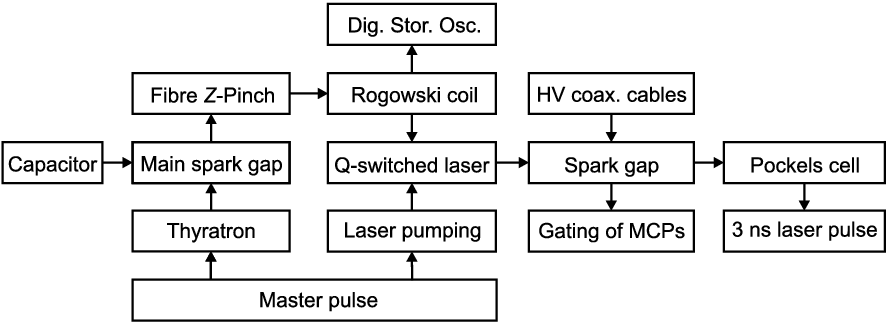}
\caption[Block diagram of synchronising a discharge and diagnostics]{Block diagram of synchronising a discharge
and diagnostics.} \label{synchro}
\end{figure}

\noindent In this section we would like to describe the controlling and synchronising a discharge and
diagnostics (cf.~Fig.~\ref{synchro}).

\subsubsection{Discharge}
\noindent The charging of the capacitor was controlled by the digital electronic device C--30 constructed
by~\citet{Kozlik02}. When the capacitor was charged, the master pulse triggered both the laser pumping and the
thyratron. After that, 30 kV thyratron pulse triggered the main spark gap and hence the $Z$--pinch discharge.

\subsubsection{Diagnostics}
\noindent All the diagnostic tools were activated by the discharge breakdown. The signal from the Rogowski coil
triggered both oscilloscopes and the Q--switched Nd:YAG laser. After the laser action, the laser beam triggered
the spark gap that discharged high voltage cables. The high voltage pulses from these cables then triggered both
microchannel plates and also the Pockels cell. The Pockels cell shortened the laser pulse width to 3 ns. The
delay of 10 ns between the gating of adjacent MCP quadrants/stripes was carried out by 2 m long coaxial cables.
There was also 10 ns time delay between two schlieren images. In order to take these two schlieren photographs
simultaneously with two pinhole images and two gated XUV spectra, the laser beam pass length (from the laser
output mirror to the $Z$--pinch plasma) was adjusted. The time delay between the current breakdown and recording
of images was controlled by the length of a coaxial cable between the Rogowski coil and the laser. The jitter
approached the value of 5 ns.

\section{Digital Storage Oscilloscopes}
\noindent Two digital phosphor oscilloscopes Tektronix TDS3054 and TDS654C were used to store data from the
Rogowski coil, voltage monitor, PIN diodes, photomultiplier tube, and gating of microchannel plates. Both
oscilloscopes have 4 channels, 500 MHz bandwidth and 5 GS/s maximum sample rate. More information can be found
on the website~\citep{Tektronix05}.

%% file: spectrograph.tex
\newpage
\subsection{XUV spectrograph}
\label{spectrograph} \noindent In this subsection, let us present the spectrograph we used in our experiment.
Since XUV spectroscopy was one of the most important contribution of the author to the experiment, the part
below covers the subject in more detail. In addition to that, principles of spectroscopy and methods applied in
the thesis can be found in appendix~\ref{spectromethods}.

\subsubsection{XUV and X--ray spectral range} \noindent Due to the fact that the temperature of $Z$--pinches varies from several eVs up
to tens of keV, a predominant fraction of the electromagnetic radiation is emitted in the XUV and X--ray
spectral region. The boundaries of XUV and X--ray ranges are not rigid, but as such traditionally accepted in
the fields of X--ray spectroscopy, plasma physics, astrophysics, X--ray laser etc. For the purpose of this
paper, we use the wavelength ranges displayed in table~\ref{ranges}.
\begin{table}[!h]
\centering
\begin{minipage}[t]{0.5\linewidth}
\centering
\begin{onehalfspacing}
   \begin{tabular}{|l| l|}
\hline \hline Spectral region& Wavelength  \\
\hline
hard X--ray & 0.01~nm $\div$ 0.2 nm \\
\hline
soft X--ray & 0.2~nm $\div$ 2.5 nm \\
\hline\hline \end{tabular}
\end{onehalfspacing}
 \end{minipage}%
 \begin{minipage}[t]{0.5\linewidth}
 \centering
\begin{onehalfspacing}
  \begin{tabular}{|l| l|}
\hline \hline Spectral region& Wavelength  \\
\hline
XUV & 2.5~nm $\div$ 40 nm \\
\hline
VUV & 40~nm $\div$ 200 nm \\
 \hline \hline
 \end{tabular}
\end{onehalfspacing}
 \end{minipage}
\caption[Definition of wavelength ranges used in our work]{Definition of wavelength ranges used in our
work~\citep{Pikuz01}.
 \label{ranges}}
\end{table}

\vspace{-0.4cm}
\subsubsection{Basic principles of XUV spectrographs}
\noindent Our XUV spectrograph was used to give information about the spectral characteristics of radiation
emitted from a~fibre $Z$--pinch. Since we were interested in K-- and L--shell emission of carbon ions in
particular, we were limited mainly to the XUV spectral region from 2 to 50 nm, i.e.~approximately from 20 to 600
eV. The construction of a spectrograph in this spectral region is determined by three important factors.

Firstly, a~high resonance absorption in all materials must be considered. It means that spectrographs have to be
evacuated and transmission through whatever materials should be restricted. In other words, reflecting elements
are preferred.

Secondly, in order to obtain high dispersion and, by extension, high spectral resolution, it is necessary to
have the characteristic period of a dispersing element of the order of the wavelength, i.e.~several nanometres.
Unfortunately, crystal spectrographs can be hardly used above 5 nm because the period of natural crystal lattice
is of the order of \AA, i.e.~0.1~nm. The usage of grating spectrographs with the normal incidence is also
limited since it is difficult to produce gratings with more than $10000$ grooves per 1 mm. Nevertheless, it is
possible to employ grazing incidence (the angle of incidence is close to 90\degree) to have the effective groove
spacing higher than $10000$ grooves per 1 mm. Moreover, the small grazing angle ensures higher reflectivity of a
grating.

Thirdly, we must take into account that we cannot use lenses to focus spectral lines onto the detector. The
common way of focusing is the usage of a concave grating with constant groove spacing. The disadvantage of such
technique are image defects, astigmatism in particular. Some defects can be reduced in the case of a toroidal
grating with varying groove spacing~\citep{Harada80,Kita83}. In our project we employed ``classical'' and
cheaper way of the detection: a concave spherical reflection grating in the Rowland circle grazing incidence
LPS--VUV1 spectrograph.

\subsubsection{Rowland mounting}
\noindent In 1881 H.\,A.~Rowland, professor of physics at the Johns Hopkins University, ruled the concave
grating, a device of spectacular value in modern spectroscopy. A concave reflection grating can be considered as
a concave mirror that disperses; it can be thought to reflect and focus light by virtue of its concavity, and to
disperse light by virtue of its groove pattern~\citep{Handbook01}.
\begin{figure}[h]
\centerline{\includegraphics{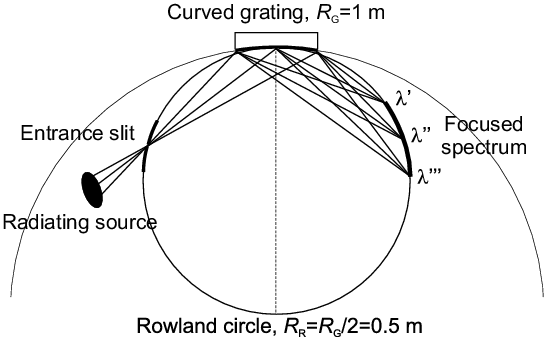}} \caption[Rowland circle spectrograph]{Rowland circle
spectrograph.}\label{Rowland}
\end{figure}

As regards the Rowland mounting, both the entrance slit and the focused spectrum lie on this circle
(cf.~Fig.~\ref{Rowland}). Light of three wavelengths $\lambda'$, $\lambda''$, $\lambda'''$ is shown focused at
different points on the Rowland circle. The diameter of the circle $R_\mathrm{R}$ equals to the tangential
radius of curvature of the grating $R_\mathrm{G}$ which passes through the grating centre.

\subsubsection{LPS--VUV1-3S--M spectrograph}
\begin{figure}[h]
\centerline{\includegraphics*{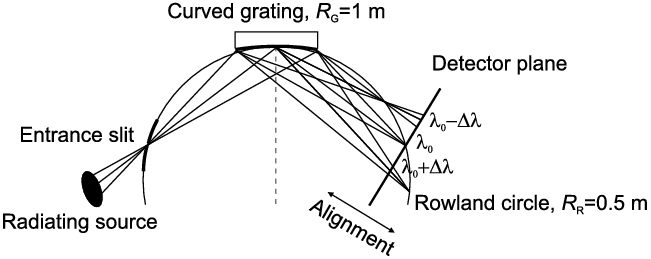}} \caption[Registration by a plane detector]{Registration by a
plane detector.}\label{LPS1}
\end{figure}

\noindent The LPS--VUV1 grazing incidence spectrograph was constructed by the Institute of Spectroscopy
(Laboratory of Plasma Spectroscopy, Russian Academy of Science)  in cooperation with Schulz Scientific
Instruments and European Pulsed Power Research Association~\citep{Schulz02}. It is a very adaptable
proximity--focused, 1--m grazing incidence spectrograph using an off Rowland circle registration scheme (see
Fig.~\ref{LPS1}). \label{schema}
\begin{figure}[h]
\centerline{\includegraphics{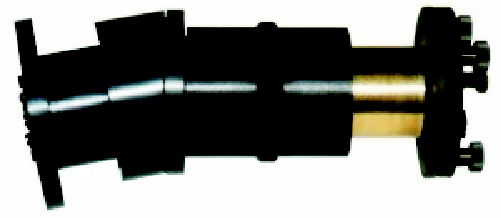}} \caption[LPS-VUV1-3S-M spectrograph]{LPS-VUV1-3S-M
spectrograph.}\label{LPS}
\end{figure}

In this registration scheme, the exact focusing takes place only for one wavelength $\lambda_0$, which
corresponds to the intersection point of the detector plane with the Rowland circle. But if the numerical
aperture of the spectrograph is small\footnote{For instance, if the plasma source is localised.}, then it is
possible to record the spectrum with sufficient spectral resolution
 in a wide spectral range $\lambda_0 \pm
\Delta \lambda$. The alignment to a different $\lambda_0$ is enabled by changing the distance between the
detector plane and the grating. The two main advantages of this scheme are its easy setup and low--cost.

The useful spectral ranges of the spectrograph $2 \div 14$~nm, $7 \div25 $~nm, and $25 \div 60$~nm were achieved
with the gratings of 1200, 600, and 300 gr./mm, respectively. The $2\div14$~nm spectral range was significant
because of the presence of carbon K--shell spectral lines and thus it was preferably used. The basic parameters
of the spectrograph and gratings are summarised in Tab.~\ref{LPSparameters} and~\ref{gratingparameters}.
\begin{table}[h!]
\centering
\begin{onehalfspacing}
 \begin{tabular}{|l|l|}
  \hline
   \hline
  Grazing angle $\phi_{0}$&4\degree \\
   \hline
  Gratings & 300, 600 and~1200 grooves / mm \\
   \hline
  Spectral range & $2 \div 80$ nm \\
   \hline
  Resolving power & $\lambda/\Delta\lambda > 100$ \\
   \hline
  Angle of incidence on detector & 45\degree or 90\degree \\
   \hline
  Angle between detector plane & \\
  and~spectrograph axis $\alpha$&10\degree \\
   \hline
  Entrance slit & $30\div50$ $\muup$m \\
   \hline
  Dimension & $\varnothing$  140 mm\,\verb|x|\,$170\div350$ mm (length can be adjusted)\\
   \hline
  Detector & Russian UV film UF-4 or MCP and CCD detector\\
  \hline
   \hline
 \end{tabular}
 \end{onehalfspacing}
\caption[Basic parameters of the LPS--VUV1--3S--M spectrograph]{Basic parameters of the LPS--VUV1--3S--M
spectrograph.}\label{LPSparameters}
\end{table}

\begin{table}[h!]
\centering
\begin{onehalfspacing}
\begin{tabular}{|l|c|c|c|c|}
\hline
 \hline
Grating no.~&1&2&3&4\\
 \hline
Radius of curvature &1000 mm&1000 mm&1000 mm&1000 mm\\
 \hline
Dimensions [mm]&30\,\verb|x|\,40\,\verb|x|\,10&30\,\verb|x|\,40\,\verb|x|\,10&30\,\verb|x|\,40\,\verb|x|\,10&30\,\verb|x|\,40\,\verb|x|\,10\\
 \hline
Ruled area [mm]&28 \verb|x| 30&28 \verb|x| 30&28 \verb|x| 30&28 \verb|x| 30\\
 \hline
Grooves / mm &1200&1200&600&300\\
 \hline
Groove shape & triangular &triangular &triangular &triangular\\
\hline
Blaze angle &1\degree&2\degree&2\degree&3\degree\\
 \hline
Blaze wavelength &2.5 nm&6.2 nm&12.2 nm&42.5 nm \\
 \hline
Coating&Au 60 nm&Au 60 nm&Au 60 nm&Au 60 nm\\
 \hline
Recommended &$2\div7$ nm&$4\div14$ nm&$7\div24$ nm&$24\div80$ nm\\
spectral range& &$2\div7$ nm (2$^\mathrm{nd}$ ord.)& & \\
 \hline
\hline
\end{tabular}
\end{onehalfspacing} \caption[Basic parameters of the gratings]{Basic parameters of
gratings~\citep{Koshelev99}.}\label{gratingparameters}
\end{table}

\newpage
\subsubsection{Estimation of wavelengths}
\noindent We can estimate wavelengths of spectral lines from the grating equation and from the geometry of the
spectrograph.

\begin{figure}[h]
\centerline{\includegraphics{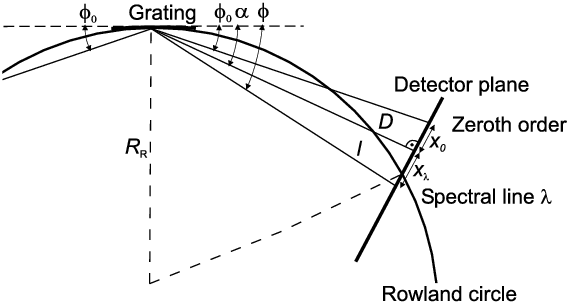}} \caption[LPS--VUV1 spectrograph geometry of detection]{LPS--VUV1
spectrograph geometry of detection.}\label{LPS2}
\end{figure}

The Rowland circle spectrograph follows the grating diffraction
equation~\citep{Michette86}
\begin{equation}
d(\cos\phi_{0}-\cos\phi)=m\lambda\sqrt{1+\frac{y^2}{s^2}}
\end{equation}
where $d$ is the groove spacing, $\phi_{0}$ is the grazing angle of incidence, $\phi$ is the grazing angle of
diffraction, $m$ is the order of diffraction, $\lambda$ is the wavelength, $y$ is the location of an imaged
point on the entrance slit\footnote{The value of $y=0$ belongs to the centre of the entrance slit.}, and $s$ is
the distance between the entrance slit and the grating centre. In the case of $y\ll s$ we obtain the well--known
equation for the plane grating
\begin{equation}\label{gratingequation}
d(\cos\phi_{0}-\cos\phi)=m\lambda
\end{equation}
Fig.~\ref{LPS2} shows the detection geometry of the spectrograph.

\begin{figure}[h]
\centerline{\includegraphics{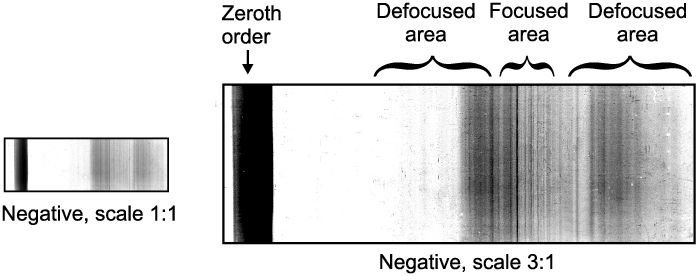}} \caption[Example of a spectrum recorded by the UV--film
UF--4]{Example of a spectrum recorded by the UV--film UF--4. Grating no. 3. }\label{ukazkaspektra}
\end{figure}

The position of the spectral line $\lambda$ can be derived by using the following symbols:
\begin{eqnarray}
\nonumber D& - & \textrm{the distance between the grating centre
and the detector plane}\\ \nonumber l&-& \textrm{the distance
between the grating centre and the spectral line}
\\ \nonumber
R_\mathrm{R}&-& \textrm{the radius of the Rowland circle}\\\nonumber x_{0}&-&\textrm{the distance between the
zeroth order and the perpendicular}
\\\nonumber
& & \textrm{from the grating centre upon the detector}\\\nonumber x_{\lambda}&-&\textrm{the distance between the
spectral line $\lambda$ and the perpendicular} \\\nonumber && \textrm{from the grating centre upon the
detector}\\\nonumber \alpha&-& \textrm{the angle between the grating ``plane'' and~the spectrograph
axis}\\\nonumber \phi_{0}&-& \textrm{the grazing angle of incidence}\\\nonumber \phi&-& \textrm{the grazing
angle of diffraction}
\end{eqnarray}
From Fig.~\ref{LPS2}, the $x$ distance between the spectral line $\lambda$ and the zeroth order can be
determined
\begin{equation}\label{odvozenilambdy}
x=x_{0}+x_{\lambda}=D[\tan(\alpha-\phi_{0})+\tan(\phi-\alpha)]
\end{equation}
By expressing the angle $\phi$ from equation (\ref{odvozenilambdy}) and substituting it in the grating equation
(\ref{gratingequation}), we obtain
\begin{equation} \label{dispersion}
\lambda=\frac{d}{m}\Bigg[ \cos\phi_{0}-\cos\Bigg(
\alpha+\arctan\bigg( \frac{x}{D}-\tan(\alpha-\phi_{0}) \bigg)
\Bigg) \Bigg]
\end{equation}
The construction parameters such as the angles $\alpha=10\degree$, $\phi_0=4\degree$ and the groove spacing $d$
are fixed and known with sufficient accuracy~\citep{Klir02}. Thus, if the distances $x$ and $D$ are known, we
can easily estimate the spectral wavelength $\lambda$, or better to say $m\lambda$. However, the zeroth order is
usually out of focus and therefore it is not possible to accurately determine the $x$ distance (see
Fig.~\ref{ukazkaspektra}).
 Also, the exact value of the distance $D$ can be hardly measured. For that reason we
determine wavelengths with respect to two reference lines\footnote{When two wavelengths are known, we can choose
$x$ and $D$ for which the dispersion curve Eq.~\ref{dispersion} best fits the experimental data.} for which the
wavelengths are known. It is clear then, that it is essential to surely identify several spectral lines in each
spectrum.

\begin{figure}[h]
\centerline{\includegraphics{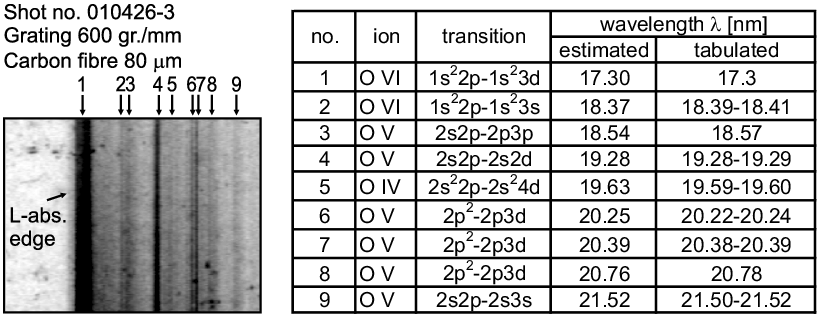}} \caption[Calibration of wavelengths by an aluminium
filter]{Calibration of wavelengths by an aluminium filter.}\label{calibration}
\end{figure}

The really certain identification was only practicable after gaining experience with several XUV sources, e.g.
a~carbon fibre $Z$--pinch, a plasma focus~\citep{Kubes03}, a~$Z$--pinch initiated from an aluminium
wire--array~\citep{Klir04a, Klir05}, etc. In the 7$\div$24 nm region, the reliable identification of lines was
made easier when a 0.8 $\muup$m thick aluminium filter was used. This technique was based on a sharp aluminium
L--absorption edge\footnote{To be more precise, it is L$_{2,3}$--absorption edge, L$_{2}$ at 17.01 nm and
L$_{3}$ at 17.1 nm~\citep{Henke93}.} at approximately 17.1 nm, i.e.~at 72.55 eV (cf.~Fig.~\ref{filters}). One
example of a~spectrum when an Al filter was applied, is shown in Fig.~\ref{calibration}. The most intensive
spectral line in the spectrum was near the L--absorption edge at 17.1 nm and it was identified as the transition
2p $^2$P -- 3d $^2$D of Li--like oxygen ion O VI at 17.3 nm. Other observed spectral lines were identified with
the aid of the dispersion curve Eq.~\ref{dispersion}.

\subsubsection{Identification of transitions}
\noindent Once wavelengths of observed spectral lines are determined, it is possible to assign them to
transitions. For that purpose we use the following spectral databases available online:
\begin{enumerate}
\item[--]Kelly's Atomic and Ionic UV/VUV Linelist~\citep{Kelly05} provided by Harvard--Smithsonian Center for
Astrophysics, \item[--]Atomic Spectra Database~\citep{NIST05} provided by the National Institute of Standards
and Technology (NIST), \item[--]Atomic line database for Astrophysics compiled by~\citet{Verner05}
and~\citet{Hoff05}.
\end{enumerate}

In our case, Kelly's compilation of experimental databases is preferred but when transition probabilities are
needed, the NIST
 database is used.

\subsubsection{Intensity of spectral lines}
\noindent Not only the wavelength but also the intensity of a spectral line is of great interest since it
enables the estimation of plasma parameters.There are six points we would like to make in this connection.

First thing to be noted is that our spectrograph is not \emph{absolutely} calibrated and we have to get along
with \emph{relative} intensities.

Second, the sensitivity of our spectrograph is wavelength dependent. In two separate paragraphs, which bring the
following pages, we touch upon the spectral dependence of grating and detector efficiency in particular.
Figures~\ref{gratingefficiency}, \ref{AUreflexe}, \ref{FILMcalibrations}, and \ref{MCPefficiency} will show that
it is not accurate to compare spectral lines with substantially different wavelengths. Nevertheless, if the
desired accuracy is not extremely high, we can compare spectral lines in the region where the grating and
detector efficiency do not change substantially.

Third, the linearity of a detector is limited, especially in the case of a UV film
(cf.~Fig.~\ref{FILMcalibration}). Consequently the comparison of spectral lines with \emph{similar} intensities
is highly appropriate.

The fourth fact we would like to mention is the saturation of spectral lines. In several cases it is possible to
check whether a spectral line is near saturation (either by plasma opacity or by the saturation of a detector).
Imagine, for instance, that we can find two spectral lines whose intensity ratio is not essentially influenced
by plasma parameters and it is almost invariable. If the measured ratio is higher in favour of the weaker line,
the stronger spectral line is most likely saturated.

The fifth note is related to the width of a spectral lines. As was already said above, the exact focusing of a
spectral line takes place only for one wavelength $\lambda_0$. Other lines are defocused to various extents.
Therefore, intensities cannot be estimated from the peak value but they have to be calculated by integrating the
spectral line profile. In some cases, especially when a spectral line is not intensive, the integration is
doubtful since the spectral line could not be separated from background and adjacent lines.

Finally, we make a remark about digitizing a UV film. Our UV films are digitised with a scanner Epson
Perfectionist 2400 Photo (4800 dpi, 16 bit grey scale) which was calibrated by the wedge of known optical
density.

\subsubsection{Grating efficiency}
\noindent For a reflection grating, relative efficiency is defined as the energy flow of monochromatic light
diffracted into the order (of our interest) \emph{relative to} the energy flow of specular reflection from a
polished mirror blank coated with the same material~\citep{Handbook01}. The determination of grating efficiency
in the XUV region is obviously a tricky problem because a spectrograph needs to be calibrated by a suitable XUV
source such as a synchrotron. Since such source has not been available and since the producer of our
spectrograph has not provided information about grating efficiency we were left with approximate theoretical
calculation. Below we present how we went about that.

For a grating with a very low blaze angle $<5\degree$, the efficiency behaviour is almost perfectly scalar; that
is, polarization effects can be neglected. The diffraction efficiency is the highest for the so--called blaze
wavelength $\lambda_\mathrm{B}$ that is diffracted by the grating in the same direction as it would be reflected
by the groove facets. Above fifty--percent efficiency is obtained from roughly $0.67\lambda_\mathrm{B}$ to
$1.8\lambda_\mathrm{B}$~\citep{Handbook01}.

\begin{landscape}
\begin{figure}
    \centering
     \subfigure{
    \includegraphics[scale=0.95]{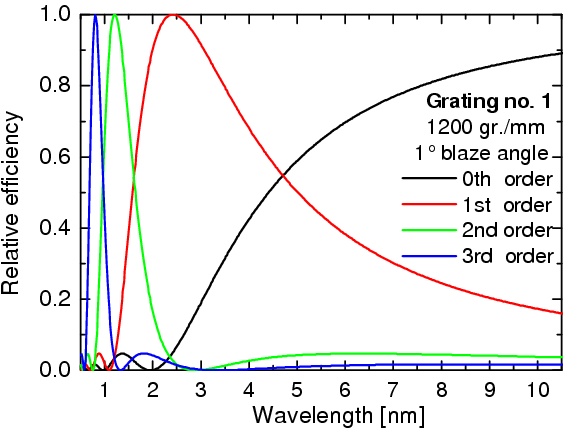} }
    \subfigure{
    \includegraphics[scale=0.95]{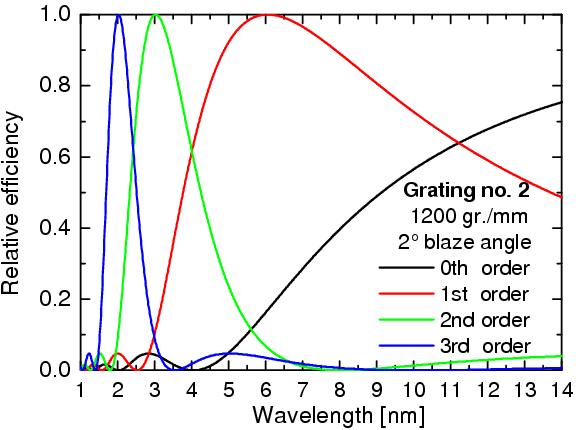} }
    \subfigure{
    \includegraphics[scale=0.95]{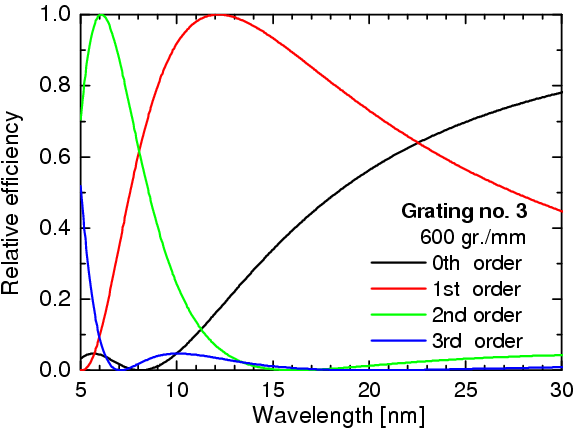} }
    \subfigure{
    \includegraphics[scale=0.95]{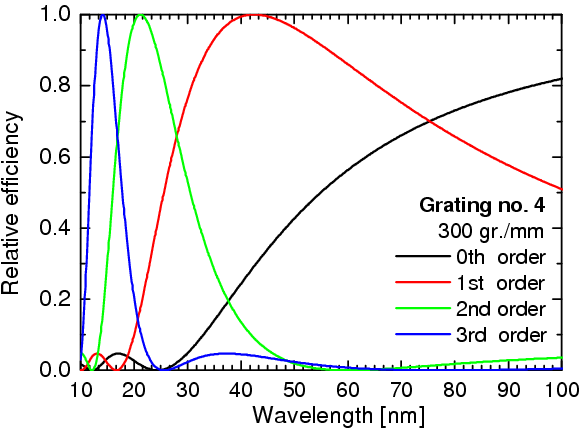} }
\caption[Diffraction efficiency of gratings no.~1--4]{Diffraction efficiency of gratings no.~1--4.}
\label{gratingefficiency}
\end{figure}
\end{landscape}

More detail dependence was calculated by the MATLAB programme provided by V\'{a}clav Neme\v{c}ek~\citep{Nemecek02}. The
calculation was based on the thin grating approximation~\citep{Hutley82}. When we used the groove spacing,
grazing angle of incidence, and blaze angle of our gratings, we obtained efficiency curves that are displayed in
Fig.~\ref{gratingefficiency}. Except the grating curvature, the MATLAB programme did not include X--ray
interaction with matter. Therefore, the reflectivity of gold coating is discussed further.

\begin{figure}[h]
\centerline{\includegraphics{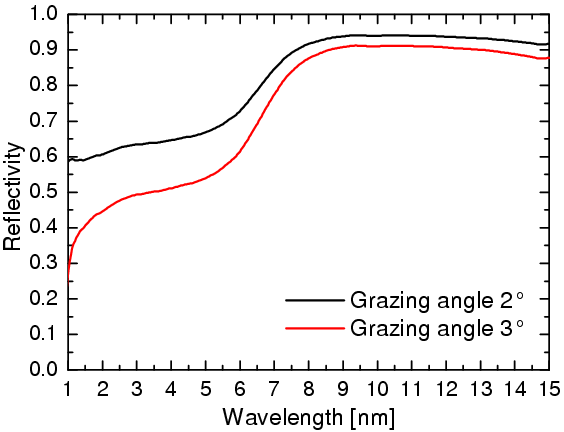}} \caption[Reflectivity of a gold thick layer for unpolarised
light]{Reflectivity of a gold thick layer for unpolarised light~\citep{CXRO}.} \label{AUreflexe}
\end{figure}

The influence of gold coating on absolute efficiency in the $1\div15$ nm region can be seen in
Fig.~\ref{AUreflexe}. In this figure, the reflectivity of a gold layer was calculated for two grazing angles of
incidence. It should be noted that the angle of incidence on the groove facets $\phi_0=4\degree$ is decreased by
the blaze angle\footnote{For blaze angles see Tab.~\ref{gratingparameters}.}. Therefore, the angle of $2\degree$
is relevant for gratings no.~2, 3, whereas the angle of $3\degree$ is relevant for grating no.~1.

Last but not least, we would like to emphasise that experimental calibration could not be fully replaced by
theoretical calculations. The theoretical calculation was essential only to exclude possible (unwanted) edges in
grating efficiency. Thus the graphs in figures~\ref{gratingefficiency} and~\ref{AUreflexe} are displayed only
for the illustration of how grating efficiencies could vary with wavelength.

\subsubsection{Detectors}
\begin{figure}[h]
\centerline{\includegraphics{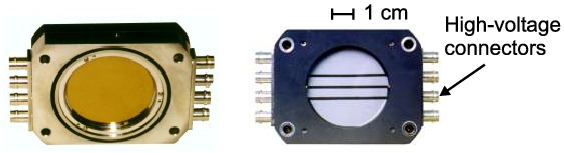}} \caption[Gold coated MCP detector with 4 stripes]{Gold coated
MCP detector with 4 stripes~\citep{Schulz02}.}\label{MCPFOTO}
\end{figure}

\noindent The LPS--VUV1 spectrograph  was designed for an easy operation in different plasma systems with
options for both time--integrated detection using a film and time--resolved recording using active detectors,
namely a microchannel plate (MCP, see Fig.~\ref{MCPFOTO}) together with charge coupled device (CCD). The
advantage of an ultraviolet film is its higher spectral resolution, whereas MCP detectors offer the nanosecond
exposure time and linear response (provided detectors are not saturated). However, we should not forget that the
gain of a microchannel plate is very sensitive to applied voltage. Moreover, the gain could vary even between
two places in the same MCP section.

As regards our experiment, gating pulses were produced by
discharging 20 cm long cables via a laser spark gap. The
high--voltage power supply was provided by the Institute of Plasma
Physics and Laser Microfusion, Warsaw.

\begin{figure}[p!]
 \begin{minipage}[t]{0.5\linewidth}
   \centering
   \includegraphics{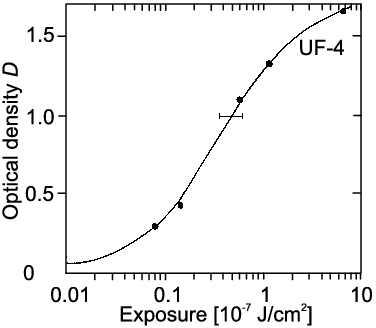}
    \caption[Characteristic
curve of the UF--4 film]{Characteristic curve of the UF--4 film~\citep{Shevelko96} ($\lambda=18$~nm, 20\degree
C, developer D-19 (1:1) -- 3 min., fixer -- 3 min.}
   \label{FILMcalibration}
 \end{minipage}%
 \begin{minipage}[t]{0.5\linewidth}
   \centering
    \includegraphics{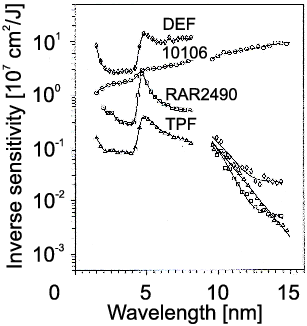}
   \setlength{\captionmargin}{0.5cm}
   \caption[Inverse sensitivity of films in the $2\div14$ nm region]{Inverse sensitivity of films in the $2\div14$ nm region.}
    \label{FILMcalibrations}
 \end{minipage}
 \begin{minipage}[t]{1\linewidth}
   \centering \subfigure[X--ray region]{
\includegraphics{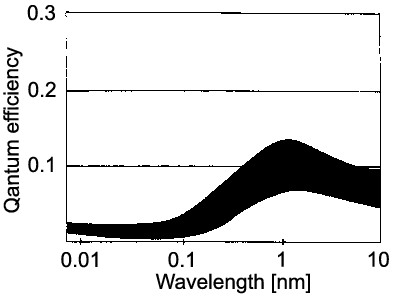} }
\hspace{1cm} \subfigure[VUV region]{
\includegraphics{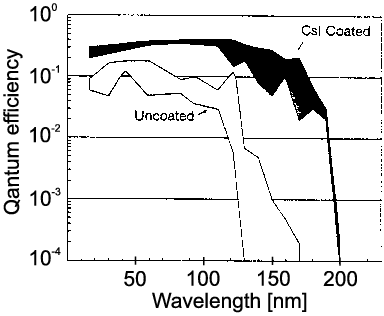} }
\caption[Typical spectral dependence of quantum efficiency of MCPs]{Typical spectral dependence of quantum
efficiency of MCPs.} \label{MCPefficiency}
 \end{minipage}
  \end{figure}

We summarised some important parameters of used detectors in tables~\ref{film}, \ref{MCP}, \ref{CCD} and
figures~\ref{FILMcalibration}, \ref{FILMcalibrations}, \ref{MCPefficiency}.
\begin{table}[h!]
\centering
\begin{onehalfspacing}
\begin{tabular}{|l|l|}
 \hline
 \hline
Width & 35 mm\\
  \hline
Recommended spectral region & $2\div200$ nm\\
  \hline
Sensitivity $\cal S_\lambda$ &  $4.10^{-8}$ J/cm$^2$\\
  \hline
Gamma $\gamma_{\lambda}$ & 1.2\\
  \hline
Thickness of sensitive layer & $3\div5$ $\muup$m\\
  \hline
Fog density $D_0$ & 0.05\\
  \hline
Resolution & $>150$ lines per mm\\
  \hline
Developer & D--19, dilution 1:5, 20\degree C, 7--8 min. \\
  \hline
Fixer & Fomafix univ. BW, 20\degree C, more than 10 min. \\
\hline
\hline
\end{tabular}
\end{onehalfspacing} \caption[Properties of the UF--4 ultraviolet film]{Properties of the UF--4 ultraviolet
film.} \label{film}
\end{table}
\begin{table}[h!]
\centering
\begin{onehalfspacing}
\begin{tabular}{|l|l|}
\hline \hline
MCP type & LPS--MCP/4S--D56 \\
 \hline
Photocathode &Au \\
\hline
Spectral region & $0.1\div200$ nm \\
\hline
Number of channel plates & 1 \\
\hline
Usable diameter & 52 mm\\
\hline
MCP channel diameter & 12 $\muup$m\\
\hline
MCP channel to channel distance & 16 $\muup$m\\
\hline
Pulse duration& $2\div3$ ns\\
\hline
Gating voltage & $4\div6$ kV\\
\hline
Gain of secondary electron emission&  $10\div1000$\\
\hline
Phosphor type& ZnS(Cu), $\lambda\approx 520$ nm \\
\hline
Fibre optic channel diameter & 6 $\muup$m\\
\hline
Fibre optic channel to channel distance & 8.5 $\muup$m\\
\hline
Conversion efficiency of electrons to photons (energy)& $\approx50\%$ \\
\hline
Pressure & $<10^{-2}$ Pa\\
\hline \hline
\end{tabular}
\end{onehalfspacing} \caption[Parameters of MCP used as an active detector in our spectrograph]{Parameters of the MCP
used as an active detector in our spectrograph.} \label{MCP}
\end{table}
\begin{table}[p!]
\centering
\begin{onehalfspacing}
\begin{tabular}{|l|l|}
\hline \hline
CCD type& digital camera Nikon D1X \\
\hline
Sensor size  &23.7 \verb+x+ 15.5 mm\\
\hline
Pixels & 5.3 million (4028\verb/x/1324)\\
\hline
A/D converter& 12 bit \\
\hline
ISO speed& $125\div800$ \\
\hline Objective & Nikkor ($35\div70$ mm/2.8D) \& Extension tube\\\hline \hline
\end{tabular}
\end{onehalfspacing}
 \caption[Basic parameters of the CCD detector]{Basic parameters of the CCD detector.}
\label{CCD}
\end{table}

\newpage For the time--integrated detection we applied the UF--4 film. The problem is that this film is not
used on a mass scale and so there is not enough information available about its sensitivity. Its absolute
calibration was made by \citet{Shevelko96}. Unfortunately, the spectral variation of the sensitivity can be only
guessed from the dependence of other films used in this spectral region. On the basis of
Fig.~\ref{FILMcalibrations} we can assume that the spectral sensitivity of the UF--4 film does not contain any
significant edges in the $2\div4$~nm region where K--shell lines of carbon ions occur.

As for as a microchannel plate is concerned, Fig.~\ref{MCPefficiency} shows typical photon detection
efficiency~\citep{Channelbook}. In our calculations we considered the efficiency in the $2\div7$ nm region
unvarying. Regarding the vignetting of our CCD camera, its influence was not significant.

\subsubsection{Resolving power}
\noindent The \emph{resolving power} $\mathcal{R}$ of a spectrograph can be expressed as the dimensionless
quantity
\begin{equation}
\mathcal{R}=\frac{\lambda}{\delta \lambda}
\end{equation}
where $\delta \lambda=\lambda_1-\lambda_2$ is the difference in
wavelength between
 two adjacent monochromatic spectral lines $\lambda\approx\lambda_1\approx\lambda_2$
 that can be separated and distinguished.

The maximum attainable resolving power for a grating is
\begin{equation}
\mathcal{R_\mathrm{max}}=\frac{2W}{\lambda}
\end{equation}
where $W$ is the width of the grating in the dispersion plane. Because of short wavelengths in the XUV and
X--ray region, the theoretical $\mathcal{R_\mathrm{max}}$ is extremely high. The resolving power is then limited
by other factors than a grating itself. For instance in our grating spectrograph, among main limitations belong
the dimension and location of the entrance slit and detector elements. The geometric width of the image of the
entrance slit (at the plane of a detector) was approximately 30~$\muup$m. In the case of the UF--4 film, this is
the limiting factor since its grain is less than 7 $\muup$m. In the case of MCP and CCD detectors, we must take
into account the convolution of the width of an entrance slit (30~$\muup$m), MCP channel to channel distance (16
$\muup$m), fibre optic channel to channel distance (8.5 $\muup$m), and CCD pixel size (25 $\muup$m in the plane
of MCP). Apart from that, the blurring of a point source takes place also between the MCP and fibre optics and
between fibre optics and the CCD detector. Altogether, we measured the width of actual instrument function of
100 $\muup$m. This width then determines the experimental resolving power.

The experimentally determined resolving powers for several wavelengths and gratings are shown in
table~\ref{resolvingpower}. These resolving powers correspond to the focus area of a~spectrum. In the defocused
part of a spectrum, the resolving power is even smaller and is decreasing with the increasing distance of a
focused line from the Rowland circle.
\begin{table}[h]
\centering
\begin{onehalfspacing}
\begin{tabular}{|l|l|l l|l l|}
\hline \hline
Wavelength & Grating &Film UF--4 & &MCP and CCD& \\
\hline $\lambda=4$ nm &no.~1 & $\delta \lambda=0.03$ nm,& $\mathcal{R}=130$ & $\delta
\lambda=0.15$ nm,& $\mathcal{R}=30$\\
\hline $\lambda=8$ nm &no.~2 & $\delta \lambda=0.03$ nm,& $\mathcal{R}=250$ &  $\delta \lambda=0.15$
nm,& $\mathcal{R}=50$\\
\hline $\lambda=18$ nm &no.~3& $\delta \lambda=0.05$ nm,& $\mathcal{R}=360$ & $\delta \lambda=0.15$
nm,& $\mathcal{R}=120$\\
\hline \hline
\end{tabular}
\end{onehalfspacing} \caption[Resolving power of our spectrograph for some wavelengths, gratings and
detectors]{Resolving power of our spectrograph for some wavelengths, gratings and detectors.}
\label{resolvingpower}
\end{table}

\vspace{-0.3cm}
\subsubsection{Spatial resolution}
\noindent The common way of obtaining the spatial resolution is to give an additional slit perpendicularly to
the entrance slit of a grating spectrograph. Then the spectrum is dispersed in one direction, whereas the second
direction could provide the spatial information of an emitting source. Unfortunately, this technique cannot be
easily used in the case of a~spherical grating. We are thus restricted to three other possibilities:

First, the spectrum from one locality could be obtained by collimating the radiation falling onto the entrance
slit of a spectrograph. In that case, the recorded spectrum contains information from one locality only, to be
more precise along the line of sight.

\begin{figure}[h!]
\centerline{\includegraphics{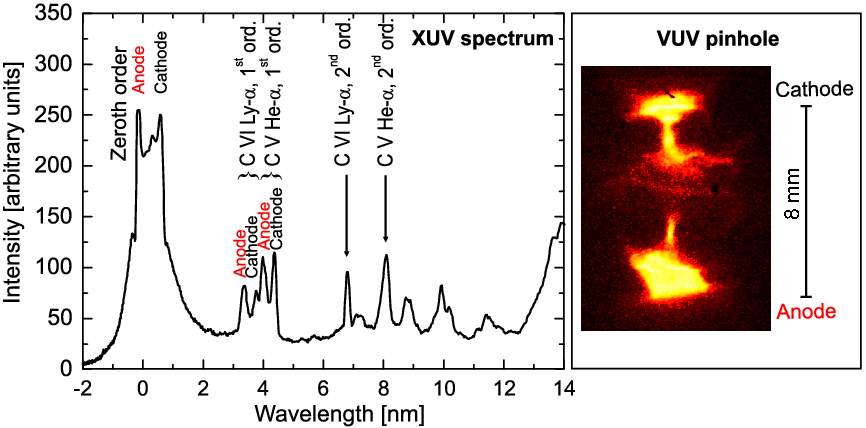}} \caption[XUV spectrum together with a pinhole
image]{XUV spectrum together with a pinhole image at 320 ns, shot no.~040225-2.}\label{spatial}
\end{figure}

Second, we can take use of the fact that there is only one spectral line that is exactly focused. Other recorded
spectral lines do not lie on the Rowland circle and are inherently defocused according to the spatial
extensiveness of an emitting source (see Fig.~\ref{LPS1} on page~\pageref{LPS1}). One exemplary spectrum is
displayed in Fig.~\ref{spatial}. In that shot, the slit of the spectrograph was oriented perpendicularly to the
fibre axis. The second diffraction order of carbon K--shell lines was in the focused area, whereas the first
order of the Ly--$\alpha$ and He--$\alpha$ lines was visibly defocused. In accordance with this defocusing, it
was possible to distinguish two regions radiating in carbon K--shell lines. One was located near the anode and
the other was near the cathode. This result agrees also with the gated VUV pinhole image. The disadvantage of
such procedure is that the spatial profile is combined with the spectrum and hence it cannot be applied to every
case and to every spectral line.

Third, the spatial information of soft X--ray emission can be also acquired by a pinhole camera. When using a
proper choice of filters, a pinhole camera provides both spatial and spectral information of a radiating source.
On the one hand, spectral information of a pinhole camera is not so in--depth as a spectrograph could give, but
on the other hand, 2D spatial resolution is achieved.

In the following subsections, we describe two pinhole cameras, time--integrated X--ray camera and time--resolved
VUV camera, that were employed in our experiments.

%% file: results.tex
\chapter{Experimental Results}

\noindent During the past four years, more than 500 shots have been observed with various diagnostic tools. As
a~result, we obtained a~large number of experimental data and their interpretations. To present these results
from each diagnostic tool separately would overburden this thesis. Instead of that, we decided to put results
from various diagnostics together so as to delineate gross dynamics of our fibre $Z$--pinch. We shall illustrate
this dynamics on carefully chosen individual shots that most evidently show general characteristics of the
pinch.

\begin{figure}[h!]
\centerline{\includegraphics{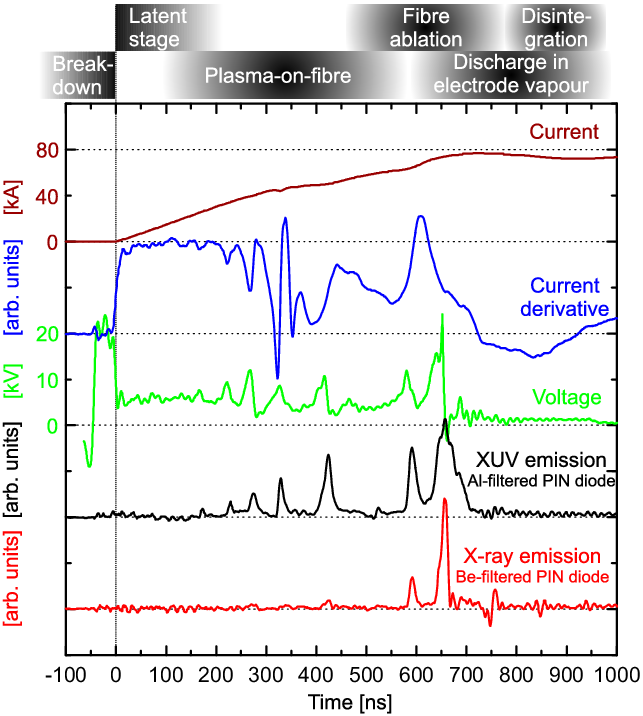}} \caption[Waveforms of current, current derivative, voltage
and PIN diode signals recorded in discharge no.~050128-1]{Waveforms of current, current derivative, voltage and
PIN diode signals recorded in discharge no.~050128-1. A~carbon fibre of 15~$\muup$m diameter and 8 mm length,
conical electrodes made of bronze, charging voltage of 20 kV.} \label{Z150scope}
\end{figure}

To make reading easier, the chapter is divided according to different stages that were distinguished in the
discharge, namely the breakdown, latent stage, plasma\,--\,on\,--\,fibre, fibre ablation, discharge in electrode
vapour and $Z$--pinch disintegration (see the upper part of the graph displayed in Fig.~\ref{Z150scope}). The
time separation between neighbouring phases was chosen somewhat arbitrary since the transition from one stage of
the discharge to another was not sharp and varied from shot to shot. All times described in this chapter refer
to the start of the current when $t=0$.

Typical waveforms recorded in our $Z$--pinch are shown in Fig.~\ref{Z150scope}. In this particular shot, the
$Z$--pinch discharge was initiated from a 15~$\muup$m diameter carbon fibre with the charging voltage of 20 kV.
The interesting and characteristic features that we observed were short pulses detected with filtered PIN
diodes. The short XUV and X--ray pulses were emitted regardless of what charging voltage, material of electrodes
and the diameter of a~carbon fibre we tried. Therefore, in the following sections we would like to elucidate the
origin of these pulses and to describe what happens in each stage of the discharge. The results of each stage
will be also supported by a short interpretation and straightforward discussion, whereas the overall discussion
of our $Z$--pinch is reserved for the next chapter.

\vspace{-0.1cm}
\section{Breakdown}
\begin{figure}[!h]
\centerline{\includegraphics{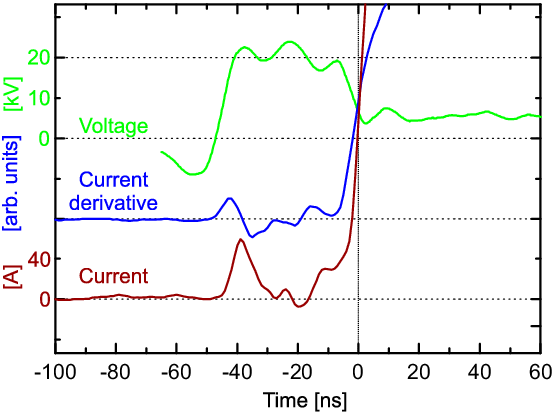}} \caption[Breakdown phase in discharge no.~050128-1]{Breakdown
phase in discharge no.~050128-1. A~15~$\muup$m diameter, 8 mm long carbon fibre, conical electrodes made of
bronze, charging voltage of 20 kV.} \label{breakdown}
\end{figure}

\vspace{-0.2cm} \noindent It is well known that the breakdown phase is very much influenced and determined by
initial physical properties of a $Z$–-pinch load. In the case of a~carbon fibre, the characteristic initial
property of the material is its high resistivity. The measured resistance of a~10 mm long, 15~$\muup$m diameter
fibre was about 7 k$\Omega$ at room temperature. It is so high resistance that a~carbon fibre behaves almost as
an insulator and it takes some time for the breakdown to occur. The delay between the applied voltage pulse and
the start of the current, i.e.~fibre breakdown, is given by the duration of avalanche ionization in strong
electric field that is present in gases/vapours desorbed from the fibre surface. The voltage does not collapse
until a lower resistance and low density plasma is formed around a~fibre. The delay between voltage and current
derivative can be seen in Fig.~\ref{breakdown}. In the case of a 20 kV charging voltage, we obtained the delay
of $50\pm10$ ns.

However, the resistance is not the only fact that influences electrical breakdown. \citet{Sarkisov02} reported
the significant effect of the radial electric field and its polarity. Because of a small fibre diameter and its
microroughness, the radial electric field could exceed the axial electric field by a few orders of magnitude. In
our case\footnote{The part of a~fibre near the anode is positive and chamber walls are negative. We expect the
radial electric field near a~fibre $E_r\approx10^8$ V\,m$^{-1}$.}, the polarity of radial electric field was
positive, i.e.~$E_r>0$. Since the electric field increased the potential barrier for electronic emission from
the fibre surface, the breakdown started from the cathode when the electric field was the lowest. As a~result,
the deposited energy towards the cathode was lower because the direct heating was prevented by a current shunted
from a~fibre.

It is also interesting to note that breakdowns of 15~$\muup$m diameter carbon fibres were quite reproducible,
whereas we had serious problems with 5 $\muup$m tungsten wires (initial resistance of about 25 $\Omega$) as well
as with fibres from deuterated polyethylene (insulator). The reason for that lies neither in resistivity nor in
the polarity of electric field, but most probably in the surface properties of a~carbon fibre. In this
connection we should perhaps add that we saw something like a ``luminescing'' column around a~carbon fibre
before the trigger pulse was applied to a spark--gap\footnote{At that moment, there was a 40 $\muup$A~current
flowing through a~fibre (see the electric scheme in Fig.~\ref{aparaturaschema}).}. We attributed this phenomenon
to a gas/vapour desorbed from the fibre surface. But whatever the explanation is, this phenomenon could
effectively facilitate the breakdown process and also prevent higher energy deposition into a~fibre core. We
measured that the Joule heat produced before the voltage collapse was about 20 mJ. It is much lower than the
energy required for atomization of a 15~$\muup$m carbon fibre which is about 200 mJ. Moreover, the energy of 20
mJ was deposited rather to ambient vapour than to a~fibre. Hence, the diameter of a~carbon fibre remained almost
unchanged after the breakdown.

\section{Latent Stage}
\noindent The ``latent'' stage of the discharge represents the time after the breakdown and before the
appearance of XUV pulses, i.e.~from 0 ns to about 150 ns for 15 $\muup$m diameter fibres.

\begin{figure}[!h]
\centerline{\includegraphics{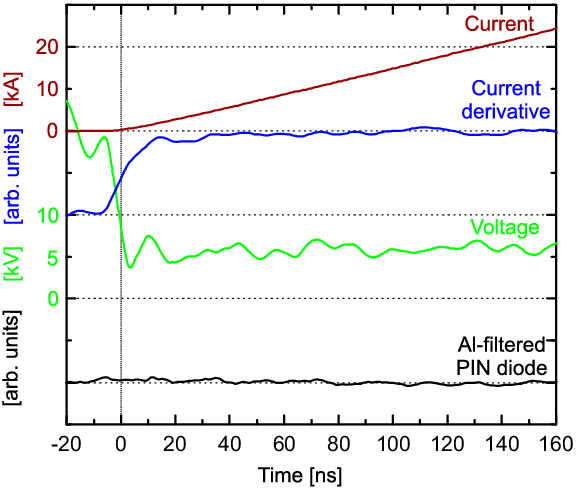}} \caption[Latent phase in discharge no.~050128-1]{Latent phase in
discharge no.~050128-1. A~15~$\muup$m diameter, 8 mm long carbon fibre, conical electrodes made of bronze,
charging voltage of 20 kV.} \label{latent}
\end{figure}

This stage has not been explored in much detail. The main reason was the lack of serious interest in this stage.
It was supported by the fact that this phase was difficult to detect. During the first 100 nanoseconds the
plasma emitted a small amount of radiation and the electron density was too low to be detected by the schlieren
system. Besides that, all the diagnostic tools were triggered mostly by the discharge breakdown and then we were
not able to obtain gated images before 150 ns. Our knowledge of the latent stage came from the non--systematic
investigation only. However few experimental data we obtained, we do have certain notion about the plasma
development after the breakdown.

According to theoretical predictions\footnote{The line density of a plasma (see the line density of
$8\times10^{17}$ m$^{-1}$ on page~\pageref{density}) is of the order of the critical density
$N_\mathrm{c}=1.3\times10^{18}A(1+\bar{z})/\bar{z}^3$~\citep{Haines96b}, where $A$ stands for the atomic mass
number and $z$ for the ionic charge. Therefore, microinstabilities could be triggered (see
page~\pageref{turbulences}). Consequently, the plasma pressure exceeds the magnetic pressure and the plasma
expands. But even if the Bennett relation holds, the current would have to follow the Haines--Hammel
curve~\citep{Hammel83} to prevent the expansion. As our current rise time of about $10^{11}$ A$\,$s$^{-1}$ is
slow, a plasma column has to expand early after the breakdown.} and experimental observations~\citep{Lorenz98}
we can assume that a low density plasma expanded to a radius of several millimetres. The expansion velocity was
limited by the sound speed that is typically $10^4$ m$\,$s$^{-1}$. The expansion was most likely slowed down by
an increasing current. At which time this happened depended on the current rise and the properties of a plasma
column.

Further information could be inferred from Fig.~\ref{latent}, especially from the voltage and $\diff I/\diff t$
waveforms. It is evident that the plasma development was not so dramatic as in the ``plasma\,--\,on\,--\,fibre''
stage (cf.~Fig.~\ref{Z150scope}). Nevertheless, it is still possible that similar processes might occur in both
stages, only less powerful in the latent stage due to smaller current. When we subtracted the inductive
contribution, the voltage across the plasma was less than 1 kV. For 10 kA~current we obtained $<0.1$~$\Omega$
resistance.

\begin{figure}[!p]
\centerline{\includegraphics{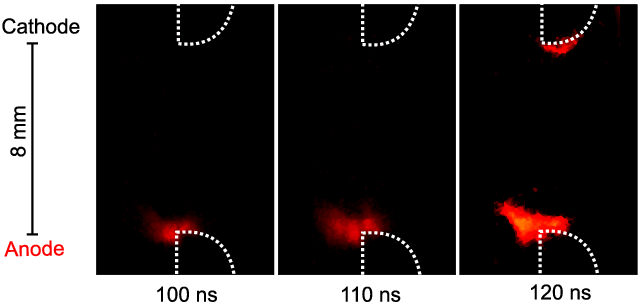}} \caption[VUV pinhole images recorded in the latent phase]{VUV
pinhole images recorded in discharge no.~020215-3. A~carbon fibre of a 15~$\muup$m diameter and 8~mm length,
``bisectional'' electrodes made of brass, charging voltage of 20~kV.} \label{0202153}
\end{figure}

\begin{figure}[!p]
\centerline{\includegraphics{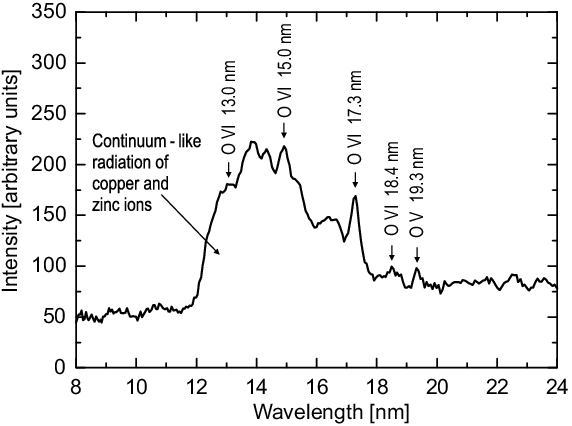}} \caption[XUV spectrum detected at 110 ns, shot
no.~011016-1]{XUV spectrum detected at 110 ns, shot no.~011016-1. A~80 $\muup$m diameter, 8 mm long carbon
fibre, ``bisectional'' electrodes made of brass, charging voltage of 23 kV.} \label{011016sp}
\end{figure}

Unfortunately, first images were recorded late in the latent stage. Fig.~\ref{0202153} shows that at 100 ns most
of the VUV radiation came from the vicinity of electrodes, especially the anode. This observation agrees with
the XUV spectrum displayed in Fig.~\ref{011016sp} where the spectral lines of oxygen, copper and zinc ions were
identified. The electron temperature estimated from Li-- and Be--like ions O VI was $15\div25$ eV. The
temperature measurement was done by the FLY code (see appendix~\ref{appendixFLY}). On the one hand, the
temperature was below 25 eV since the O VI spectral line at 11.6 nm (2s--4p and 2p--5d transitions) was too weak
in comparison with the spectral line at 17.3 nm (2p--3d transition). But on the other hand, because the O VI
spectral line at 17.3 nm was at least five times stronger than the transition 2s2p--2s3d of Be--like oxygen O V,
the temperature was higher than 15 eV. The merging of spectral lines and continuum--like radiation caused that
it was problematic to separate spectral lines. In addition, we do not know what the plasma density was. That was
the reason why more accurate estimation of the temperature was not achieved.

\section[Plasma--on--Fibre]{Plasma--on--Fibre\footnote{This term was derived by analogy from the expression
``plasma\,--\,on\,--\,wire'' which was used by \cite{Wessel92} to name the imploding aluminium-plasma jet onto a
coaxial wire. However, the implosion onto a metal wire differs from the implosion onto a~fibre with a very low
conductivity, and so we decided to use the term ``plasma\,--\,on\,--\,fibre'' instead of
``plasma\,--\,on\,--\,wire''.}}

\input{pof}

\section{Fibre Ablation}
\label{ablace} 
\begin{figure}[h!]
\centerline{\includegraphics{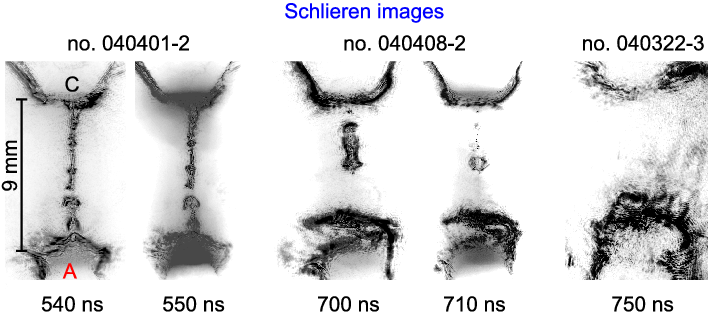}} \caption[Schlieren images that recorded the ablation of
a~fibre]{Schlieren images that recorded the ablation of a~fibre. A~15~$\muup$m diameter and 9 mm long carbon
fibre, conical electrodes made of brass, 20 kV charging voltage.} \label{ablation}
\end{figure}

\noindent Approximately 500 ns after the breakdown, gaps in the fibre indicated that several
parts of a~fibre had been already ablated, especially the part near the anode. In Fig.~\ref{ablation}, one can
see the $m=0$ behaviour which most likely increased the ablation rate. In the case of conical electrodes, the
central part of a~fibre persisted up to 700~ns and then formed a dense and cold plasma ``island''. The entire
length of a~fibre was seen totally ablated 750~ns after the current breakdown. After that, the radiation was
still detected with the Al-- and Be--filtered PIN diodes.

 Something like a shock--wave was seen near the
anode in shot no.~040408-2 in figure~\ref{ablation}. The following figure~\ref{shock} shows a radiating, sharp
density gradient that was propagating from the cathode.

\begin{figure}[h!]
\centerline{\includegraphics{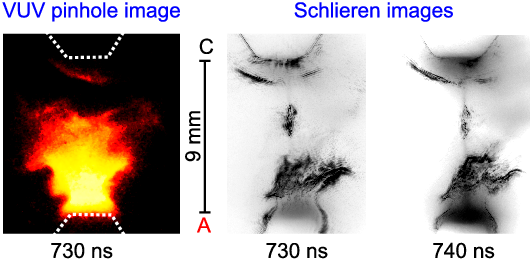}} \caption[Sharp density gradient near the cathode]{Sharp density
gradient near the cathode in shot no.~040408-1. A~15~$\muup$m diameter and 9~mm long carbon fibre, conical
electrodes made of brass, 20 kV charging voltage.} \label{shock}
\end{figure}

\section{Discharge in Electrode Vapour}
\begin{figure}[h!]
\centerline{\includegraphics{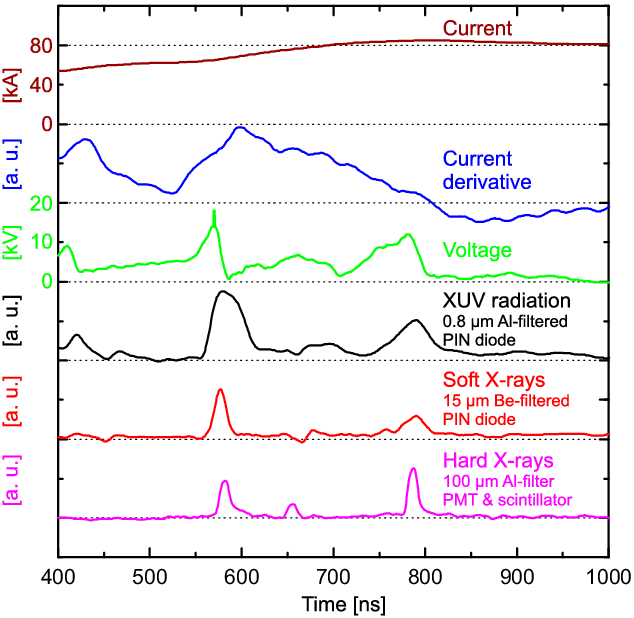}} \caption[Discharge in electrode vapour]{Discharge in
electrode vapour, shot no.~050201-3. A~15~$\muup$m diameter and 8 mm long carbon fibre, conical electrodes made
of brass, 20 kV charging voltage.} \label{hxr_scope}
\end{figure}

\noindent Fig.~\ref{hxr_scope} shows typical waveforms of the current derivative $\diff I/\diff t$, voltage, PIN
diode signals and PMT signal between 400 and 1000 ns.

\vspace{0.2cm}
\vspace{0.15cm}

In this phase, emitted pulses with various FWHMs were accompanied with rises of voltage\footnote{We expect that
the electrode vapour spread over a~significant part of the chamber. This could reduce the current flowing
through the Rogowski coil and consequently stop the sensitive detection of fast changes of a plasma impedance.
As a~result, the coincidence between an X--ray pulse and a dip in $\diff I/\diff t$ disappeared after 400 ns.}
up to 30 kV (similarly to the plasma\,--\,on\,--\,fibre phase). However, in comparison with the first 500 ns of
the discharge, these pulses were more energetic. The energy of $>1$ keV photons exceeded 10 mJ in one pulse and
the abundance of $>6$ keV photons was detected behind a 100 $\muup$m thick Al filter\footnote{We detected
radiation filtered also with a 750 $\muup$m thick Al foil. Thus the energy of several photons was higher than 10
keV (see Fig.~\ref{ALfilter}).}. This fact brought us naturally to the question of what led to the emission of
X--ray pulses. It was evident that the important role was played by the material of electrodes. The following
subsections shall add to the explanation of that.

\subsection{Evaporation of material from electrodes}
\begin{figure}[h!]
\centerline{\includegraphics{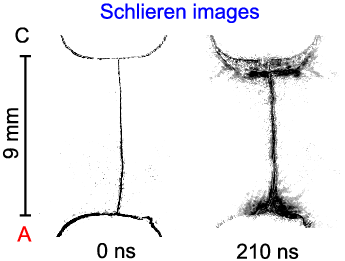}} \caption[Material evaporated from electrodes in shot
no.~021024-2]{Material evaporated from electrodes, shot no.~021024-2. A~15~$\muup$m diameter and 9 mm long
carbon fibre, bisectional electrodes made of brass, 20 kV charging voltage.} \label{021024-2}
\end{figure}

\noindent  The emission from the electrodes as well as the expansion of the electrode vapour were apparent early
in the discharge (see Fig.~\ref{0202153} and~\ref{021024-2}). With the growing time the influence of material
from electrodes was increasing. The observed velocity of evaporated materials from electrodes approached $10^4$
m\,s$^{-1}$, therefore the discharge occurred more or less in the vapour of electrodes later than~400~ns after
the current breakdown. This expectation was proven experimentally when we observed that spectral lines of copper
and tin ions started to dominate in XUV spectra after 500 ns (cf.~Fig.~\ref{040402-1}). By the way, the peak of
tin ions at 13.5 nm has become important for EUV lithography.

\subsection{Magnetic pinching of evaporated material}
\begin{figure}[h!]
\centerline{\includegraphics{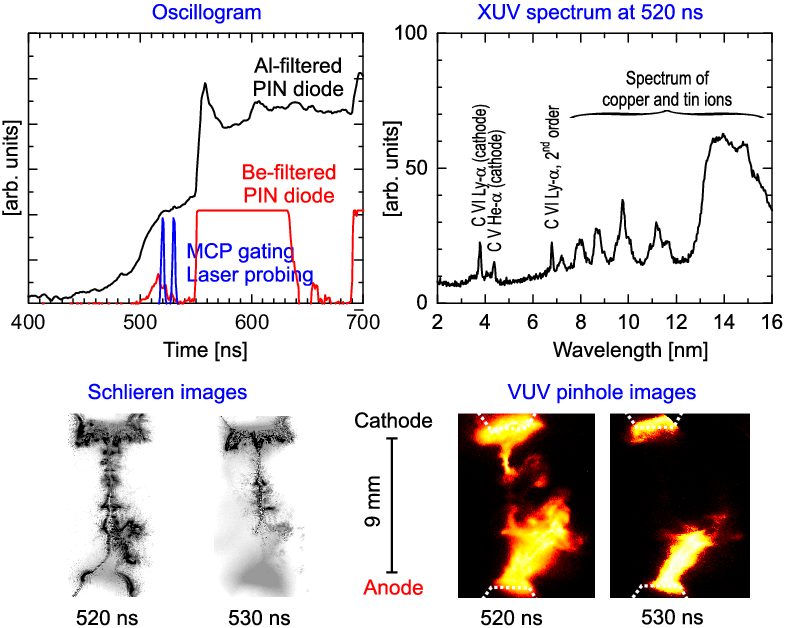}} \caption[Implosion of material evaporated from
electrodes]{Implosion of material evaporated from electrodes, shot no.~040402-1. A~15~$\muup$m diameter and 9 mm
long carbon fibre, conical electrodes made of bronze, 20 kV charging voltage.} \label{040402-1}
\end{figure}

\noindent Fig.~\ref{040402-1} displays the off-axis implosion of electrode material near the anode. The
implosion velocity was about $3\times10^4$ m\,s$^{-1}$. At about 560 ns, i.e.~at the time when we could expect
the end of the implosion, an X--ray pulse was generated. Schlieren images indicated that more mass participated
in the implosion. The electron line density above $10^{20}$~m$^{-1}$ was two orders of magnitude higher than in
the plasma\,--\,on\,--\,fibre phase.

\newpage
\subsection{Development of plasma instabilities}
\begin{figure}[h!]
\addtolength{\belowcaptionskip}{-0.3cm} \addtolength{\belowcaptionskip}{-0.3cm}
\centerline{\includegraphics{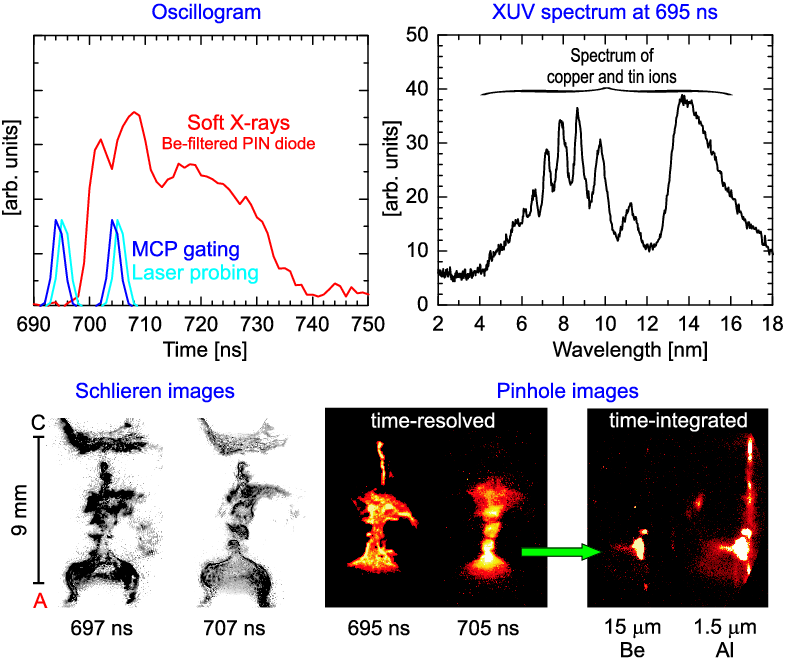}} \caption[$Z$--pinch plasma during the onset of an~intensive X--ray
pulse]{$Z$--pinch plasma during the onset of an~intensive X--ray pulse, shot no.~040414-3. A~carbon fibre of
15~$\muup$m diameter and 9 mm length, conical electrodes made of bronze, 20 kV charging voltage.} \label{rtg}
\end{figure}

\noindent The final stage of the implosion was detected in shot no.~040414-2 (see Fig.~\ref{rtg}). In this
particular shot, images were obtained just before and during the X--ray pulse of a~2~ns rise time. The schlieren
and pinhole images caught the development of the $m=0$ instability with three ``blobs''. Furthermore, the most
intensive radiation (as could be seen in the time--integrated pinhole images) came from one off--axis elongated
bright spot which corresponded to the constriction of one $m=0$ neck. These bright spots occurred mainly near
the anode and their number was shot\,--\,to\,--\,shot dependent.

At this point we shall turn to the plasma\,--\,on\,--\,fibre stage. In Fig.~\ref{rtg}, the entire length of
a~carbon fibre is displayed in the time--integrated Al--filtered pinhole image. The radiation from a~fibre was
absorbed by the Be-filter. Such result agrees with the observation of XUV pulses (and no X--ray pulses) during
the plasma\,--\,on\,--\,fibre phase. Next, the spatially non--uniform exposure along the fibre gave another
evidence that the radiation came from that part of a~fibre where an $m=0$ instability developed
(cf.~page~\pageref{pofinstabilities}).

\subsubsection{Temperature of bright spots}
\vspace{-0.2cm} \noindent The plasma parameters within bright spots has not been explored in detail. The density
of the bright spot displayed in Fig.~\ref{rtg} can be only guessed. As far as the temperature is concerned, the
X--ray pinhole camera and photomultiplier signal evidently demonstrated that more energetic photons were emitted
whenever bright spots were observed. Nevertheless, these photons could be produced by ``non--Maxwellian''
electrons accelerated in high electric field\footnote{The plasma voltage was up to 30 kV. The peak voltage could
exceed the charging voltage of 20 kV because the large inductance of the driver $L_0$ kept current flowing
through a plasma when the plasma resistance rapidly increased.}, whereas the plasma temperature could be low.
The only thing which can be used is the temperature estimated from an XUV spectrum.

\begin{figure}[h!]
\addtolength{\belowcaptionskip}{-0.6cm} \centerline{\includegraphics{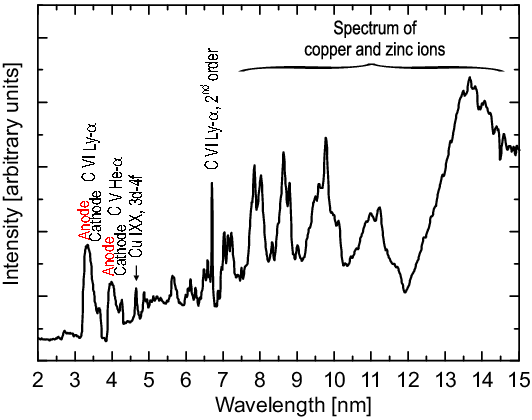}}
\caption[Time--integrated XUV spectrum, shot no.~050427-2]{Time--integrated XUV spectrum, shot no.~050427-2.
A~carbon fibre of 15~$\muup$m diameter and 9 mm length, conical electrodes made of brass, 20 kV charging
voltage.} \label{050427-2}
\end{figure}

Fig.~\ref{050427-2} presents the time--integrated spectrum in the 2\,--\,15 nm region. In this spectrum, the
first order of carbon K--shell lines did not lie on the Rowland circle and therefore the Ly--$\alpha$ and
He--$\alpha$ lines of carbon ions were defocused. This large defocusing indicates that these K--shell lines
originated from an extensive source. To be more specific, although the spectral lines in question were emitted
from the entire length of a~fibre, it is particularly true for the region near the anode. Contrary to these
spectral lines, the line at about 4.7 nm was relatively narrow and therefore it must have originated from a
smaller region. This spectral line was identified as the transition 3d $^2$D -- 4f $^2$F of a Na--like copper
ion Cu IXX at 4.74 nm. It means that the Na--like stage of copper occurred in a~plasma and
thus\footnote{According to Eq.~\ref{odhadteploty} on page~\pageref{odhadteploty} the temperature was estimated
from Cu XVIII ionization energy of about 650 eV~\citep{Attwood99}.} the temperature was well above 100 eV. The
peak temperature could be of course even higher.

\subsection{Experimental data interpretation}
\vspace{-0.2cm}
 \noindent Now, we shall discuss occurrence of X--ray pulses accompanied with rises of voltage.
It could be done similarly to the interpretation of the plasma\,--\,on\,--\,fibre stage.

The peak current at this phase was about 80 kA, therefore the increase of voltage above 20 kV leads to the
$R_\mathrm{P}+\dot{L}_\mathrm{P}$ term
 of about 0.3 $\Omega$. We believe that such value was achieved within plasma
instabilities that were also presumably sources of energetic photons. These X--ray photons were  generated for
several reasons. Firstly, instabilities could develop substantially because the fibre had been already
ablated\footnote{This was also the reason for off-axis implosion.}. Secondly, the current at this phase was
higher than in the plasma\,--\,on\,--\,fibre stage. Thirdly, the plasma contained ions with higher atomic number
$Z$ than carbon has. As a~result, not only continuum emission but also line radiation occurred in the X--ray
region.


\section{$Z$--pinch Disintegration}
\noindent The disintegration of the $Z$--pinch occurred at about 1 $\muup$s. At that time, discharge chamber
walls were most likely reached by ejecta of a plasma. Also, X--ray and UV flux emitted from a~$Z$--pinch could
envelop the surface of the insulator with a plasma. Due to that, the significant part of the current was flowing
outside the ``$Z$--pinch region'' and the decreasing amount of radiation was detected (see
Fig.~\ref{visiblePIN}).

\section{Energetics of $Z$--pinch}
\noindent The previous sections have described individual phases of our $Z$--pinch, from the breakdown to its
disintegration. In this section we shall deal with the overall energetics of our fibre $Z$--pinch. After that we
will discuss the influence of the electrodes and finally we will conclude this chapter by saying a few words
about differences between fibres of various diameters.

The measurement of plasma voltage and current enabled us to estimate the energy deposited into our $Z$--pinch.
The deposited energy was spent on the ablation/evaporation of a~fibre/electrodes, ionization and heating of a
plasma, thermal diffusion, advection, etc. Only a small part of the deposited energy was radiated from the
plasma. The overall energetics of our fibre $Z$--pinch is summarised in Tab.~\ref{energetika}.
\begin{table}[h!]
\centering
\begin{onehalfspacing}
\begin{tabular}{|l|l|}
\hline \hline
Physical quantity& Energy   \\
\hline
Energy stored in a capacitor bank $\frac{1}{2}C_0U_0^2$ & 600 J ($U_0=20$ kV)\\
 \hline
Energy deposited into a plasma $\int_0^\infty (U_\mathrm{P}-L_\mathrm{P}\dot{I})I\diff t$ &  $50\div100$ J\\
\hline
Energy required for the atomization of a~fibre & 200 mJ\\
\hline
Energy required for the fibre ionization to the He--like stage & 4.5 J\\
\hline
Energy required for the fibre ionization to bare nuclei & 30 J\\
\hline
Emitted energy &  $\approx5$ J\\
\hline
Energy emitted in the XUV region (30--70 eV) &  $\approx 300$ mJ\\
\hline
Energy emitted in soft X--rays ($>1$ keV) & $\approx 20$ mJ \\
\hline\hline
\end{tabular}
\end{onehalfspacing} \caption[Energetics of our fibre $Z$--pinch]{Energetics of our fibre $Z$--pinch.}
\label{energetika}
\end{table}

\begin{figure}[h!]
\addtolength{\belowcaptionskip}{-0.3cm} \centerline{\includegraphics{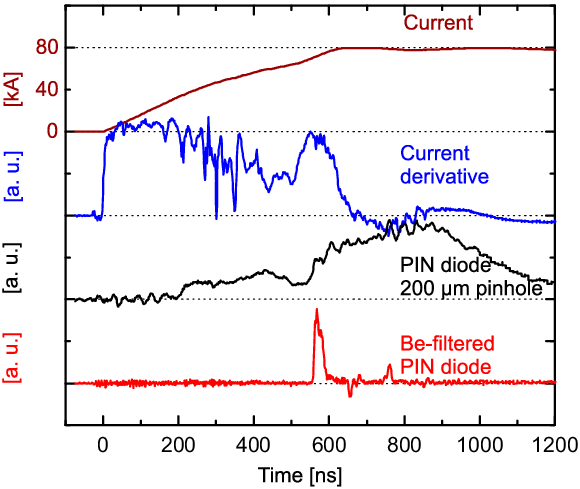}} \caption[PIN diode
signal together with the current waveform and signal of a Be-filtered PIN diode, shot no.~050512-1]{PIN diode
signal (200 $\muup$m pinhole used instead of a filter) together with the current waveform and signal of a
Be-filtered PIN diode, shot no.~050512-1. A~15~$\muup$m diameter and 9 mm long carbon fibre, conical electrodes
made of brass, 20 kV charging voltage.} \label{visiblePIN}
\end{figure}

The total emitted energy was estimated from the PIN diode used with a 200 $\muup$m pinhole, which was this time
used instead of a filter (see the blue line in Fig.~\ref{responsivities}). The results from shot no.~050512-1
are presented in Fig.~\ref{visiblePIN}.

It shows that most of the energy was radiated in less energetic photons during the long--lasting emission. This
emission was increasing with the rising current and was not so sensitive to the fast changes in $Z$--pinch
dynamics as XUV pulses were. Surprisingly, this long--lasting emission overlapped XUV and X--ray pulses which
one could expect to be more apparent.

\begin{figure}[h!]
\addtolength{\belowcaptionskip}{-0.3cm}
 \centerline{\includegraphics{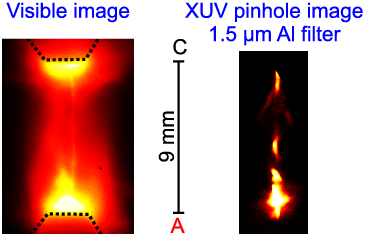}}
\caption[Time--integrated visible and X--ray images, shot no.~040422-1]{Time--integrated visible and X--ray
images, shot no.~040422-1. A~15~$\muup$m diameter and 9 mm long carbon fibre, conical electrodes made of bronze,
20 kV charging voltage.} \label{visible}
\end{figure}

As regards places of the long--lasting emission origin\footnote{Time--integrated visible images were recorded by
the schlieren system when the Nd:YAG laser did not start up.}, they are displayed in the Fig.~\ref{visible}.
Clearly, low energy photons were emitted mainly from regions near electrodes, whereas XUV photons originated
from a~fibre.

\section{Influence of Electrodes}
\begin{figure}[h!]
\centerline{\includegraphics{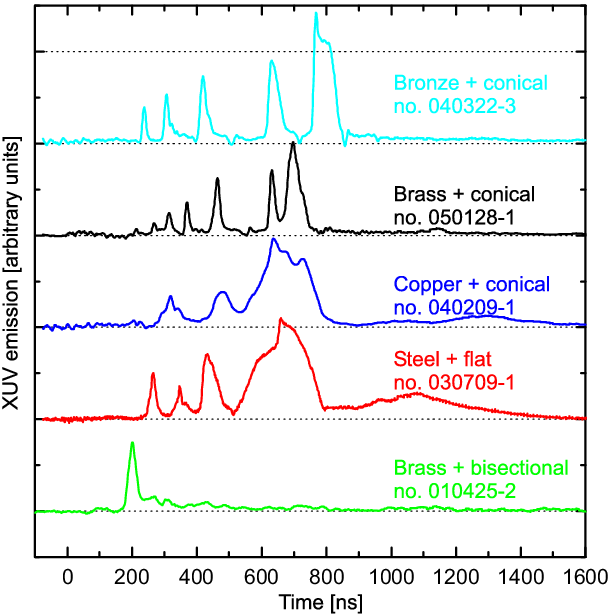}} \caption[PIN diode signals for different materials and
shapes of electrodes]{Al filtered PIN diode signals for different materials and shapes of electrodes.}
\label{electrodesXUV}
\end{figure}

\noindent Describing our fibre $Z$--pinch, it is impossible to leave out the influence of electrodes on the
plasma dynamics. For instance, Fig.~\ref{electrodesXUV} shows PIN diode signals for various electrodes. One can
see how the number, intensities, and widths of XUV pulses changed with the choice of electrodes. In this section
we present the influence of material and shape of electrodes.

\begin{figure}[t!]
\centerline{\includegraphics{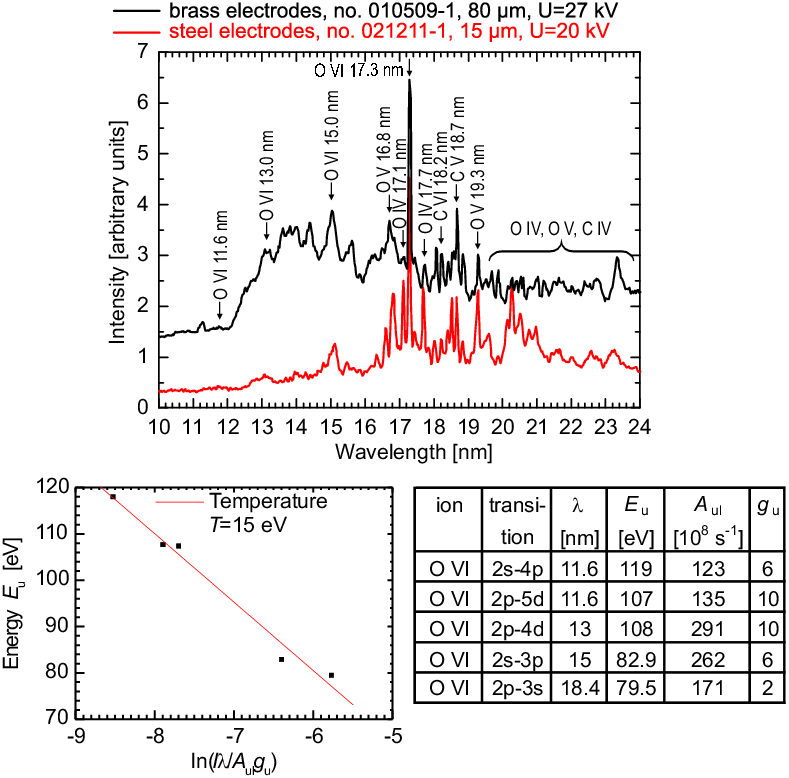}} \caption[Comparison of two time--integrated XUV
spectra when brass and steel electrodes were used]{Comparison of two time--integrated XUV spectra when brass and
steel electrodes were used. Temperature estimated from spectral lines of O VI ions in shot. no 021211-1.}
\label{electrodeSPECTRUM}
\end{figure}

As far as the choice of the electrode material is concerned, it influenced mainly emission spectra. The
difference between two XUV spectra in the case of brass and stainless steel electrodes is shown in
Fig.~\ref{electrodeSPECTRUM}. In the 10--24 nm spectral region the spectral lines of Li-and Be-like oxygen ions
were dominant. The intensities of O VI lines corresponded to the Boltzmann distribution of excited states. The
appropriate ``integral'' electron temperature was 15 eV (cf.~Fig.~\ref{electrodeSPECTRUM} and Eq.~\ref{ratio} on
page~\pageref{ratio}). The lines of H- and He-like carbon ions were also present. Some of the lines corresponded
to higher--$Z$ ions. In the case of brass electrodes, the characteristic feature was continuum-like radiation
between 12 and 18 nm. After replacing brass electrodes by stainless steel ones, the continuum in question
disappeared. Hence, the material of electrodes determined (i) the spectral region where the emission occurred
and consequently (ii) intensities and widths of XUV pulses. We did not, however, observed any significant change
of gross $Z$--pinch dynamics with the choice of electrode material. It was rather the shape of electrodes what
seemed to affect deeply the discharge.

\begin{figure}[h!]
\addtolength{\belowcaptionskip}{-0.2cm} \centerline{\includegraphics{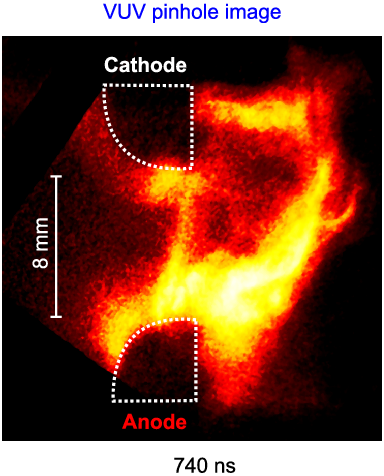}} \caption[VUV
pinhole image in the discharge with bisectional electrodes made of stainless steel]{VUV pinhole image in the
discharge with bisectional electrodes made of stainless steel, shot no.~030331-3. A~15~$\muup$m diameter and 8
mm long carbon fibre, 25 kV charging voltage.} \label{bisectional}
\end{figure}

The shape of electrodes determined not only intensities, widths and number of XUV pulses but it also influenced
the zipper effect, fibre ablation, X--ray emission, etc. We found out, for example, that a~fibre was not fully
ablated and there were no X--ray pulses observed in the case of ``bisectional'' electrodes. The reason for that
could be the non--symmetry of the electrodes that caused the early deflection of a coronal plasma from a~fibre
to chamber walls (cf.~Fig.~\ref{bisectional}).

In the case of conical electrodes, a~fibre was ablated and X--ray pulses were observed. Nevertheless, as it has
been mentioned above, the non--symmetric implosion occurred separately at the anode, at the cathode, and at the
centre, and thus a~higher number of pulses was observed.

\vspace{-0.3cm}
\section{Various Fibre Diameters}
\vspace{-0.1cm} \noindent Most of our results presented above came from experiments that were carried out with
carbon fibres of a 15~$\muup$m diameter. Similar results\footnote{As examples, we could mention the occurrence
of XUV pulses, dips in $\diff I/\diff t$, and plasma\,--\,on\,--\,fibre.} were observed also with 6 $\muup$m
and~80 $\muup$m diameter fibres. As regards differences between fibres,
 the discharge with a 6~$\muup$m diameter fibre was quite irreproducible,
 K--shell emission of carbon ions was less
intensive and the influence of electrode material seemed to be more apparent than with a 15~$\muup$m diameter
fibre. In the case of a 80~$\muup$m diameter fibre, XUV pulses lasted longer and the fibre was probably not
fully ablated. However, the comparison between different fibre diameters was dependent on the shape of
electrodes. In addition, shot\,--\,to\,--\,shot variation prevented us from detail quantitative comparison. The
discussion of this shot\,--\,to\,--\,shot variation goes beyond the scope of our experiment because we were
mainly focused on the basic description of $Z$--pinch dynamics and it was not our ambition at that time to study
and, by extension, to increase the discharge reproducibility.

%% file: pof.tex
\begin{figure}[!h]
\centerline{\includegraphics{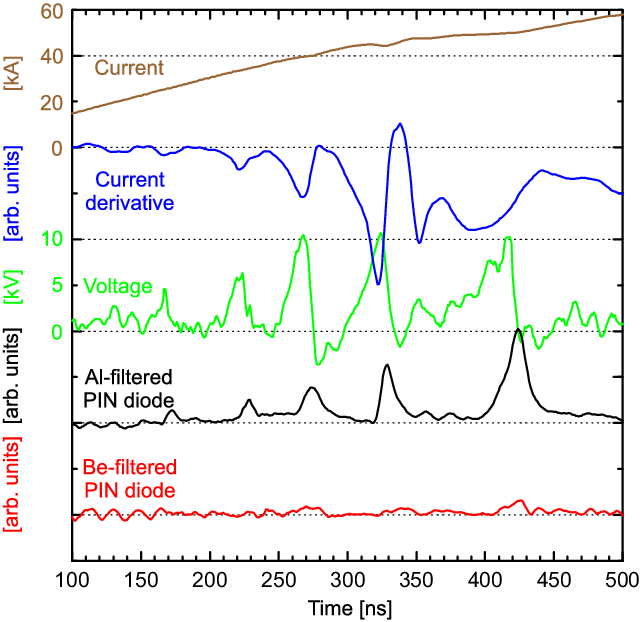}} \caption[Plasma\,--\,on\,--\,fibre phase in discharge
no.~050128-1]{Plasma\,--\,on\,--\,fibre phase in discharge no.~050128-1. A 15~$\muup$m diameter and 8 mm long
carbon fibre, conical electrodes made of bronze, charging voltage of 20 kV.} \label{pof}
\end{figure}

\noindent Fig.~\ref{pof} shows typical waveforms of the current derivative $\diff I/\diff t$,
voltage\footnote{In this case, we subtracted the induction contribution of the voltage. It means
$U^{'}_\mathrm{P}=U_\mathrm{P}-L_\mathrm{P}\dot{I}=(R_\mathrm{P}+\dot{L}_\mathrm{P})I$. To be more precise, we
had to subtract also the induction contribution of the electrodes and cable connected with the voltage probe.}
and PIN diode signals between 100 and 500 ns.

This phase was characterised by short XUV pulses of $10\div30$ ns FWHMs\footnote{Despite the fact that the
shot\,--\,to\,--\,shot variation resulted in a~different number of pulses, their intensities, and widths, it was
possible to identify characteristic features of our fibre $Z$--pinch.}. When the collimated PIN diode was used,
the pulse width was sometimes smaller than~3 ns. The first XUV pulses appeared at about 150 ns. The intensity of
these XUV pulses was increasing with the growing time. Because the signal of the Be-filtered PIN-diode was very
small, the energy of detected photons was mainly in the $20\div70$ eV region (cf.~spectral responsivity of
Al-filtered PIN diode in Fig.~\ref{responsivities} on page~\pageref{responsivities}). For instance, the energy
of $20\div70$ eV photons emitted during the last XUV pulse in Fig.~\ref{pof} was about 100 mJ, whereas the
energy of $>1$ keV photons approached 1 mJ.\label{energyXUV}

The notable feature of the radiation was the strong correlation with the voltage and the circuit $\diff I/\diff
t$. The gradual fall of $\diff I/\diff t$ as well as the growth of the voltage up to 10 kV can~be clearly seen
before each pulse in Fig.~\ref{pof}. These phenomena were typical for our fibre $Z$--pinch regardless of the
diameter of fibre or the shape and material of electrodes that were used. Hence it was our primary interest to
find out how it happened that short XUV pulses occurred.

\subsection{Implosion}
\begin{figure}[h!]
\centerline{\includegraphics{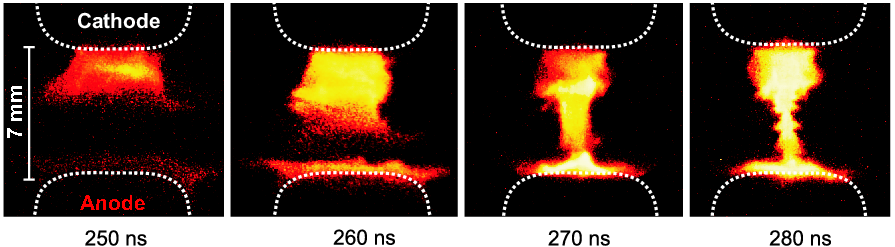}} \caption[VUV pinhole images detected in shot no.~030828-1. The
collapse of a coronal plasma onto a~fibre with the velocity of $2\times10^5$ m$\,$s$^{-1}$]{VUV pinhole images
detected in shot no.~030828-1. The collapse of a coronal plasma onto a~fibre with the velocity of $2\times10^5$
m$\,$s$^{-1}$. The last image corresponded to the peak of the 20~ns XUV pulse. A 15~$\muup$m diameter and 7 mm
long carbon fibre, electrodes made of steel, charging voltage of 20 kV.} \label{imploze1}
\end{figure}

\noindent
 The only diagnostic tool which provided images of what happened before XUV pulses was the gated VUV
pinhole camera. For example, the behaviour before the first XUV pulse is displayed in Fig.~\ref{imploze1}. In
this shot, during the gradual fall of $\diff I/\diff t$, VUV images show the relatively stable\footnote{The
implosion seemed to be stable. The development of instabilities was observed first at the end of the implosion,
during the stagnation. This stability could be ascribed to snowplough stabilization (see page
\pageref{snowplow}). However, the skin--depth of several millimetres was comparable with the radius of the
imploding plasma. Thus a diffuse $Z$--pinch was formed and ``typical'' snowplough did not occur.} implosion of
the coronal plasma onto a~fibre. The implosion velocity approached the value of $2\times10^5$~m$\,$s$^{-1}$. The
compression ratio was about 10. Because the compression caused the increase of the plasma inductance and
resistance, we observed the drop of $\diff I/\diff t$.

\subsubsection{Zipper effect}
\begin{figure}[h!]
\centerline{\includegraphics{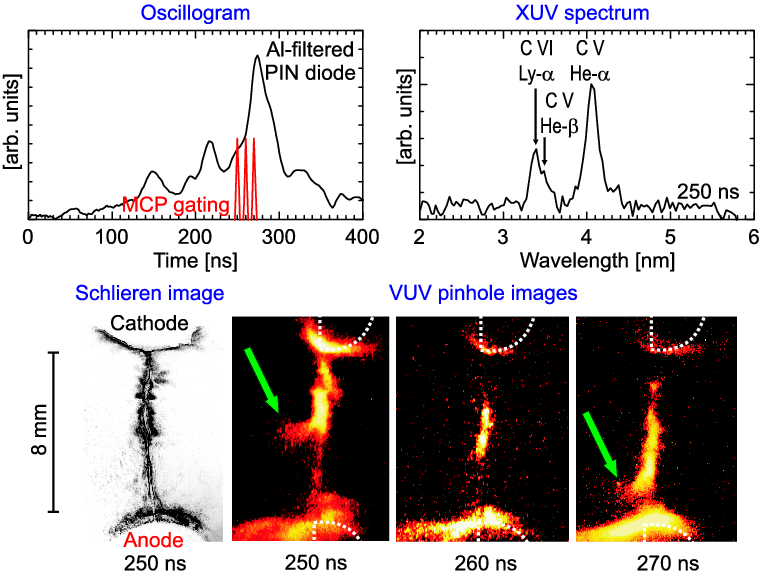}} \caption[Zippering from the cathode with the velocity of
$10^5$ m$\,$s$^{-1}$]{Zippering from the cathode with the velocity of $10^5$ m$\,$s$^{-1}$, shot no.~030321-1. A
15~$\muup$m diameter and 8 mm long carbon fibre, bisectional electrodes made of steel, charging voltage of
20~kV, schlieren sensitivity $n_\mathrm{e}\approx5\times 10^{18}$ cm$^{-3}$. } \label{imploze2}
\end{figure}

\noindent In most cases the implosion was non--symmetric\footnote{It could be the consequence of the polarity
effect during the breakdown. As it was mentioned above, the direction of the radial electric field influences
the energy deposition~\citep{Sarkisov02}. According to this model, the implosion should stagnate first close to
the cathode.}. As a~result, the radiation was ``zippering" along the fibre with the velocity of up to
$1\times10^5$~m$\,$s$^{-1}$. This phenomenon could be seen in the images in Fig.~\ref{imploze2}. The zipper was
spreading mainly from the cathode, however, zippering from the anode was also observed.
\begin{figure}[h!]
\centerline{\includegraphics{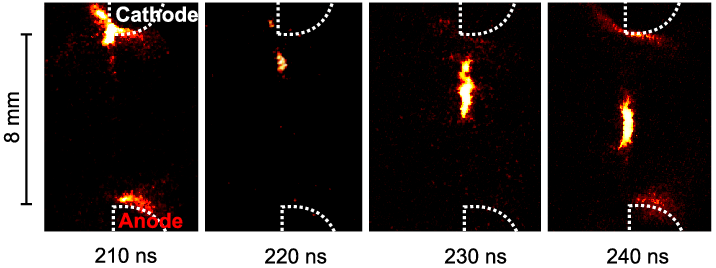}} \caption[XUV pinhole images in shot
no.~021031-1. Zippering with the velocity of $10^5$ m$\,$s$^{-1}$]{Zippering with the velocity of $10^5$
m$\,$s$^{-1}$. The last image corresponded to the peak of the 20 ns XUV pulse. XUV pinhole images (0.8 $\muup$m
thick Al filter was used) in shot no.~021031-1. A carbon fibre of 15~$\muup$m diameter and 8 mm length,
bisectional electrodes made of brass, 12 $\muup$F capacitors, charging voltage of 20 kV.} \label{imploze3}
\end{figure}

At this point we would like to point out that it was necessary to detect a low density corona in order to
recognise the implosion. Therefore, we have never seen the implosion in schlieren images. As regards the pinhole
camera, when an Al--foil was used to filter radiation (Fig.~\ref{imploze3}), we observed only the zipper effect
without any evidence of the implosion. Also, we would like to draw the attention to the fact that in this shot
the radiation from a~fibre exceeded the emission from regions near electrodes. That was observed although an
Al--filter did not permit K--shell spectral lines of carbon ions.

\subsubsection{Plasma density}
\label{density} \noindent The plasma density during the implosion could be derived from the implosion velocity
which approached $2\times10^5$~m\,s$^{-1}$. This value should be of the order of the Alfv\'{e}n velocity,
i.e.~$v_{\mathrm{imp}}\approx\sqrt{\mu I^2_{0}/\pi N_\mathrm{i} M_\mathrm{i}}$ (see Eq.~\ref{alfven} on page
\pageref{alfven}).  For the electric current $I=40$~kA and for the mass of carbon ion $M_{\rm{i}}\doteq
2\times10^{-26}$~kg, we obtain the line density of ions in the coronal plasma
$N_{\rm{i}}=8\times10^{17}$~m$^{-1}$. According to Tab.~\ref{properties1}, we take the density of carbon atoms
in a~fibre $n_{\rm{a}}\approx1.1\times10^{29}$~m$^{-3}$. Then the line density of a~15~$\muup$m diameter fibre
is $N_{\rm{a}}\approx2\times10^{19}$~m$^{-1}$. From these values we may conclude that roughly\footnote{This is a
rough estimation because on the one hand, atoms from electrodes were present in the coronal plasma and therefore
the number of ablated carbon ions could be lower, but on the other hand, however, we could expect that not all
the ablated carbon ions were participating in the implosion. By the way, we do not know whether 100\% of the
current followed the implosion to the axis.} a few per cents of the total mass of a~fibre were ablated at that
time.

However, the line density of $N_{\rm{i}}=8\times10^{17}$~m$^{-1}$ determines not only the mass of a~fibre
ablated but also the average plasma density. First, if we assume that the diameter of a plasma column at the
beginning of the implosion is of the order of one centimetre (cf.~Fig.~\ref{imploze1}), we obtain the ion
density
 of about $n_\mathrm{i}=N_\mathrm{i}/\pi r^2\approx10^{16}$ cm$^{-3}$. Second, at the end of plasma implosion we
observed the plasma diameter of the order of millimetre and hence the ion density was about $10^{18}$ cm$^{-3}$.
\label{hustota} This value corresponds to the plasma density estimated from schlieren diagnostics (see
page~\pageref{hustota2}).

\subsubsection{Plasma temperature}
\noindent Before we present results of temperature measurement, it should be emphasised that the plasma column
was inhomogeneous in temperature. It means that the determination of temperature depends on the choice of
spectral lines which are used for calculation. In our case, the electron temperature could be estimated from (i)
Li-- and Be--like oxygen ions, (ii) Li--like carbon ions and (iii) K--shell spectral lines of carbon ions. The
first two choices gave information mainly about colder regions\footnote{Using the ratio of C IV, O V and O VI
lines, the electron temperature was about 20 eV regardless the phase of the discharge.} where Li--like ions
occur in an abundance. That is why we decided to use mainly the lines of H-- and He--like carbon ions.

\begin{figure}[h!]
\centerline{\includegraphics{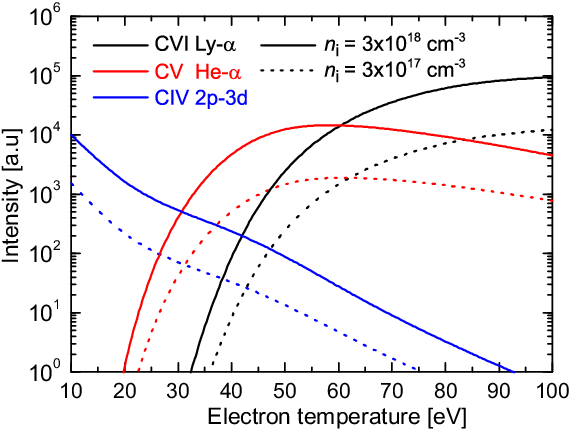}} \caption[Temperature dependence of intensities of selected
lines]{Temperature dependence of intensities of selected lines calculated by the FLY code for
$n_\mathrm{i}=2\times10^{17}$ cm$^{-3}$ (dotted lines) and $n_\mathrm{i}=2\times10^{18}$ cm$^{-3}$. }
\label{lines}
\end{figure}

In most cases, K--shell lines of carbon ions were not observed at the beginning of the implosion, i.e.~before
150 ns. According to Fig.~\ref{lines}, it was a~result of a low ion density and temperature. The electron
temperature was likely below 40 eV.

\begin{figure}[h!]
\centerline{\includegraphics{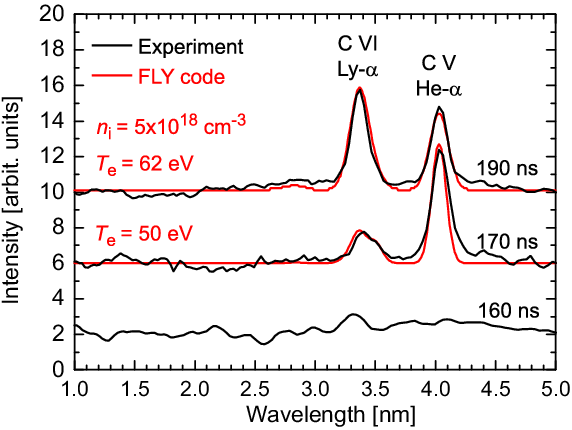}} \caption[XUV spectrum in shot no.~011031-3. Synthetic
spectra simulated with FLY code for optically thin plasma]{XUV spectrum in shot no.~011031-3. Synthetic spectra
simulated with the FLY code for optically thin plasma. The last spectrum was detected 10 ns prior to the peak of
the XUV pulse of a 20 ns FWHM. A carbon fibre of 15~$\muup$m diameter and 8 mm length, bisectional electrodes
made of brass, 12 $\muup$F capacitors, charging voltage of 23 kV.} \label{011131-3}
\end{figure}

The He--$\alpha$ line was observed first during the implosion, but especially shortly before the stagnation. The
evolution of Ly--$\alpha$ and He--$\alpha$ lines before the stagnation is demonstrated in Fig.~\ref{011131-3}.
If the plasma is optically thin, the Ly--$\alpha$ to He--$\alpha$ ratio is almost density independent and can
serve as a convenient method for temperature measurement. It can be clearly seen in Fig.~\ref{011131-3} that the
Ly--$\alpha$ to He--$\alpha$ ratio was increasing during the implosion and thus the temperature was growing.
This agrees with the increasing kinetic pressure that was inferred from the deceleration of the implosion
velocity before the stagnation.
\begin{figure}[h!]
\centerline{\includegraphics{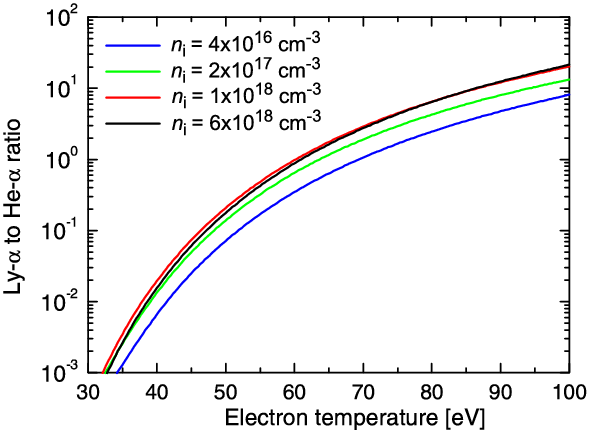}} \caption[Ratio of Ly--$\alpha$ and He--$\alpha$ lines of carbon
ions for optically thin plasma]{Ratio of Ly--$\alpha$ and He--$\alpha$ lines of carbon ions for optically thin
plasma, FLY.} \label{LyToHe}
\end{figure}


\subsection{Stagnation}
\noindent At the end of the implosion, a coronal plasma reached a~fibre and stagnated at the axis for a brief
period of time. This time corresponded to the peak of the short intensive pulse of XUV radiation. This pulse
originated predominantly from the corona around a~fibre. Sometimes, however, we observed that the signal of
Al--filtered PIN diode did not correspond to intensities of K--shell lines of carbon ions. This discrepancy
could be explained by the fact that Al--foil absorbs carbon K--shell lines. Hence, the intensity of XUV pulse is
determined by L--shell lines of carbon and/or by the amount of impurities.
\begin{figure}[!h]
\centerline{\includegraphics{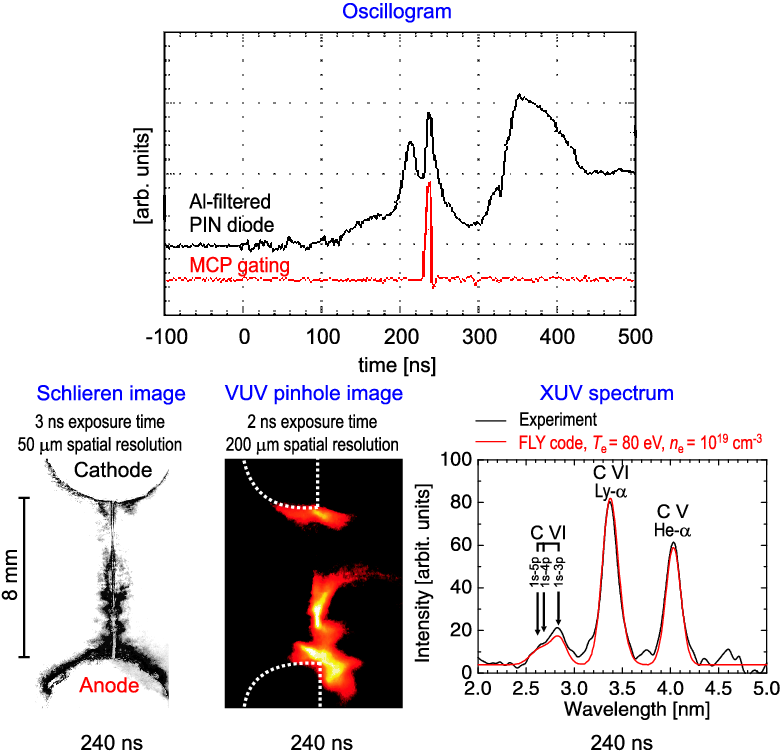}} \caption[Simultaneous XUV spectrum, schlieren and pinhole
images exposed at the peak of the XUV pulse]{Simultaneous XUV spectrum, schlieren and pinhole images exposed at
the peak of the XUV pulse, shot no.~030219-4. A 15~$\muup$m diameter and 8 mm long carbon fibre, bisectional
electrodes made of brass, 12 $\muup$F capacitors, charging voltage of 20 kV, schlieren sensitivity
$n_\mathrm{e}\approx5\times 10^{18}$ cm$^{-3}$.} \label{stagnation}
\end{figure}

The simultaneous diagnostics in Fig.~\ref{stagnation} demonstrates that the radiation was emitted from that part
of a~fibre where MHD instabilities were developed. These perturbations were usually axisymmetric indicating an
$m=0$ instability. Schlieren images corresponded to pinhole images and thus the schlieren system and pinhole
camera gave similar value of perturbation wavelengths $\lambda$. In the case of 15~$\muup$m carbon fibres, the
wavelength $\lambda$ and pinch radius (or to be more precise, the diameter of instabilities) $R$ varied from 0.3
to 1 mm and 0.5 to 1.5 mm, respectively. The $kR$ product, where $k=2\pi/\lambda$ is the axial wave number, was
$2\div10$ and seemed to increase with the time. However, accurate evaluation was impossible since only a few
lobes developed along the fibre. Moreover, the changing wavelength $\lambda$ showed the presence of a mixture of
instabilities (see the schlieren image in Fig.~\ref{imploze2}).

\vspace{-0.5cm}
\subsubsection{Plasma parameters}
\vspace{-0.1cm} \noindent The graph in Fig.~\ref{stagnation} shows the experimental and simulated spectrum. The
simulated spectrum was calculated by the collisional--radiative code FLY. As far as the input parameters for the
steady--state simulation are concerned, the electron temperature $T_{\rm{e}}$ and density $n_{\rm{e}}$ were 80
eV and $10^{19}$~cm$^{-3}$, respectively. Optical depth effects were taken into account. The estimation of the
electron density $n_\mathrm{e}$ was ambiguous owing to its dependence on the choice of the optical path length.

Therefore, plasma density was obtained from schlieren images assuming a parabolic density profile. The highest
sensitivity of our schlieren system determined the minimum electron density of $5\times 10^{18}$ cm$^{-3}$ (see
page~\pageref{schliersensitivity}). \label{hustota2} When we used a~larger stop in the schlieren system we
observed plasmas with electron densities above $10^{19}$ cm$^{-3}$. On the basis of our measurements, we assumed
that the average electron density during the stagnation was of the order of $n_\mathrm{e}\approx10^{19}$
cm$^{-3}$. For the average ion charge of 5, we get the ion density of
$n_\mathrm{i}\approx2\times10^{18}$~cm$^{-3}$ which is two times higher than the density estimated from the
implosion velocity (see page~\pageref{hustota}). It is a reasonable agreement especially if we consider that
both methods give only a rough estimation and that the density during the stagnation could be increased by the
particles ablated from a~fibre.

\vspace{-0.3cm}
\subsubsection{Bennett equilibrium}
\vspace{-0.1cm} \noindent Assuming the steady--state equilibrium of a~$Z$--pinch during the stagnation, it is
possible to compare our experimental results with the Bennett equilibrium.

During the stagnation the current $I$ flowing through the pinch was about 40 kA, the temperature $T$ was about
70 eV. The ionization state was primarily H--like and He--like carbon thus the ion charge $\overline{z}$ was
about 5.5. Then the Bennett equation~\ref{Bennet} on page~\pageref{Bennet} gives the ion line density
$N_\mathrm{i}=10^{18}$ m$^{-1}$. This value is of the order of the density determined from the implosion
velocity (see page~\pageref{hustota}).

\subsection{Expansion}
\begin{figure}[h!]
\centerline{\includegraphics{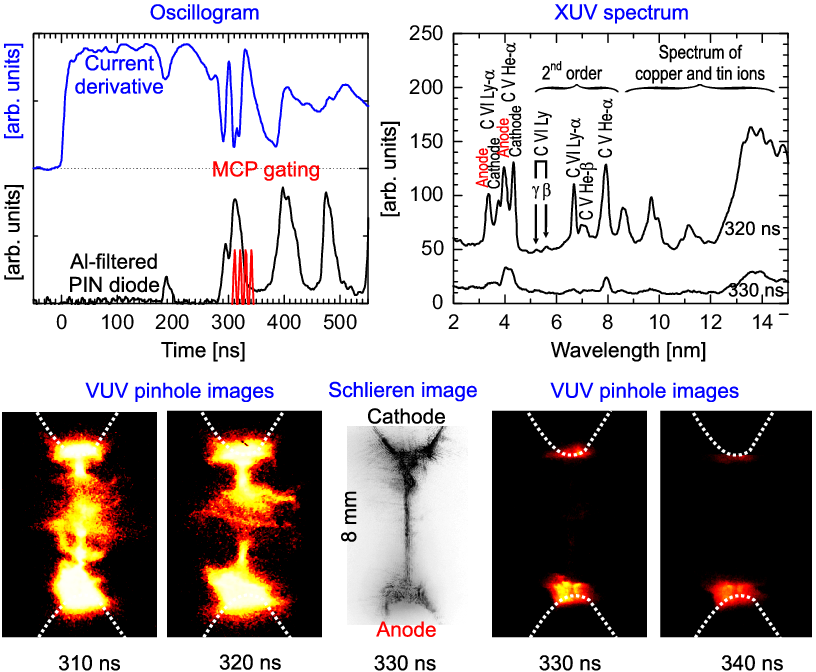}} \caption[Cooling after stagnation in discharge
no.~040225-2]{Cooling after the stagnation in discharge no.~040225-2. A 15~$\muup$m diameter and 8 mm long
carbon fibre, conical electrodes made of bronze, 3 $\muup$F capacitor, 20 kV charging voltage, schlieren
sensitivity $n_\mathrm{e}\approx10^{19}$ cm$^{-3}$.} \label{expansion}
\end{figure}

\noindent The stagnation lasted from a several nanoseconds to a few tens of nanoseconds. After the stagnation we
observed the expansion of a plasma column. The rapid expansion was inferred from the schlieren images. For
instance in Fig.~\ref{expansion}, the density of a plasma column was below the sensitivity of the schlieren
system already 20 ns after the peak of XUV radiation. To determine the expansion velocity was somewhat
problematic because of the drop of plasma density and emitted power. However, it was clear that the expansion
velocity was high enough to cause the rapid increase of $\diff I/\diff t$. The expansion was accompanied by the
gradual cooling. In Fig.~\ref{expansion}, the Ly--$\alpha$ to He--$\alpha$ ratio fell down during 10 ns.
According to Fig.~\ref{LyToHe}, the temperature dropped from 60 eV to less than~45~eV.

\begin{figure}[h!]
    \centering \subfigure[Expanded dense core at 275 ns, it occurred 10
ns after the peak of 10 ns pulse in discharge no.~030409-4.]{ \label{core}
\includegraphics{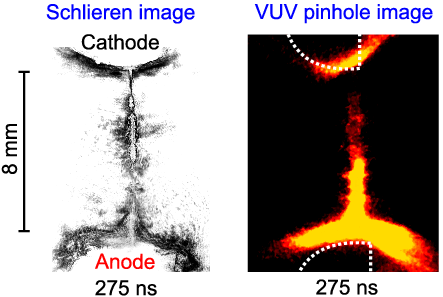}}
\hspace{0.8cm} \subfigure[Schlieren image of the helix observed at 300 ns, it occurred 50 ns after the peak of
XUV pulse in the discharge no.~\mbox{030410-2.}]{ \label{helix}
\includegraphics{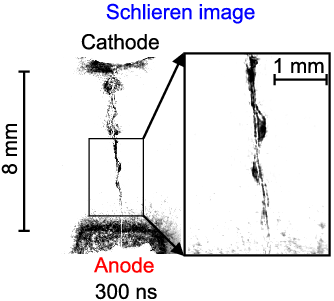}}
\caption[Expanded dense core and helix observed after the expansion]{A 15~$\muup$m diameter and 8 mm long carbon
fibre, bisectional electrodes made of steel, 3~$\muup$F capacitor, 25 kV charging voltage, schlieren sensitivity
$n_\mathrm{e}\approx5\times 10^{18}$ cm$^{-3}$.}
\end{figure}

During or after the expansion, we sometimes observed cold and dense plasma regions. For example, Fig.~\ref{core}
shows the expanded dense core of a~fibre 10 ns after the series of short XUV pulses.

Once or twice we also observed the helix after the expansion. Fig.~\ref{helix} was recorded 50 ns after the XUV
pulse. At this time the plasma was expanded to a radius of several millimetres, whereas the diameter of helix
was about 50 $\muup$m. This begs the question of what kept this cold structure in existence.


\subsection{Multiple pulses}
\noindent Fig.~\ref{pof} shows a~high number of pulses that were detected by a PIN diode. The intensity of
pulses was increasing with the growing time most likely because of the rising current. It was necessary to
decide whether pulses originated from various places along a~fibre or from the whole length of a~fibre at
different moments.

We believe that both phenomena might occur. During the rise of the current it was possible that the multiple
implosion of the corona onto a~fibre took place. However, when there were more than two pulses, they did not
originate from the whole length of a~fibre. Especially in the case of conical electrodes, the non--symmetric
implosion occurred separately at the anode, at the cathode, and at the centre (cf.~Fig.~\ref{expansion}). As
a~result, we observed higher number of pulses than in the case of flat electrodes.


\subsection{Experimental data interpretation}

\subsubsection{Gradual fall of current derivative and rise of voltage}
\noindent First observation which needs to be interpreted is the gradual fall of circuit $\diff I/\diff t$  and
the rise of voltage before each XUV pulse (see Fig.~\ref{pof}).

 The starting point of our explanation is Fig.~\ref{RLC}
on page~\pageref{RLC}. According to it we can express
\begin{eqnarray}
U(t)&=&U_0-\int_0^t \frac{I(t')}{C_0}\diff t'=
\frac{\diff}{\diff t}\bigg((L_0+L_{\rm{P}})I\bigg)+(R_0+R_{\rm{P}})I\\
\frac{\diff I}{\diff t}&=&\frac{U-(R_0+R_{\rm{P}}+\dot{L}_{\rm{P}})I}{L_0+L_{\rm{P}}}
\end{eqnarray}
where $C_0$, $L_0$, $R_0$ are parameters of a short circuit and $L_{\rm{P}}$, $R_{\rm{P}}$ stand for the
inductance and resistance of a~pinch plasma, respectively. On the one hand, the inductance of the pinching
plasma ($L_{\rm{P}}<10$~nH) was lower than~the inductance of the~rest of circuit ($L_0\approx100$~nH) and, in
addition to that, the inductance depends on the pinch radius only logarithmically. Therefore, it did not affect
the current derivative to an excessive degree. On the other hand, because of a low resistance of the current
generator ($R_0=25$~m\,$\Omega$), the current waveform was very sensitive to the plasma resistance
$R_\mathrm{P}(t)$ and to the change of inductance $\dot{L}_\mathrm{P}(t)$.  That is why the observed fall of
$\diff I/\diff t$  can~be attributed to the rise of the impedance $(R_{\rm{P}}+\dot{L}_{\rm{P}})$ during the
implosion and stagnation. The increase of $(R_{\rm{P}}+\dot{L}_{\rm{P}})$ also caused the voltage growth since
\begin{equation}
U_{\rm{P}}  = \frac{\diff}{\diff t}(L_{\rm{P}}I)+R_{\rm{P}}I = L_{\rm{P}}\frac{\diff I}{\diff t}+\Bigg(\frac{\diff L_{\rm{P}}}{\diff t}+R_{\rm{P}}\Bigg)I\\
\end{equation}

\begin{figure}[h!]
\centerline{\includegraphics{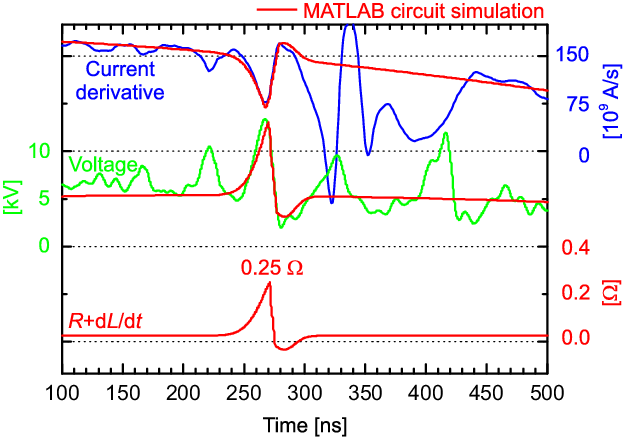}} \caption[Experiment vs. MATLAB simulation of the
$R$--$L$--$C$ circuit]{Experiment vs. MATLAB simulation of the $R$--$L$--$C$ circuit, shot no.~050128-1.
A~carbon fibre of 15~$\muup$m diameter and 8 mm length, conical electrodes made of brass, 3~$\muup$F capacitor,
20 kV charging voltage.} \label{simulation}
\end{figure}

In order to obtain quantitative results, we simulated the $R$--$L$--$C$ circuit with a time--dependent
$R_\mathrm{P}+ \dot{L}_\mathrm{P}$ term (see appendix \ref{appendixMATLAB}). An example of our simulation is
shown in Fig.~\ref{simulation} where the experimentally observed waveforms are presented together with MATLAB
simulation. Fig.~\ref{simulation} shows that the observed waveforms in the period between 250 ns and 300~ns
could be explained by the rise of the $R_\mathrm{P}+\dot{L}_\mathrm{P}$ term up to 0.25~$\Omega$ and by its
subsequent drop to -0.05 $\Omega$. An inevitable question is what part of this value could be attributed to the
resistance $R_{\mathrm{P}}$ and what part belongs to the change of the inductance $ \dot{L}_\mathrm{P}$?

\subsubsection{Resistance versus time--varying inductance} \noindent
To solve this question we present the following calculation assuming the current flowing through a cylindrically
symmetric $Z$--pinch\footnote{We assume the uniform current density since the skin depth is of the order of
millimetres.}.

We can~express $R_{\rm{P}}$ and $L_{\rm{P}}$ as
\begin{eqnarray}
R_{\rm{P}}&=&\frac{l}{\pi R^2 \sigma}\\
L_{\rm{P}}&=&\frac{\mu l}{2\pi}\ln\frac{b}{R} \label{indukcnost}
\end{eqnarray}
where $l$, $R$, $\sigma$ stand for the length, radius, and conductivity of a~plasma column, respectively. The
quantity $b$ represents the radius of the experimental chamber.

\noindent Calculating the derivative
\begin{equation}
\label{lpdot} |\dot{L}_\mathrm{P}|=\frac{\mu lv_\mathrm{imp}}{2\pi R}
\end{equation}
we obtain the ratio
\begin{equation}
 \label{pomerLaR}
\frac{|\dot{L}_{\rm{P}}|}{R_{\rm{P}}} =\frac{1}{2}\mu \sigma R v_\mathrm{imp}
\end{equation}
For quantities observed during the implosion
\begin{displaymath}
R=2\hspace{2mm}\mathrm{mm},\quad v_\mathrm{imp}=10^5\hspace{2mm}\mathrm{m}\,\mathrm{s}^{-1},\quad
T_\mathrm{e}=40\hspace{2mm}\mathrm{eV},\quad \bar{z}=5,\quad \ln \Lambda\doteq10
\end{displaymath}
\begin{equation}
\label{values}
 \sigma_{\rm{Spitzer}}\doteq 7.7\times10^{-3} T^{3/2}/(\bar{z} \ln \Lambda)\doteq5\times10^{4}\hspace{2mm} \mathrm{S}\,\mathrm{m}^{-1}
\end{equation}
the ratio in question is
\begin{equation}
\frac{\dot{L}_{\rm{P}}}{R_{\rm{P}}}\doteq6
\end{equation}

This led us to conclusion that during the implosion the energy was coupled from the generator through a
time--varying inductance. The electrical energy was mainly converted (i) into the radial motion of ions, (ii)
into the (adiabatic) compression of a plasma column and (iii) into the increase of magnetic energy\footnote{From
$\int_{t_0}^tU_\mathrm{P}I\diff t=\int_{t_{t_0}}^tR_\mathrm{P}I^2\diff t+\int_{t_0}^t\dot{L}_\mathrm{P}I^2\diff
t+\int_{t_0}^tL_\mathrm{P}\dot{I}\,I\diff t=\int_{t_0}^tR_\mathrm{P}I^2\diff t+\frac{1}{2}\int_{t_0}^t
\dot{L}_\mathrm{P}I^2\diff t+\frac{1}{2}\int_{t_0}^t \dot{L}_\mathrm{P}I^2\diff t+\frac{1}{2}\int_{t_0}^t
L_\mathrm{P} (\diff I^2/\diff t)\diff t=\int_{t_0}^tR_\mathrm{P}I^2\diff t+ \frac{1}{2}\int_{t_0}^t
\dot{L}_\mathrm{P}I^2\diff t + [L_\mathrm{P}I^2]_{t_0}^t$ follows that the energy
$\int_{t_0}^t(R_\mathrm{P}+\frac{1}{2}\dot{L}_\mathrm{P})I^2\diff t$ is fed into the plasma and
$[L_\mathrm{P}I^2]_0^t$ goes into the rise of magnetic field.}. Shock heating of ions might also occur, however,
we have not observed any shock--wave during the implosion of a coronal plasma onto a~fibre.

Late in the implosion, as the radius $R$ and velocity $v_\mathrm{imp}$ were decreasing, the influence of plasma
resistance was growing and the resistive heating started to play an important role. Speaking quantitatively, for
values in Eq.~\ref{values} and for $l=3$ mm (a part of a~fibre participating in the implosion when conical
electrodes were used), we obtain a time-varying inductance
\begin{equation}
\dot{L}_\mathrm{P}=0.03\hspace{2mm}\Omega
\end{equation}
which is lower than 0.25 $\Omega$.

As regards the plasma resistance, for reasonably estimated values during the stagnation \vspace{-0.2cm}
\begin{equation}
R=0.5\hspace{2mm}\mathrm{mm},\quad l=3\hspace{2mm}\mathrm{mm},\quad T_\mathrm{e}=70\hspace{2mm}\mathrm{eV},\quad
\bar{z}\ln \Lambda\doteq60,\quad \sigma_{\rm{Spitzer}}\doteq 10^{5}\hspace{2mm} \mathrm{S}\,\mathrm{m}^{-1}
\end{equation}
we get
\begin{equation}
R_{\rm{P}}\doteq0.04\hspace{2mm}\Omega
\end{equation}
Such value is also lower than 0.25~$\Omega$. In order to reach 0.25 $\Omega$, we can naturally reduce the radius
of a plasma column. However, the effect of instabilities seems to be more sensible explanation.

\subsubsection{Plasma instabilities}\label{pofinstabilities} 
\begin{figure}[!h]
\centerline{\includegraphics{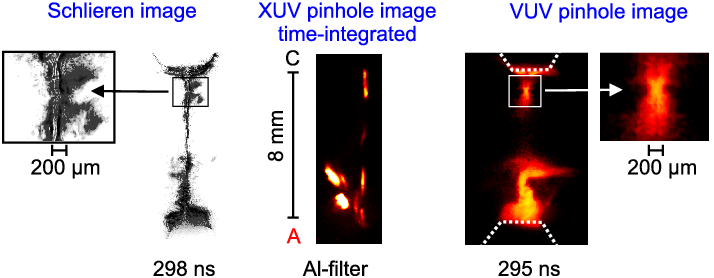}} \caption[Observed $m=0$ instability in shot no.~040421-1.
Off-axis bright spots originated from the later stage of a discharge]{Observed $m=0$ instability in shot
no.~040421-1. Off--axis bright spots originated from the later stage of a discharge. A 15~$\muup$m diameter and
8 mm long carbon fibre, conical electrodes made of brass, 20 kV charging voltage.} \label{m0nestabilita}
\end{figure}

\noindent
The substantial contribution to plasma column resistance during the stagnation could be ascribed to
instabilities. The strong argument for that are voltage spikes which corresponded to XUV pulses. At the same
time, these XUV pulses did not originate from the whole length of a~fibre but from several bright spots. These
bright spots corresponded to the interaction of the necks of an $m=0$ instability with the dense core
(cf.~Fig.~\ref{m0nestabilita}).

\noindent For values observed in Fig.~\ref{m0nestabilita}
\begin{equation}
\label{values3} R=0.1\hspace{2mm}\mathrm{mm},\quad l=0.3\hspace{2mm}\mathrm{mm} ,\quad
T_\mathrm{e}=70\hspace{2mm}\mathrm{eV},\quad \bar{z}\ln \Lambda\doteq60, \quad \sigma_{\rm{Spitzer}}\doteq
10^{5}\hspace{2mm} \mathrm{S}\,\mathrm{m}^{-1}
\end{equation}
we obtain the plasma resistance
\begin{equation}
R_{\rm{P}}\doteq0.1\hspace{2mm}\Omega
\end{equation}
\noindent If we consider both the wide plasma column and several necks of $m=0$ instability, the total plasma
resistance during the stagnation could achieve 0.25 $\Omega$.

However, these values represent only rough estimations since the radius of a plasma channel was limited by the
spatial resolution of the order of 100 $\muup$m. Moreover, we do not know where exactly the current was flowing
and what was the role of anomalous resistivity within instabilities.

We must not also forget that instabilities contribute not only to plasma resistance but also to the
$\dot{L}_{\rm{P}}$ term. An $m=0$ instability develops with the Alfv\'{e}n velocity of $10^5$ m\,s$^{-1}$ and thus
according to values in Eq.~\ref{values3}

\begin{equation}
\dot{L}_\mathrm{P}=\frac{\mu lv_\mathrm{A}}{2\pi R}\doteq0.05\hspace{2mm}\Omega
\end{equation}
The power $\frac{1}{2}\dot{L}_\mathrm{P}I^2$ is then transferred to ions.

 As regards an $m=1$ instability, its contribution to the $\dot{L}_{\rm{P}}$ term depends on the
rate of how fast it develops. We often observed helical structures in pinhole images, whereas in schlieren
images they were less apparent. Similar results were observed by \citet{Lorenz98}. The complex helical
structures develop usually during an expansion phase after stagnation (cf.~Fig.~\ref{helix} and
\citealp{Spielman94, Spielman01}). Recently, the influence of an $m=1$ instability upon $Z$--pinch dynamics has
been quantitatively discussed by \citet{Chittenden04}. He found that the plasma is going $m=1$ unstable when the
Lorentz force starts to compress the stagnated plasma. Then the onset of an $m=1$ instability provides an
efficient mechanism for dissipation of the magnetic energy surrounding the stagnated pinch. Not only does the
formation of a helix allow the magnetic field to do work in driving an expansion but it also enhances the ohmic
heating by elongating the current path length\footnote{The energy conservation in wire--array $Z$--pinches has
become a debated issue because radiated energies could be more than $3\div4$ times greater than the ion kinetic
energy from the implosion. Already 20 years ago, however, \citet{Riordan81} pointed out that the observed
radiation yield was significantly higher than the kinetic energy input. Besides anomalous resistivity in the
trailing mass~\citep{Peterson98}, the obvious extra energy source available is the magnetic energy that is much
greater than the kinetic energy of implosion. The magnetic energy could be dissipated by instabilities through
the $\dot{L}_\mathrm{P}$ term. \citet{Rudakov97, Rudakov00} and \citet{Velikovich00} proposed the model of
current reconnections around an $m=0$ instability and subsequent collapse of toroidal magnetic flux tubes.
\citet{Haines04} considered fine-scale instabilities and viscous heating of ions to play the major role.
Nevertheless, a~lot of plasma physicists see $m=1$ behaviour in the current flow to be the most likely
explanation of a~``paradox'' heating in question (cf.~\citealp{Kubes02, Chittenden04}).}.

On the basis of preceding paragraphs we can conclude that it is not easy to unambiguously determine what
contributed to the $R_\mathrm{P}+\dot{L}_\mathrm{P}$ term. Nevertheless, we can say that a time--varying
inductance was important during the implosion, whereas at the stagnation the influence of plasma instabilities
and resistive heating seemed to be the dominant one.

\vspace{-0.2cm}
\subsubsection{Energetics during plasma\,-\,on\,-\,fibre stage}
\noindent The previous paragraphs clarified rises of voltage. During these spikes, the energy from the current
generator was fed into a plasma. The excess of energy was almost immediately (within 5 ns) radiated and this is
why we observed XUV pulses. Since we measured the current and voltage, it was possible to estimate the
energetics of our $Z$--pinch. We could calculate the ohmic heating and the kinetic energy delivered to ions in
the period between 250 ns and 300 ns (cf.~Fig.~\ref{simulation}). We shall start with the kinetic energy first.

The total kinetic energy delivered to the axis could be estimated from a time--varying inductance as follows:
\begin{equation}
W_\mathrm{imp}=\int_{t_0}^t \frac{1}{2}\dot{L}_\mathrm{P}I^2 \diff t\approx \frac{1}{2}I^2\int_{t_0}^t
\dot{L}_\mathrm{P}\diff t=\frac{1}{2}[L_\mathrm{P}(t)-L_\mathrm{P}(t_0)]I^2
\end{equation}
According to Eq.~\ref{indukcnost} we have
\begin{equation}
W_\mathrm{imp}=\frac{\mu l }{4\pi}\ln\Bigg(\frac{R(t_0)}{R(t)}\Bigg)I^2
\end{equation}
Using
\begin{equation}
I=35\hspace{2mm}\mathrm{kA},\quad l=3\hspace{2mm}\mathrm{mm} ,\quad \frac{R(t_0)}{R(t)}\doteq10\hspace{2mm}
\end{equation}
the result is
\begin{equation}
W_\mathrm{imp}=1.5\hspace{2mm}\mathrm{J}
\end{equation}

Even more energy seemed to be delivered by collisional processes during the stagnation phase. Moving to the
ohmic heating, using values discussed above, it can be estimated as
\begin{equation}
\int_0^t R_\mathrm{P}I^2 \diff t\approx R_\mathrm{P}|_\mathrm{max}I^2\cdot t_\mathrm{FWHM}\doteq
0.2\cdot(35\times10^3)^2\cdot15\times10^{-9}\hspace{2mm}\mathrm{J}\doteq3.5\hspace{2mm}\mathrm{J}
\end{equation}
The overall energy deposited to the plasma during one XUV pulse was thus about 5 J. This energy was used for
ablating of a central fibre, ionization\footnote{The energy of 4.5 J is needed for the whole fibre ionised to
the He--like stage, whereas 30 J are necessary for the ionization to bare nuclei.} and heating of a plasma.
Another part of the input energy was spent by advection, thermal diffusion of a plasma, plasma expansion, etc.
The radiated energy (100 mJ in the XUV region, see page~\pageref{energyXUV}) represented only a small fraction
of the overall deposited energy.

\subsubsection{Expansion velocity}
\noindent Another value that can be evaluated from Fig.~\ref{simulation} is the expansion velocity. Because the
$R_\mathrm{P}+ \dot{L}_\mathrm{P}$ term fell to -0.05 $\Omega$, the change of inductance was higher than
0.05~$\Omega$. From Eq.~\ref{lpdot} and for $l=3$~mm, $R=500$~$\muup$m (cf.~Fig.~\ref{expansion}), we obtain the
expansion velocity $v_{\rm{exp}}$ above $4\times10^4$~m\,s$^{-1}$. 


%% file: discussion.tex
\newpage
\fancyhead[R]{\slshape 5. Discussion}
\chapter{Discussion}
\label{discussion} \noindent In this chapter we compare our results with other $Z$--pinch experiments. First, we
shall summarise the most important results of other fibre experiments. Then, we will point out results which are
similar or different to ours. The particularity of our fibre $Z$--pinch is the implosion of a coronal plasma
onto a~fibre and thus it will be treated of in more detail. Finally, we will deal with the most energetic phase
of our discharge which occurred in electrode vapour.

\section{Fibre $Z$--pinch Experiments}
\noindent Fibre $Z$--pinch experiments were carried out on modern high--voltage pulsed--power generators at Los
Alamos National Laboratory, Naval Research Laboratory, Imperial College in London, Kernforschungszentrum
Karlsruhe, and Kurchatov Institute in Moscow~(see section \ref{fibres} and appendix \ref{appendixS300}). Even
though each experimental group observed somewhat different behaviour of a~fibre $Z$--pinch, the gross dynamics
of a~fibre pinch was described. We shall sum up the most important findings of these research groups in the
following text.

We start with the global expansion of a coronal plasma surrounding the solid core. The expansion was observed
although the Pease--Braginskii current was reached and the current rise was fast enough to cause the radiative
collapse. This rapid expansion could happen for several reasons and we touch upon them briefly. One of the
reasons might be anomalous resistivity in a low density coronal plasma which, if it occurs, causes the plasma
kinetic pressure to exceed the magnetic pressure~\citep{Chittenden95}. As a~result, the Bennett equilibrium does
not hold and the pinch expands. Another reason for the expansion may lie in MHD instabilities. During the
development of an $m=0$ instability, the bulges of the instability expand and the constricted regions cause
enhanced non-uniform heating of the core~\citep{Sheehey92, Riley96}. An $m=0$ instability was apparent very
early in the discharge in shadowgrams and schlieren images.  These instabilities exhibited highly dynamic
behaviour~\citep{Chittenden97}, but they were not disruptive (most likely because of the dense core which lasted
for a relative long time). During the interaction of the coronal plasma with the remaining dense core, bright
spots occurred. After the dynamic bright--spot phase, the disruption accompanied with the neutron production and
X--ray emission was observed as soon as the fibre was completely ablated.

All these results, but mainly the rapid expansion and the early development of plasma instabilities, effectively
decrease the plasma density and eliminate the possibility of using a~fibre $Z$--pinch as a fusion reactor.

\section{Implosion of Coronal Plasma onto Central Fibre}
\noindent  Our work with a~carbon fibre $Z$--pinch brought some results that were to a certain extent similar to
those of above mentioned experiments despite the fact that peak currents and rise times were substantially
different. Similarly to these experiments, we made the following observations:
\begin{itemize}
\item[--] A low density coronal plasma carried almost all the current, a cold dense core persisted for a long
time. The interaction of a coronal plasma with this dense core then substantially influenced the plasma
dynamics. \vspace{-0.2cm}\item[--] The XUV emission came mainly from bright spots which were produced by
instabilities that developed in a low density corona. However, these instabilities were not disruptive as long
as the dense core survived unionised. \vspace{-0.2cm}\item[--] After the fibre ablation, the intense X--ray
radiation was produced.
\end{itemize}

Nevertheless, our $Z$--pinch differed in several issues of a great importance too. Fibre $Z$--pinches driven by
fast pulsed power generators are usually considered to be in pressure balance because the current rises rapidly
and the plasma pressure could be balanced against the magnetic forces in every moment. Contrary to that, our
fibre $Z$--pinch driven by microsecond capacitive discharge expanded first to the diameter of about one
centimetre. When the current built up, the implosion of a coronal plasma onto a central fibre occurred. This
means that our fibre $Z$--pinch behaved as a dynamic pinch for a lapse of time. Even though the current was
about 40 kA, the implosion velocity approached the value of $2\times10^5$ m$\,$s$^{-1}$ which implies the
kinetic energy of carbon ions of about 2.5 keV.

It is true that noticeable implosion could be observed also on pulsed power generators, however, the current
prepulse\footnote{Recently, a current prepulse applied before the main pulse has been suggested to evaporate a
wire and to allow it to expand. The reason for that is to create a homogenous plasma for subsequent
implosion~\citep{Choi05}.} must be applied. Such experiment was carried out on a~pulsed power generator IMP
($I_\mathrm{max}=200$ kA, $t_{10\%-90\%}=60$ ns, Imperial College, London, \citealp{Lorenz98}).
   In this experiment, the prepulse delivered 10 kA current with a 50 ns
quarter period into a 7 $\muup$m diameter carbon fibre. The breakdown of the fibre occurred when the voltage
reached about 20 kV. The prepulse generated a low density coronal plasma of $N_{\rm{i}}=3\times10^{17}$ m$^{-1}$
(cf.~with our value of $8\times10^{17}$~m$^{-1}$, page~\pageref{density}) which expanded to a radius of the
order of a millimetre. After the switch of the main discharge current, they observed, similarly as we did,
implosion of a~low density plasma onto a~fibre\footnote{In Lorenz's experiment, the faster current rise--time
implied lower skin depth and thus the \emph{snowplough} implosion occurred.}, the zipper from the cathode
towards the anode, the soft X--ray pulse, and subsequent rapid expansion.

Another similar feature between Lorenz's and our experiment was the observation of $m=0$ and $m=1$ instabilities
during the stagnation of the corona at the fibre. For Alfv\'{e}n velocity of $2\times10^5$~ms$^{-1}$, the pinch
diameter of about 1 mm, and the conductivity of $10^5$~S\,m$^{-1}$, the ideal MHD model predicts the
characteristic time of instability growth $\tau=R/v_\mathrm{A}=2.5$ ns. Despite a low Lundquist number
$S=\mu\sigma v_\mathrm{A}R\approx10$, enhanced stability was not confirmed and the presence of a~fibre did not
seem to significantly suppress MHD instabilities. However, these were not excessively disruptive.

The development of instabilities caused the generation of XUV pulses from several bright spots. Because of a low
current and, by extension, a relative high resistivity, the main heating mechanism was most likely Joule
heating. Nevertheless, the $\dot{L}_\mathrm{P}$ term cannot be ruled out. In that case, plasma instabilities
could provide an effective mechanism for dissipating the magnetic energy adjacent to the pinch.

The coronal plasma in our experiment and in the experiment of \cite{Lorenz98} was formed from a~fibre. Another
possibility would be to create a plasma from a wire--array or gas--puff which subsequently implodes onto a
central fibre. We shall look at these two approaches in the following sections. As regards the difference
between the coronal plasma created from a~fibre and the plasma formed from a wire--array or gas--puff, the
higher mass was imploded onto a~fibre in experiments with wire--arrays or gas--puffs. Clearly, the energetics of
these $Z$--pinches is dominated by the implosion energy. In our fibre experiment, the implosion energy was
relatively low. Much more important phenomena was probably the transfer of the current in the vicinity of
a~fibre when the corona imploded.

\section{Wire--Array and Gas--Puff Implosion onto Central Fibre} \noindent The implosion of a wire-array $Z$--pinch onto
a deuterated fibre was studied on the S--300 device (4 MA, 700 kV and 100 ns; RRC Kurchatov Institute, Moscow).
These experiments were performed at a peak current of about 2 MA with a rise time of about 100
ns~\citep{Klir04a, Klir05}. An aluminium wire--array of 1 cm in diameter was used as a load. Each wire--array
consisted of $30\div60$ aluminium wires of 15~$\muup$m in diameter and 1 cm in length. The deuterated
polyethylene (CD$_2$)$_\mathrm{n}$ fibres with diameters between 80 and 120 $\muup$m were placed on the axis of
the array. The purpose of these experiments on the S--300 generator was to study the generation of XUV and
X--ray radiation together with the neutron production. It was found out that the implosion of a wire--array had
the positive influence on the neutron yield. The neutron yield was one order in magnitude higher in the case of
an imploding wire--array onto a~fibre than without a wire--array (cf.~\citealp{Klir04a, Klir05} and appendix
\ref{appendixS300}).

 Two shots with 240 aluminium wires imploding
onto a (CD$_2$)$_\mathrm{n}$ fibre were also carried out on $Z$--machine in 1998.
 However, more neutrons of $(2.8\pm0.2)\times10^{12}$ were
generated on Saturn during the implosion of a hollow deuterium gas--puff\footnote{This configuration was very
similar to the deuterium plasma--sheath imploding onto a (CD$_2$)$_\mathrm{n}$ fibre on the PF--1000 plasma
focus~\citep{Kubes04}.} onto a 250 $\muup$m deuterated polyethylene fibre located on the pinch
axis~\citep{Spielman91}. Recently, the highest neutron yield from the D-D reaction of $6\times10^{13}$ has been
achieved with double--shell deuterium gas--puff, i.e.~without any fibre~\citep{Velikovich05}.

On the basis of our results and observations of other research groups, we believe that the implosion onto
a~fibre offers the possibility of current transfer near the fibre. However, when one considers higher neutron
yield, solid fibres are probably less suitable than deuterium gas--puffs.

\section{Plasma\,--\,on\,--\,Wire} \noindent The peculiarity of a solid fibre is its initial
non--conductivity that causes the transfer of the current from a~fibre and thus makes the energy coupling into
a~fibre difficult. Different situation occurs when a conducting wire is used instead of a~fibre. The implosion
of an aluminium jet onto a coaxial aluminium wire was first studied by~\citet{Wessel92}. They found the
resultant plasma to be more uniform and hotter than a wire--only and jet--only pinch. The pinch also
demonstrated that an imploding plasma could couple the energy from a current generator to a micron--sized wire.

\begin{figure}[h!]
\centerline{\includegraphics{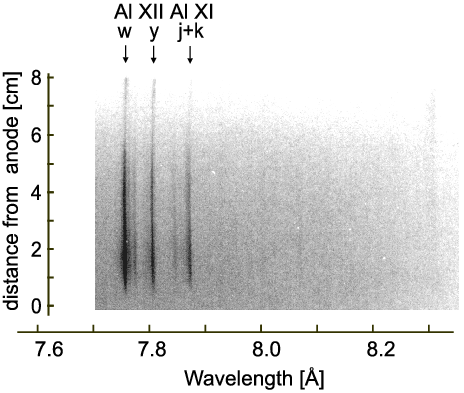}} \caption[Spatially resolved X--ray spectrum recorded on the
PF--1000 plasma focus.]{Spatially resolved X--ray spectrum recorded by V.\,M.~Romanova during the implosion of
the deuterium plasma--sheath onto an Al wire (80 $\muup$m diameter and 8 cm length), PF--1000 plasma focus
(IPPLM, Warsaw), shot no.~2564, 1.5 MA peak current, deuterium filled up to 3 Torr.} \label{varsava}
\end{figure}

Our experiments on the PF--1000 plasma focus in Warsaw showed similar behaviour. In some shots with deuterium
filling, we placed an Al wire at the axis and top of the anode having no galvanic connection to the anode.
Fig.~\ref{varsava} shows the spatially resolved spectrum recorded with the time--integrated X--ray spectrograph,
which detected resonance, intercombination and satellite lines of He--like aluminium ions. Clearly, the whole
length of an Al wire did not manifest typical large--scale instabilities which are characteristic for wire
$Z$--pinches. If we realise that the length of a wire was 8 cm, it was quite an impressive result for a pinch
plasma, which may lead to reconsideration of a wire $Z$--pinch as an X--ray laser medium.

\newpage
Having dealt with the plasma\,--\,on\,--\,fibre phase of our discharge, we shall now turn to the phase in which
electrode vapour determined plasma dynamics.

\section{Discharge in Electrode Vapour and Bright Spots}
\noindent The influence of electrode vapour on $Z$--pinch dynamics was visibly increasing with the growing time.
The dominant role of the material evaporated from the electrodes was apparent after the fibre ablation. After
that we observed bright spots and hard X--ray emission together with sharp voltage spikes and dips in $\diff
I/\diff t$. As far as the gradual rise of plasma voltage is concerned, it could be explained by the implosion of
a plasma column. However, the implosion could not cause the observed sharp increase of voltage which was related
to the development of plasma instabilities near the maximum compression.

Much alike results were also obtained in plasma foci, vacuum sparks and higher--$Z$ gas--puffs driven by low
impedance generators. This phase of the discharge produces the most energetic electrons, ions, and photons,
therefore a~large number of mechanisms were suggested to explain the production of fast particles (see, for
instance, \citealp{Anderson58, Haines83, Trubnikov86, Vikhrev86, Deutsch88, Herold88}, a recent review given by
\citealp{Haines01} and \citealp{Ryutov00}). Basically, we can distinguish three mechanisms that exist in various
modifications~\citep{Ryutov00}:
\begin{itemize}
\item[--] The first explanation from the 1950s suggests that charged particles are accelerated by a~high
inductive voltage in the neck during the development of an $m=0$ instability.\vspace{-0.2cm} \item[--] In
another mechanism, the production of fast particles is ascribed to an adiabatic compression of plasma in the
neck of an instability. \vspace{-0.2cm}\item[--] The third explanation is based on microinstabilities that could
be triggered in a low density plasma with high current density, for instance within $m=0$ instabilities.
Microinstabilities then result both in strong anomalous heating and in the formation of a non--Maxwellian tail
of the ion velocity distribution.
\end{itemize}
\noindent If the charged particles are not of thermal origin, they have to be accelerated by the voltage across
the plasma $U_\mathrm{P}=(R_\mathrm{P}+\dot{L}_\mathrm{P})I+L_\mathrm{P}\dot{I}$. In high current fibre
$Z$--pinches, there was no evidence of pinching or expansion of a dense plasma at the time of disruption
(see~\citealp{Robledo97, Mitchell98}). The rise of plasma voltage could be therefore explained by
microinstabilities triggered in a low density plasma. At this point we would like to draw the attention to the
long duration of energetic particle emission. Such long pulses (several tens of ns) can be hardly explained by
$\dot{L}_\mathrm{P}$ and $\dot{I}$ because the inductance and current would have to be changing for a very long
time. The rise of plasma resistance $R_\mathrm{P}$ seems to be much more feasible.

In our experiment, we also observed the rapid rise--time of X--rays and relatively long (several tens of
nanoseconds) duration of X--ray pulse (cf. Fig.~\ref{rtg} on page~\pageref{rtg}). However, we do not have any
experimental evidence of electron and ion beams and thus we do not discuss the mechanisms of charged particle
acceleration in greater detail. Instead of that, we shall conclude this chapter by a report on bright spots that
were observed in our experiment.

\subsubsection{Bright spots}
\noindent Bright spots are localised plasma regions intensively radiating in visible, XUV and X--ray spectral
ranges. They were thoroughly studied first in a vacuum spark~\citep{Cohen68}. Since then they have been often
observed in all types of $Z$--pinches.
 The temperature, density and life--time of bright spots varied over
several orders of magnitude and different kinds of bright spots have been reported. Even though the origin of
bright spots is still being discussed, it is clear that the nonlinear phase of the sausage instability,
radiative collapse and anomalous resistivity play a very important role~\citep{Vikhrev82, Kunze00}.

In our experiment, we have to distinguish two types of bright spots. The first group of bright spots was
produced during the plasma\,--\,on\,--\,fibre phase. These bright spots emitted less energetic photons than
bright spots created in electrode vapour. To avoid confusion, we call the latter ones ``hot'' spots. These hot
spots arose from $m=0$ instabilities. Because the discharge occurred in electrode vapour, higher--$Z$ ions
(copper, tin, etc.) increased radiation losses through line emission. After the fibre ablation, $m=0$
instabilities could develop significantly according to the radiative collapse model (see \ref{collapse} on page
\pageref{collapse}). As far as the position of hot spots occurrence is concerned, they appeared predominantly
near the anode similarly to vacuum sparks~\citep{Koshelev91}.


%% file: conclusion.tex
\chapter*{Summary, Conclusions and Prospects}
\addcontentsline{toc}{chapter}{Summary, Conclusions and Prospects}

\fancyhead[R]{\slshape Summary, Conclusions and Prospects} \noindent The following pages first summarise the
results obtained in our experiment. After that, conclusions from these results are presented. Finally, prospects
for the future are proposed.

\section*{Summary}
\noindent The primary aim of this thesis has been the study of the evolution and gross dynamics of a~fibre
$Z$--pinch. Our experiments were driven by a slow capacitive discharge of about 1~kJ energy. The current
generator delivered 80 kA with 850 ns quarter period into a~load. Most of the experiments were performed in
vacuum with carbon fibres of 15~$\muup$m diameter and about 1 cm length. The plasma was observed with a~large
number of diagnostic tools. Particularly, simultaneous diagnostics (XUV spectra, VUV pinhole images and
schlieren images) were very helpful.

The characteristic feature of our experiments were XUV and X--ray pulses accompanied with rises of voltage and
with dips in $\diff I/\diff t$. In order to find the origin of these pulses, an in--depth experimental study was
performed. During this study, we distinguished several phases of our $Z$--pinch, namely the breakdown, latent
phase, plasma\,--\,on\,--\,fibre, fibre ablation, implosion of material evaporated from electrodes, development
of $m=0$ instabilities and bright spots. We found out that XUV pulses were first emitted during the development
of instabilities in the plasma\,--\,on\,--\,fibre stage.  After the ablation of a~fibre, both XUV and X--ray
pulses were generated from hot spots which developed in higher--$Z$ material evaporated from electrodes. The
energetics of the $Z$--pinch and basic plasma parameters were estimated in each phase. We also made an attempt
to discuss relevant heating mechanisms.

\section*{Conclusions}
\noindent In conclusion it should be stressed that the dynamics of a~fibre $Z$--pinch substantially differs from
$Z$--pinches initiated from a metal wire. Even though ICF purposes caused a~carbon fibre $Z$--pinch to be of a
modest interest now, we believe that unique properties of carbon could provide valuable data not only for
$Z$--pinch physics but also for material science in which carbon is often used. Our experiment with a~carbon
fibre proved that it is possible to study a~lot of $Z$--pinch phenomena on a small device provided that a plasma
is thoroughly diagnosed. As an example we could mention the study of \emph{$Z$--pinch implosion} with the
velocity above 10$^5$ m\,s$^{-1}$ on the current level of 50 kA. Another interesting phenomenon is the implosion
\emph{onto a~fibre or wire}. First, it offers a new possibility of how to transfer the current with a sharp
rise--time in the vicinity of a~fibre or how to modify the shape of an X--ray pulse. Next, the fibre in the
centre of an imploding plasma introduces homogeneity to $Z$--pinch discharges. Last but not least, a~fibre can
be used as a target for an imploding plasma and can serve as a diagnostic tool.


 Our results also showed that dynamics of a~fibre $Z$--pinch significantly varies depending on a current
generator used. In the case of a slow current rise--time, the implosion of a coronal plasma onto a~fibre might
occur late in the time. This way, more energy stored in a capacitor bank could be transferred into a plasma.

Furthermore, the current waveform of a low impedance and low voltage capacitor is quite sensitive to $Z$--pinch
behaviour, especially to the change of the $R_\mathrm{P}+\dot{L}_\mathrm{P}$ term. Although this fact is
considered to be a disadvantage, it can serve well some diagnostic purposes. Since the voltage and current can
be quite easily measured on a low voltage generator, we can calculate a plasma resistance and energetics of
a~$Z$--pinch. Moreover, if we assume cylindrical symmetry and Spitzer resistivity, the resistance and
spectroscopically measured temperature determine the diameter of a plasma column in which the current is
flowing. Then this diameter can be compared with the one estimated from schlieren or pinhole images. Thus, we
are coming to one of the main problem of $Z$--pinches: what does the distribution of a current flowing through a
plasma look like? Of course, it is necessary to measure the current distribution locally. However, the above
mentioned technique gives at least a rough estimation of a diameter.

\section*{Future prospects}
\noindent Present investigation describes only the gross dynamics of our fibre $Z$--pinch. All of observed
phenomena could be studied in greater extent. Further experimental work depends on the subject of interest,
i.e.~which of the observed phenomena we want to study in detail. According to this interest, the diagnostics
should be improved. One may attempt to increase the current, to increase the length of a~fibre, to use more
fibres in one shot, to detect a low density corona in the latent phase, to detect charged particles, to perform
X--ray spectroscopic measurement of hot spots, etc. But what should be improved in any case, is the modelling
without which it is almost impossible to get deeper insight into $Z$--pinch phenomena.

It is well known that a~fibre $Z$--pinch does not belong to the main stream of $Z$--pinch research because it is
not such efficient radiation source as high-$Z$ wires. That is why experiments with carbon fibres are limited to
basic research, i.e.~to the study of general phenomena such as plasma implosion, instabilities, bright spots,
resistance, current distribution, etc. However, the future work is not restricted to a~carbon fibre. The
contribution of our experiments lies also in the development of a comprehensive set of diagnostics, which could
be used for various $Z$--pinch loads and configurations. \pagebreak

\fancyhead[R]{\slshape \rightmark}

%% file: references.tex
\pagestyle{fancy} \lhead{} \fancyhead[R]{\slshape Bibliography}
\fancyfoot[C]{\thepage}

\bibliographystyle{apalike}
\bibliography{thesis}


%% file: appendix.tex
\newpage

\appendix
\pagestyle{fancy} \lhead{}
\fancyhead[R]{\slshape \rightmark} \fancyfoot[C]{\thepage}
\chapter{Applied Spectroscopic Methods}
\label{spectromethods} \noindent In order to introduce fundamental terms\footnote{Partial LTE, optical depth,
rate equations, etc.} and to describe spectroscopic methods, we present the basic theory of emission
spectroscopy. As there are numerous spectroscopic methods (see, for example, \citealp{Griem64, Griem74, Griem97,
deMichelis81, Hauer91, Thorne99}) we will deal only with those that were applied in our experiment. Besides the
aforementioned papers, our approach was also inspired by several lectures of prof.~H.\,J.~Kunze and by the PhD
thesis of one of his students~\citep{Aschke00}.

\section{Introduction}
\noindent The characteristic property of a plasma is the radiation in a~large spectral range. Emission
spectroscopy is therefore a diagnostic technique that has been applied to an evaluation of appropriate plasma
parameters since the very beginning of plasma physics \citep{Griem64,deMichelis81}. By emission spectroscopy it
is possible to measure plasma composition, electron temperature and density, ion temperature and density,
electric and magnetic field, plasma motion or turbulence, plasma dynamics, ionization dynamics,
etc.~\citep{Kunze89}. However, to obtain some information about a plasma, a~lot of things have to be taken into
account to exclude misinterpretation. That is also the reason why we would like to briefly delineate the process
of how the plasma emission is formed. We shall start with radiative and collisional processes which influence
the population of energy levels and hence the local emissivity of a plasma. After that we will deal with
radiative transfer. And finally we will describe how the plasma parameters were estimated.

\section{Radiative Processes}
\noindent Radiative processes in a plasma can be divided into three main groups according to the type of
transition. We can distinguish transitions (i) between free and free electrons, (ii) between free and bound
electrons, and (iii) between bound and bound electrons.
\begin{itemize}
\item[--]\emph{Free--free transitions} are associated with a gain or loss of an electron energy in the field of
an ion\footnote{The collision between two electrons cannot produce radiation by electric or magnetic dipole
processes.}.  The loss of the electron energy into the radiation is connected with the deceleration of electrons
and that is why we speak about Bremsstrahlung. The absorption of a photon by an electron is called the inverse
Bremsstrahlung. Free states are not quantised, therefore radiation/absorption is characterised by a broad
continuum.

\item[--]\emph{Free--bound transitions} correspond to the radiative recombination and photoionization processes.
Recombination continuum extends from each bound state of an electron and is characterised by edges. Similarly,
the absorption edges occur in the spectral dependence of the photoionization cross--section.

\item[--]\emph{Bound--bound transitions} occur when an electron bounded by an ion makes a transition to another
bound state. When the final state has lower energy than the initial one, the photon of well--defined energy is
emitted. The inverse mechanism is the photoexcitation. The radiation is characterised by emission or absorption
spectral lines.

The rate of spontaneous emission from an upper energy level ``u'' to a lower level ``l'' is defined by means of
the Einstein probability coefficient $A_\mathrm{ul}$ as follows:
\begin{equation}
\frac{\diff n_\mathrm{u}}{\diff t}\Bigg|_{\mathrm{u}\rightarrow
\mathrm{l}}=-A_\mathrm{{ul}} n_\mathrm{u}
\end{equation}
where $n_\mathrm{u}$ is the number density of electrons in the upper level. The overall rate of spontaneous
radiative transitions from the upper level ``u'' is given by the sum of the rates from the upper level ``u'' to
all lower levels
\begin{equation}
\frac{\diff n_\mathrm{u}}{\diff
t}=-\sum_{\mathrm{l}<\mathrm{u}}A_\mathrm{{ul}}
n_\mathrm{u}=-A_\mathrm{u\rightarrow} n_\mathrm{u}
\end{equation}
It is quantum electrodynamics that treats of the transition probability $A_\mathrm{ul}$ and offers selection
rules. As far as our experiments with carbon fibres are concerned, we identified only (allowed) electric dipole
transitions with $\langle l|e\vec{r}|u\rangle\neq 0$. The intercombination (spin--forbidden $\triangle
\mathbb{S}\neq 0$) lines are not usually observed for low--$Z$ elements, such as carbon, where the spin--orbit
interaction is small.
\end{itemize}

\begin{figure}[!h]
\centerline{\includegraphics{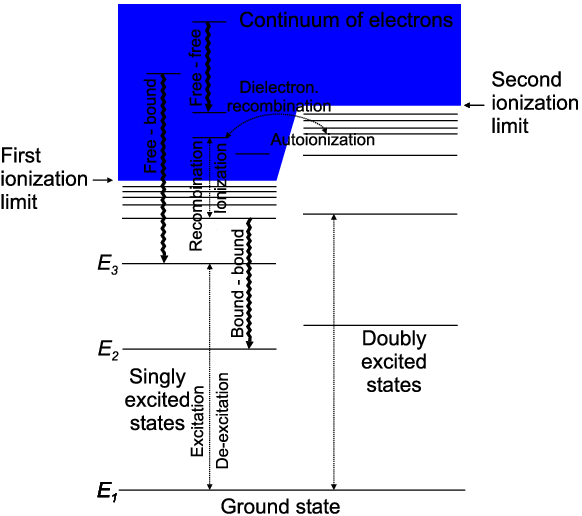}} 
\setlength{\captionmargin}{2cm} \caption[Several possible radiative and collisional processes]{Several possible
radiative and collisional processes. \small{(\mbox{Radiative processes are illustrated as wavy lines.})}}
\label{procesy}
\end{figure}

\section{Collisional Processes in Plasma}

\noindent Similarly to radiative processes, we can distinguish between \emph{collisional} excitation --
de-excitation, ionization -- recombination, etc. But on the contrary to radiative processes, collisional
transitions between two states do not follow any selection rules. The crucial collisional process in a plasma is
an \emph{interaction of a bound electron with free electrons}. The collision of a bound electron with ions is
much less significant due to the substantial difference in mass.

\newpage
When a free electron collides with an electron bounded in an ion A, excitation or ionization of a bound electron
can occur:

\vspace{0.2cm}
\begin{tabular}{l l l l}
 \hspace{1.5cm} A$^z+\mathrm{e}^-$ & $\leftrightarrow$ & A$^{z*}+\mathrm{e}^-$ &(excitation,
 de-excitation)\\[0.2cm]
 \hspace{1.5cm}  A$^z+\mathrm{e}^-$ & $\leftrightarrow$ & A$^{z+1}+\mathrm{e}^-+\mathrm{e}^-$ &(ionization, three-body
 recombination)
\end{tabular}
\vspace{0.2cm}

\noindent In three-body recombination, i.e.~the inverse of collisional ionization, an excited bound state of the
ion A$^{z}$ is formed and the excess energy is given to a free (non-captured) electron. The cross--section of
this reaction is increasing with the square of the electron density $n_\mathrm{e}$ and is larger for transitions
into highly excited levels. On the contrary to that, the radiative capture (radiative recombination) of an
electron occurs mainly into the ground state:

\vspace{0.2cm}
\begin{tabular}{l l l l}
 \hspace{1.52cm} A$^{z+1}+\mathrm{e}^-$ & $\leftrightarrow$ & A$^{z}+h\nu$ &\hspace{0.77cm} (radiative~capture, photoionization)
\end{tabular}
\vspace{0.2cm}

\noindent
 In the case of a~high--temperature and low density plasma
($n_\mathrm{e}<10^{16}\,\mathrm{cm}^{-3}$), the dominant recombination process is dielectronic recombination: a
free electron is resonantly captured into an excited level, with simultaneous excitation of one of the electrons
of the recombining ion A$^{z+1}$. A doubly--excited ion A$^{z**}$ is formed:

\vspace{0.2cm}
\begin{tabular}{l l l l}
\hspace{1.5cm} A$^{z+1}+\mathrm{e}^-$ & $\leftrightarrow$ & A$^{z**}$ &\hspace{1.3cm} (dielectronic recombin.,
autoionization)
\end{tabular}
\vspace{0.2cm}

\noindent
 As the doubly--excited state is above the first ionization limit of the recombined ion, it is
highly unstable. The two electron excited state can either undergo the inversion process (autoionization) or
radiatively decay by the emission of photons. Dielectronic recombination is responsible for the occurrence of
satellite spectral lines on the long--wavelength side of resonance lines. Satellites on the long--wavelength
side can be also emitted after the innershell excitation:

\vspace{0.2cm}
\begin{tabular}{l l l l}
\hspace{1.5cm} A$^{z}+\mathrm{e}^-$ & $\rightarrow$ & A$^{z**}+\mathrm{e}^-$ &\hspace{0.8cm} (innershell
excitation)
\end{tabular}
\vspace{0.2cm}

\noindent Lastly, we shall mention the charge--exchange recombination that happens during collision between two
ions or between an ion and atom. When the charge of the ion A is greater than that of the ion B, the electron
bounded in the ion A could be captured by the other ion B into a~highly excited level:

\vspace{0.2cm}
\begin{tabular}{l l l l}
\hspace{1.5cm} A$^{z_\mathrm{A}}+B^{z_\mathrm{B}}$ & $\rightarrow$
& A$^{z_\mathrm{A}-1}+B^{z_\mathrm{B}+1}$ &(charge--exchange
recombination)
\end{tabular}
\vspace{0.2cm}

\noindent
 The quantitative description of collisional
processes begins with a microscopic cross--section $\sigma (v)$.
The number of transitions ``a'' (of an ion A induced by electrons)
per unit time and per unit volume is given by equation
\begin{equation}
R_\mathrm{a}=-\frac{\diff n_\mathrm{A}}{\diff t}=\langle\sigma
(v)\cdot v\rangle
n_\mathrm{e}n_\mathrm{A}=C_\mathrm{a}n_\mathrm{A}
\end{equation}
where $\langle\sigma (v)\cdot v\rangle=\int\sigma (v)f_\mathrm{e}(v)v\diff v$ represents the so--called rate
coefficient. The rate coefficient is calculated by the averaging of $\sigma (v)\cdot v$ over the electron
velocity distribution $f_\mathrm{e}(v)$. For the excitation/deexcitation rate coefficient we usually use symbol
$X$. Symbol $S$ represents the ionization rate coefficient and symbol $\alpha$ stands for the recombination
coefficient. The quantity $C_\mathrm{a}$ is called the probability of collision ``a'' of one ion A with
electrons.

\section{Charge State Distribution and Population of Energy Levels}
\noindent The rate of radiative and collisional processes in a plasma determines the charge state distribution
and population of each energy level.

\subsection{Charge state distribution}
\noindent A given degree of ionization is given by ionization and recombination of two neighbouring ionization
degrees as well as by diffusion and convection of ions. In principle, the charge state distribution can be
solved from the equation of continuity
\begin{equation}
\frac{\partial n^{z}}{\partial t}+\nabla\cdot\Gamma_{z}=n_\mathrm{e}(n^{z-1}S_{z-1}-n^{z}S_{z})+
n_\mathrm{e}(n^{z+1}\alpha_{z+1}-n^{z}\alpha_{z})
\end{equation}
where $\Gamma_{z}$ stands for the flux density of ions with the charge $z$.

Some simplification can be made if transport of ions is neglected, i.e.~\begin{equation} \frac{\partial
n^{z}}{\partial t}=n_\mathrm{e}(n^{z-1}S_{z-1}-n^{z}S_{z})+ n_\mathrm{e}(n^{z+1}\alpha_{z+1}-n^{z}\alpha_{z})
\end{equation}
In $Z$--pinch plasmas, one has to usually solve a~time--independent system of equations
\begin{equation}
0=n_\mathrm{e}(n^{z-1}S_{z-1}-n^{z}S_{z})+ n_\mathrm{e}(n^{z+1}\alpha_{z+1}-n^{z}\alpha_{z})
\end{equation}
which leads to the so--called \emph{ionization equilibrium}. This set of rate equations could be solved if
plasma parameters and the recombination  and ionization rate coefficients $\alpha$, $S$ are known.

\subsection{Level populations}
\noindent Once the calculation of the charge state distribution was described, we are going to discuss the level
populations in the particular ionization stage of interest.

 In order to simplify notation, the rate of all collisions filling the upper level ``u'' will be marked as
$R^\mathrm{c}_{\rightarrow \mathrm{u}}$. Similarly, the rate of radiative processes will be marked as
$R^\mathrm{r}_{\rightarrow \mathrm{u}}$ and the rate of collisions depopulating the level ``u'' as
$R^\mathrm{c}_{\mathrm{u}\rightarrow}=C_{\mathrm{u}\rightarrow}n_\mathrm{u}=X_{\mathrm{u}\rightarrow}n_\mathrm{e}n_\mathrm{u}$.
The symbol $A_{\mathrm{u}\rightarrow}$ will stand for the radiative decay of the level ``u''. By using these
quantities we can write the following rate equation
\begin{equation}
\label{rychlostnirovnice} \frac{\diff n_\mathrm{u}}{\diff t}=R^\mathrm{c}_{\rightarrow
\mathrm{u}}+R^\mathrm{r}_{\rightarrow
\mathrm{u}}-C_{\mathrm{u}\rightarrow}n_\mathrm{u}-A_{\mathrm{u}\rightarrow}n_\mathrm{u}
\end{equation}
 Compared with atomic time constants $\tau_\mathrm{u}=1/(A_{\mathrm{u}\rightarrow}+C_{\mathrm{u}\rightarrow})$,
  $n_\mathrm{e}$ and $T_\mathrm{e}$ change usually slowly. It means that the population
density $n_\mathrm{u}$ could be assumed to be in quasi--equilibrium, i.e.~$\diff n_\mathrm{u}/\diff t=0$. From
the rate equation~\ref{rychlostnirovnice} we therefore obtain the population of upper level in
\emph{collisional--radiative equilibrium}
\begin{equation}
\label{quasiequilibrium}
 n_\mathrm{u}=\frac{R^\mathrm{c}_{\rightarrow
\mathrm{u}}+R^\mathrm{r}_{\rightarrow
\mathrm{u}}}{A_{\mathrm{u}\rightarrow}+C_{\mathrm{u}\rightarrow}}
\end{equation}
An analytic solution taking into account all processes is too complicated and so it is necessary to use
numerical simulation (for our purposes we used the time--dependent atomic physics code FLY described in
appendix~\ref{appendixFLY}). However, simplification of equation~\ref{quasiequilibrium} is still possible in two
limiting cases. The first case is a low density plasma with $n_\mathrm{e}<10^{14}$~cm$^{-3}$ where the
population follows the \emph{corona equilibrium}. In  what follows, we will restrict to the second limiting
case, to a~high density plasma \emph{in local thermodynamic equilibrium} (LTE).

\newpage
\section{Local Thermodynamic Equilibrium}
\noindent The criterion for LTE is that the collisional rates must be much larger than radiative rates. The
radiative terms in Eq.~\ref{quasiequilibrium} can be then neglected and we get
\begin{equation}
\label{LTE1}
 n_\mathrm{u}=\frac{R^\mathrm{c}_{\rightarrow
\mathrm{u}}}{C_{\mathrm{u}\rightarrow}}=\frac{n_\mathrm{e}X_{\rightarrow
\mathrm{u}}}{X_{\mathrm{u}\rightarrow}}
\end{equation}
From the principle of the detailed balance\footnote{Every process must be balanced by its exact inverse.}
follows that
\begin{equation}
\label{LTE2} \frac{X_{\mathrm{l}\rightarrow
\mathrm{u}}}{X_{\mathrm{u}\rightarrow\mathrm{l}}}=\frac{g_\mathrm{u}}{g_\mathrm{l}}\mathrm{e}^{-\frac{E_\mathrm{u}-E_\mathrm{l}}{kT_\mathrm{e}}}
\end{equation}
where $g_\mathrm{u}$, $g_\mathrm{l}$ are statistical weights for levels with energies $E_\mathrm{u}$ and
$E_\mathrm{l}$, respectively. Applying Eq.~\ref{LTE2} to any two levels in atomic species and combining it with
Eq.~\ref{LTE1}, it is possible to derive the well--known Boltzmann distribution for the level population
densities
\begin{equation}
\frac{n_\mathrm{u}}{n_\mathrm{l}}=\frac{g_\mathrm{u}}{g_\mathrm{l}}
\mathrm{e}^{-\frac{E_\mathrm{{u}}-E_\mathrm{{l}}}{kT_\mathrm{e}}}=\frac{g_\mathrm{u}}{g_\mathrm{l}}
\mathrm{e}^{-\frac{h\nu}{kT_\mathrm{e}}} \label{Boltzmann}
\end{equation}
When we extend Boltzmann relation to continuum states~\citep{Griem97} we can arrive at the Saha equation
\begin{eqnarray}
\frac{n_\mathrm{e}n^{z+1}}{n^z}&=&2\frac{Q^{z+1}(T_\mathrm{e})}{Q^z(T_\mathrm{e})} \left( \frac{2\pi
m_\mathrm{e}kT_\mathrm{e}}{h^{2}}\right)^{3/2} \mathrm{e}^{-\frac{E_\mathrm{ion}}{kT_\mathrm{e}}} \nonumber\\
\frac{n_\mathrm{e}n^{z+1}}{n^z}&=&6.10^{21}\frac{Q^{z+1}(T_\mathrm{e})}{Q^z(T_\mathrm{e})}T_\mathrm{e}^{3/2}\mathrm{e}^{-\frac{E_\mathrm{ion}}{T_\mathrm{e}}}\quad[\mathrm{eV;}\,\mathrm{cm}^{-3}]
\label{Saha}
\end{eqnarray}
which describes the total densities of adjacent ionization stages $n^z$ and $n^{z+1}$. The term $+ze$ represents
the effective charge. $E_\mathrm{ion}$ stands for an ionising energy of an ion/atom with the density $n^z$.
$Q(T_\mathrm{e})$ is the partition function defined by
\begin{equation}
Q^z(T_\mathrm{e})=\sum_ig_i^z\cdot \mathrm{e}^{-\frac{E_i-E_1}{kT_\mathrm{e}}}
\end{equation}
\vspace{-0.7cm}

The sum goes over all bounded states $i$. The number of terms that are included in the partition function
depends not only on the plasma temperature but also on the density.

Since the collision processes are dominant we can further assume that the velocities of ions and electrons are
described by a Maxwell distribution with a temperature identical with that of the Boltzman distribution.

To sum up, local thermodynamic equilibrium represents the equilibrium between particles. Then the plasma can be
described by the Maxwell, Boltzmann and Saha distribution with the same temperature $T_\mathrm{e}$. Temperature
$T_\mathrm{e}$ may vary from one place to another, hence the name ``local''. The \emph{complete} thermodynamic
equilibrium is further described by the Planck blackbody function for radiation energy with the same temperature
$T=T_\mathrm{e}$ as previous distributions. This may not to be valid for plasma in LTE where the radiation
density could be below the blackbody level.

\section{Prerequisites for Local Thermodynamic Equilibrium}
\noindent The first necessary condition of LTE is a Maxwell distribution of electron velocities.

Another condition is given by the validity of reducing equation~\ref{quasiequilibrium} to equation~\ref{LTE1}.
It means that inequality  $A_{\mathrm{u}\rightarrow}\ll C_{\mathrm{u}\rightarrow}$ has to be satisfied. Usually
a factor of ten is considered as sufficiently high, i.e.~$10\ A_{u\rightarrow}\leq C_{u\rightarrow}$.
Unfortunately, it is impossible to derive any easy criterion for \emph{complete} LTE (i.e.~for populations of
all levels of all ionic states). The review of~\citet{deMichelis81} provides the necessary, though not
sufficient, criterion
\begin{equation}
n_\mathrm{e}\geq 1.8\times10^{14}\sqrt{kT_\mathrm{e}}\Delta E^3_\mathrm{ul} \quad [\mathrm{eV; cm}^{-3}]
\end{equation}
This criterion is difficult to satisfy especially for the largest energy level difference $\Delta E_\mathrm{ul}$
in the level scheme. Nevertheless, for any value of density $n_\mathrm{e}$, the collisions still dominate
between high--lying levels. Level, above which excited states are in equilibrium with the ground state of the
next ionization stage, is called thermal limit. Then we talk about a~plasma in \emph{partial} LTE. We use the
following criterion~\citep{Griem64}
\begin{equation}
\label{kriteriumLTE} n_\mathrm{e}\geq
7\times10^{17}\frac{(z+1)^6}{n_\mathrm{TH}^{17/2}}\sqrt{\frac{kT_\mathrm{e}}{E_\mathrm{H}}} \quad [\mathrm{eV;
cm}^{-3}]
\end{equation}
that determines the minimal electron density $n_\mathrm{e}$ which is needed for partial LTE of levels above the
principal quantum number $n_\mathrm{TH}$. Here $z$~is the effective nuclear charge and $E_\mathrm{H}\doteq13,6$
eV is the ionization energy of hydrogen.

As regards our experiment, it follows from equation~\ref{kriteriumLTE} that the LTE approximation cannot be
applied for K--shell lines of carbon ions. However, if the electron density is higher than a few times of
10$^{19}$ cm$^{-3}$, PLTE may be valid for levels $n\geq2$ of Li--like oxygen ions (see
Fig.~\ref{electrodeSPECTRUM} on page~\pageref{electrodeSPECTRUM}).


\newpage

\section{Radiative Transfer}
\noindent Until now we have been interested in the influence of collisional and radiative processes on the
population of energy levels. It was important since the population of energy level $n_\mathrm{u}$ and the
transition probability of spontaneous emission $A_\mathrm{ul}$ determine the local emission in a plasma. The
local emission is described by the emissivity, the power emitted per unit solid angle per Hz by unit volume
\begin{equation}
\epsilon_\nu=\frac{\diff^3 \Phi}{\diff V\diff \Omega\diff \nu}
\end{equation}
The emissivity is given by all radiative transitions u$\rightarrow$l
\begin{equation}
\epsilon(\mathrm{u}\rightarrow \mathrm{l})=\frac{h\nu}{4\pi}A_{\mathrm{ul}}n_\mathrm{u}
\end{equation}

The importance of emissivity lies in the fact that it contains information about local properties of the plasma.
However, it is impossible to measure the emissivity directly by passive emission spectroscopy. We usually
measure the spectral radiance $L_\nu$ at the surface of a plasma: the power emitted per unit solid angle per Hz
by unit projected area
\begin{equation}
L_\nu=\frac{\diff^3 \Phi}{\diff A \cos\theta \diff \Omega\diff
\nu}
\end{equation}
 The relation between these two quantities is trivial
only in the case of an optically thin plasma where the absorption and re-emission of photons are neglected.
Otherwise we have to solve the equation of radiative transport, which may be written for the line of sight along
the $x$--axis as
\begin{equation}
\label{radiativetransfer} \diff L_{\nu}(\vec{r},\Omega)=\bigg[
\epsilon_{\nu}(\vec{r},\Omega)-\kappa(\nu,\vec{r})
L_{\nu}(\vec{r},\Omega)\bigg] \diff x
\end{equation}
Here $\kappa(\nu,\vec{r})$ represents the effective absorption coefficient that could be negative if stimulated
emission occurs. Now we shall define a quantity called the source function $S_{\nu}(\vec{r})$. It can be
expressed  as the ratio of the emissivity to the absorption coefficient
\begin{equation}
 S_{\nu}(\vec{r},\Omega)=\frac{\epsilon_{\nu}(\vec{r},\Omega)}{\kappa(\nu,\vec{r})}
\end{equation}
And by using the optical depth $\tau(\nu,\vec{r})$
\begin{equation}
\diff \tau(\nu,\vec{r})=\kappa(\nu,\vec{r})\diff x
\end{equation}
we can rewrite equation \ref{radiativetransfer} as follows:
\begin{equation}
\frac{\diff L_{\nu}(\vec{r},\Omega)}{\diff \tau(\nu,\vec{r})}=S_{\nu}(\vec{r},\Omega)-L_{\nu}(\vec{r},\Omega)
\label{radiativetransfer2}
\end{equation}
Unfortunately, this equation cannot be solved in general  because the emissivity and absorption coefficient may
have different spectral dependence. Moreover, the absorption of photons influences the population of energy
levels and thus it is necessary to solve the equation \ref{radiativetransfer2} together with the system of
equations~\ref{rychlostnirovnice}.

The solution of equation~\ref{radiativetransfer2} can be found only in limiting cases. If we suppose the
homogenous plasma with the characteristic length along the line of sight $l_\Omega$ and the spatially isotropic
source function $S_{\nu}$ we arrive at the relation
\begin{equation}
L_{\nu}=S_{\nu}(1-\mathrm{e}^{-\tau(\nu)})
\end{equation}
where $\tau(\nu)=\int\diff \tau(\nu,\vec{r})=\kappa(\nu) l_\Omega$.

\noindent One of the limiting cases occurs when the plasma is optically thin \mbox{($\tau\ll 1$)}, i.e.~without
absorption, we can use the Taylor series $\mathrm{e}^{-\tau(\nu)}\approx 1-\tau(\nu)$. Then we obtain
\begin{equation}
L_{\nu}\approx\int\epsilon_{\nu}(\vec{r},\Omega)\diff
x=\epsilon_{\nu}l_\Omega
\end{equation}
Another limiting case is when $\tau\gg 1$. Then we can derive
\begin{equation}
\label{opticallythick}
L_{\nu}=S_{\nu}=\frac{\epsilon_{\nu}}{\kappa(\nu)}=\frac{2h\nu^3}{c^2}\cdot\frac{1}{\mathrm{e}^{h\nu/kT}-1}
\end{equation}
The last equality in equation~\ref{opticallythick} is valid only in the case of the plasma in LTE. The limiting
case $\tau\gg 1$ is not reached in the entire spectral range. Since the absorption increases with $\lambda^2$,
the plasma becomes optically thick towards long wavelength~\citep{Kunze89}. As far as shorter wavelengths are
concerned, the emission lines become optically thick more readily than continuum.

\newpage
From the spectroscopic point of view, it is necessary to recognise whether a spectral line is optically thick.
For Doppler--broadened lines and the plasma in LTE, the optical thickness at the line centre $\nu_0=c/\lambda_0$
is given by
\begin{equation}
\tau(\nu_0)=r_\mathrm{B}\sqrt{\frac{\pi M_\mathrm{i}c^2}{2kT}
}\lambda_0f_{\mathrm{lu}}n_\mathrm{l} l_\Omega
\end{equation}
where $r_\mathrm{B}=e^2/4\pi\varepsilon_0m_\mathrm{e}c^2$ is the classical (Bohr) electron radius. When
substituting known constants in the previous equation, we get the relation
\begin{equation}
\tau(\nu_0)\doteq1,08.10^{19}\sqrt{\frac{M_\mathrm{i}}{2kT}} \lambda_0f_\mathrm{lu}n_\mathrm{l} l_\Omega
\end{equation}
Here the ion mass $M_\mathrm{i}$ is in atomic units $m_\mathrm{u}$, temperature $kT$ is in~eV, the population of
lower level $n_\mathrm{l}$ is in~m$^{3}$, the wavelength $\lambda_0$ is in~nm, and the length of plasma
$l_\Omega$ is in~m.

Whether the emission lines are optically thick or thin, can be checked experimentally by comparing relative
intensities of close multiplet lines. The radiance ratio of two optically thin multiplet lines is determined by
the statistical weights of lower levels $g_\mathrm{l}$ and the oscillator strengths $f_\mathrm{lu}$
\begin{equation}
\frac{L_{\nu_1}}{L_{\nu_2}}=\frac{f_{\mathrm{l}_1\mathrm{u}_1}g_{\mathrm{l}_1}}{f_{\mathrm{l}_2\mathrm{u}_2}g_{\mathrm{l}_2}}
\end{equation}
If the measured ratio is higher in favour of the weaker line, the spectral lines are not optically thin. As
a~result, the temperature cannot be easily determined by the intensity ratio of these spectral lines.

As regards our experiment, we found that the C VI Ly--$\alpha$ line and O VI 2p--3d line were almost always
optically thick.

\section{Intensity of Spectral Lines}
\noindent In the scientific literature there is often used the term \emph{intensity} of a spectral line $I$. The
fact of the matter is that the correct term is the spectral radiance $L_\nu$ (mostly spatially and/or temporally
integrated). However, as the term ``intensity'' became common, we use it in this thesis in spite of the fact
that the spectral radiance would be more appropriate.

\newpage
\section{Spectroscopic Measurements of Plasma Parameters}
\noindent The emission spectroscopy of plasma belongs to one of the fundamental passive diagnostic technique.
The emissivity of plasma contains much information about plasma properties and dynamics. The key issue is the
choice of a suitable method which enables the estimation of plasma parameters, especially the electron
temperature and density. Our choice of a method was predetermined by the properties of our grating spectrograph
(see subsection~\ref{spectrograph} on page~\pageref{spectrograph}). The spectrograph enabled us to measure
relative intensities of spectral lines only. The measurement of line profile was not possible due to limited
spectral resolution. That is why the manifold choices of spectroscopic techniques (cf.~\citealp{Griem64,
Griem97, deMichelis81}) are reduced to only several fundamental methods.

\subsubsection{Spectroscopic measurement of temperature and density} \noindent First, it is important to mention
that spectroscopic measurement of temperature is actually the measurement of a parameter determining certain
equilibrium distribution \citep{Thorne99}. We can thus distinguish between the electron temperature
$T_\mathrm{e}$ and the ion temperature $T_\mathrm{i}$. Moreover, we can distinguish between the electron
temperature occurring in the Maxwell velocity distribution, in the Boltzmann distribution~\ref{Boltzmann}
(so-called population or excitation temperature), in the Saha--Boltzmann equation~\ref{Saha} (ionization
temperature), etc. Unless any equilibrium distribution exists, the temperature cannot be defined and the plasma
has to be described by distribution functions themselves. Most of the spectroscopic temperature measurements
primarily yield electron temperature $T_\mathrm{e}$. The same was true in our case. The kind of the electron
temperature (excitation, ionization, etc.) follows from the method used and it is not specified in our work.

In this thesis we applied the following spectroscopic methods of the temperature and density measurement:
\begin{itemize}
\item[--] Very easy and undemanding method of a rough temperature measurement is the identification of ionic
states in a plasma. It holds that an ionic state with ionization potential $E_\mathrm{ion}$ is reached in a
plasma when the electron temperature obeys
\begin{equation}
\label{odhadteploty}
 T_\mathrm{e}>C_\mathrm{i} E_\mathrm{ion}
\end{equation}
with $C_\mathrm{i}$ varying approximately between 0.1 and 0.3 \citep{Hauer91}.

More accurate value of temperature can be obtained from calculation of ionization equilibrium (the temperature
dependence of population of various ionic states). The free parameter in this calculation is the electron or ion
density. To calculate ionization equilibrium we used atomic physics code FLY.

\item[--] Another common method is based on relative line intensities of the same ion. Provided that the plasma
is in LTE or at least in partial LTE, the population of energy levels follows the Bolzmann
distribution~\ref{Boltzmann}. Then the ratio of two optically thin spectral lines is given by
\begin{equation}
\label{ratio}
\frac{L_{\nu}}{L_{\nu'}}=\frac{A_\mathrm{ul}g_\mathrm{u}\lambda '}
{A_\mathrm{u'l'}g_\mathrm{u'}\lambda}\cdot
\mathrm{e}^{-(E_\mathrm{u}-E_\mathrm{u'})/kT_\mathrm{e}}
\end{equation}
It can be clearly seen from Eq.~\ref{ratio} that, except several constants, the ratio is dependent only on the
electron temperature. The sufficient sensitivity is achieved for $E_\mathrm{u}-E_\mathrm{u'}> kT_\mathrm{e}$.
This method can be extended to a~higher number of lines than two. We can make a graph with $\ln (L_{\nu}
\lambda/g_\mathrm{u} A_\mathrm{ul})$ on the $x$--axis and with the energy of upper level $E_\mathrm{u}$ on the
$y$--axis. The slope of the line then equals $-kT_\mathrm{e}$. This method was used in cases when the code FLY
could not be applied, e.g.~for Be--like ions.

\item[--] In the regions where (partial) LTE approximation is no more valid, the ratio of two spectral lines is
not only temperature--, but also density--dependent. Under such circumstances a detailed accounting must be made
of the ionization, recombination, excitation, deexcitation and radiative transfer. As far as this work is
concerned, we used the code FLY not only to calculate the ratio of line intensities in non--LTE approximation
(usually in collisional--radiative equilibrium) but also to reconstruct the whole synthetic spectrum for a
particular temperature and density. Non--LTE requires more complicated treatment than LTE approximation, but on
the other hand, it could provide the estimation of density which was very needed since the other methods of
density measurement\footnote{For instance, Stark broadening, depression of ionization potential, Inglis--Teller
limit, the ratio of resonance and intercombination lines of He--like ions.} cannot be applied.
\end{itemize}

\chapter{Code FLY}
\label{appendixFLY}

\section{Basic Parameters}
\noindent FLY~\citep{Lee95,Lee96} is a commercially available suite of computer codes which was developed for
analysing experiments where K--shell spectra\footnote{K--shell spectroscopy is engaged in the radiative
transitions to the most inner shell (K--shell) of ions.} are observed. The FLY suite is capable of simulating
K--shell emitters with the atomic number $Z$ from 2 to 26, i.e. from helium to iron. The suite allows the user
to calculate the population of the states in an LTE or non--LTE approximation by assuming steady--state or a
time dependent evolution. The kinetics model FLY provides details of the level populations of \mbox{hydrogen--,}
helium-- and lithium--like ions. Less ionised species are represented as a ground state only. FLY is a
``zero--dimension" suite of codes, thus the plasma is specified with the local conditions. The optical depth
effects are approximated for a homogenous plasma ``column" only. The most important characteristics of the suite
can be summarised in the following items:
\begin{itemize}
\item[--] $0$-dimension post-processing of MHD. \vspace{-0.2cm}\item[--] Atomic number $Z$ from 2 to 26, i.e.
from helium to iron. \vspace{-0.2cm}\item[--] K--shell spectroscopy: detailed information on H-, He- a Li- like
ion stages. K-shell satellite lines and also continuum radiation are included. \vspace{-0.2cm}\item[--]
Steady-state (either LTE, or non-LTE) as well as time--dependent solution of rate equations.
 \vspace{-0.2cm}\item[--] Line profile synthesis: Doppler profile. Stark broadening
for specific transition only. \vspace{-0.2cm}\item[--] Effect of optical depth within the escape factor
approximation. \vspace{-0.2cm}\item[--] Possibility of external radiation field. \vspace{-0.2cm}\item[--] Easy
to use. Comparison of the synthetic spectrum with the experimental one.
\end{itemize}

\section{Description}
\noindent The suite of code consisted of three parts\ -- FLY, FLYPAPER and FLYSPEC codes.

\section{FLY}
\noindent The first code in the suite, FLY, calculates the level populations of hydrogen--, helium-- and
lithium--like ions. The population of energy levels can be solved either in non--LTE or LTE case. The rate
equations for populations include the most relevant ionising, recombining, exciting and deexciting processes.
The necessary input is the atomic number of interest together with the history of magnetohydrodynamic
parameters. The other possible input is the grid of temperature and density for which the populations will be
calculated. When the time history of plasma parameters is known, the code calculates a~time--dependent evolution
of the populations. When a grid of temperatures and densities is the input, the code computes a steady state
approximation of population levels. The output file of the FLY code serves as an input into the other two codes
of the suite --- FLYPAPER and FLYSPEC.

\section{FLYPAPER}
\noindent FLYPAPER code can be used to evaluate:

\begin{itemize}
\item[--] populations of levels, \vspace{-0.2cm}\item[--] the ratio of line intensities,
\vspace{-0.2cm}\item[--] optical depth
\end{itemize}

as a function of the time or plasma parameters (density and
temperature).

\section{FLYSPEC}
\noindent FLYSPEC code allows the reconstruction of the synthetic spectrum for a particular time or for
particular temperature and density. The line profiles can be calculated for certain transitions: the Lyman and
Balmer series for hydrogen--like ions, 1s$^{2}$ $^{1}$S$_{0}$ -- 1s\,np $^{1}$P transitions for helium--like
ions, and transitions to the ground levels 2s, 2p in the case of lithium--like ions. The recombination emission
is calculated for mentioned series only. In addition, the user can perform integrations of spectra over time.

\section{Code Limitation}
\noindent Firstly, it should be stressed that the FLY suite is a $0$--dimension code. Therefore, the~correct
treatment of radiative transfer (which is a non--local problem especially where a~density gradient occurs) is
not possible. Moreover, only Doppler broadening is incorporated in the FLY code and therefore the optical depth
effects are overestimated.

Secondly, if we need to obtain synthetic spectra of L-- or M-- shell emitters we also need to use another code
than FLY. This is true also for detail description of a Li--like ionization state because a lower excited state
(Be--like) is described only by a~single ground state\footnote{This restriction was overcome in FLYCHK, the
extension of FLY~\citep{Chung03}.}. Since the FLYSPEC code includes recombination for certain transition only,
the continuum spectrum is underestimated.

Thirdly, the appropriate use of any post--processors assumes that preceding MHD modelling of plasma evolution is
not influenced by the fact that the detail radiative processes are not included. This condition may not be
fulfilled in a strongly radiating plasma consisting of high--$Z$ elements.

Fourthly, another serious problem represents non--Maxwellian distribution of electrons since it often occurs
along the $z$--axis in a~$Z$--pinch plasma and cannot be described by means of electron temperature.

Lastly, it should be mentioned that the FLYSPEC programme provides insufficient instrumental broadening. In
order to compare the synthetic spectra with the experimental ones, we have to therefore process the synthetic
spectra with a spreadsheet programme.

\section{Platform}
\noindent The kernel of the suite FLY was written in FORTRAN77 language. Running the code does not require any
atomic and spectroscopic data since they are internally included. Programme runs on any UNIX/Linux workstation
as well as on desktop PC under MS-DOS.

\chapter{Deuterated Fibre Experiment on S--300 facility}
\label{appendixS300}
\input{s300}

\chapter{MATLAB Simulation of $R$--$L$--$C$ Circuit \mbox{with Time--Dependent Resistance}}
\label{appendixMATLAB} \noindent The equivalent circuit of our $Z$--pinch is shown in Fig.~\ref{RLC}. We can
substitute this circuit by the $R$--$L$--$C$ circuit where $C$ stands for $C_0$, $L$ stands for
$L_0+L_\mathrm{P}(t)$, and $R(t)$ stands for $R_0+R_\mathrm{P}(t)+\dot{L}_\mathrm{P}(t)$. Because the plasma
inductance $L_\mathrm{P}$ is much lower than $L_0$ and depends  on the pinch radius logarithmically, we consider
the inductance $L$ to be constant. It is therefore possible to solve the $R$--$L$--$C$ circuit with a
time--dependent resistance only.
\begin{figure}[!h]
\centerline{\includegraphics{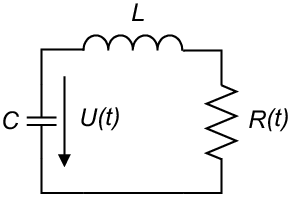}} \caption[$R$--$L$--$C$ circuit with a time--dependent
resistance]{$R$--$L$--$C$ circuit with a time--dependent resistance.} \label{RLCsubstitute}
\end{figure}

 \noindent
  Then the voltage $U(t)$ is given by
\begin{equation}
U(t)=U_0-\int_0^t \frac{I(t')}{C}\diff t'=L\frac{\diff I(t)}{\diff
t}+R(t)I(t)
\end{equation}
By differentiating this equation we obtain
\begin{eqnarray}
- \frac{I(t)}{C} &=& L\ddot{I}(t)+R(t)\dot{I}(t)+\dot{R}(t)I(t)\\
0&=&\ddot{I}(t)+\frac{R(t)}{L}\dot{I}(t)+\Bigg(\frac{\dot{R}(t)}{L}+\frac{1}{CL}\Bigg)I(t)
\end{eqnarray}
This ordinary differential equation was solved in MATLAB by the \verb"ode45" function for various dependencies
of resistance $R(t)$. We were looking for such resistance $R(t)$ which produced $U_\mathrm{P}$ and $\diff
I/\diff t$ similar to experimentally measured waveforms.

%% file: s300.tex
\noindent We have studied fibre $Z$--pinch not only on a small device $Z$--150 but also on a~large pulsed power
facility S--300 in Moscow. This appendix presents the results that were accepted for publishing in Physica
Scripta in 2006.

\section{Introduction}
\noindent The dense $Z$--pinches initiated from cryogenic deuterium fibres were investigated in the 80's and
90's in connection with the research of controlled thermonuclear fusion and radiative collapse \citep{Scudder85,
Sethian87, Decker89, Lebedev98}. The plan was to heat and ionise the fibre from frozen deuterium and to confine
the high density and high temperature plasma column within a small diameter. The fibres from deuterated polymer
were also employed \citep{Stephanakis72, Young77, Kies91, Mitchell98} because their discharge behaviour was
roughly the same (cf.~\citealp{Lebedev98}) and at the same time they were easily available and could be handled
much easier than the frozen deuterium ones. Unfortunately, the development of MHD instabilities and global
expansion of a pinch column were observed from the very beginning of the discharge and so the idea of the fibre
$Z$--pinch as a fusion reactor was abandoned.

The purpose of our fibre experiments on the S--300 generator was (i) to compare results with the wire--array
implosion onto a central fibre~\citep{Klir04a, Klir05} and (ii) to study the generation of XUV and X--ray
radiation together with the neutron production which gives insight into the acceleration of ions and hence into
the processes taking place in $Z$--pinches.

\section{Apparatus and Diagnostics}

\subsection{Current generator and $Z$--pinch load}
\noindent So far most experiments with dielectric fibres have been pursued on high impedance (about 1 $\Omega$)
pulsed power generators since they are better optimised to drive currents into a~high impedance load, which a
dielectric fibre surely is (cf.~\citealp{Mitchell96}). Despite this, however, we carried out fibre $Z$--pinch
experiments on a low impedance S--300 generator (3.5 MA, 100 ns, 0.15 $\Omega$, \citealp{Chernenko96}) at the
Kurchatov Institute in Moscow. The experiments were performed at a current level of 2 MA with a rise time of
about 150 ns. The $Z$--pinch was formed in a vacuum chamber from a~deuterated polyethylene fibre of 100 $\muup$m
diameter and 1 cm length.

\subsection{Diagnostics}
\begin{figure}[!h]
\centerline{\includegraphics{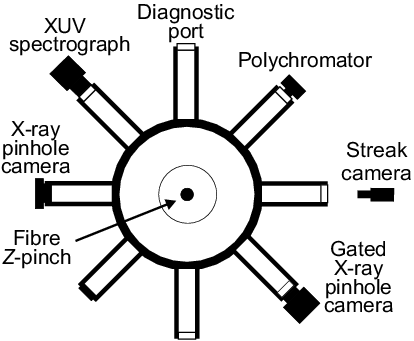}} \caption[Diagnostic set--up on the S--300 generator,
Moscow]{Diagnostic set--up on the S--300 generator, Moscow.} \label{S300setup}
\end{figure}

\noindent In order to study dynamics of $Z$--pinch plasma, an extensive set of diagnostic tools was used. The
diagnostic set--up is displayed in Fig.~\ref{S300setup}.

First, to provide time and space resolved information about visible emission, an optical streak camera was used.
The streak camera was performed in the radial mode, i.e.~with a slit parallel to the $Z$--pinch axis.

Second, X--ray radiation was detected with two X--ray pinhole cameras, XUV grazing incidence spectrograph, and
11-channel soft X--ray polychromator. One X--ray pinhole camera, which was gated, recorded 4 frames with 2 ns
exposure and 10 ns inter-frame separation. All four frames were filtered with 24 $\muup$m thick beryllium. The
other pinhole camera, time--integrated and differentially filtered (without a filter, and with 5 $\muup$m and 24
$\muup$m mylar), was used to observe the plasma in various spectral ranges with the spatial resolution of 100
$\muup$m. Even better spectral information was obtained by a time--integrated XUV grazing incidence spectrograph
which recorded carbon K--shell lines. Time resolved studies of soft X--ray emission was conducted using an
11-channel polychromator. The various combinations of glancing incidence mirrors, transmission filters, and
semiconductor detectors enabled the detection of photons in channels of 50, 80, 120, 180, 270, 365, 600, 800,
1000, 1200 and 2200 eV.

And finally, as far as neutrons are concerned, a time--of--flight analysis of neutrons was made possible by two
axially positioned SSDI--8 scintillators at the distance of 2.70 m and 7.45 m (downstream). The neutron yield
was measured by an indium activation counter. Since the activation counter has not been calibrated in situ and
since we have some experimental evidence that the yield was underestimated by one order of magnitude, it is
possible that the neutron yields should be multiplied by 10 throughout this article and
references~\citep{Klir04a, Klir05}.

\section{Experimental Results}

\subsection{Current waveform and X--ray emission}
\begin{figure}[!h]
\centerline{\includegraphics{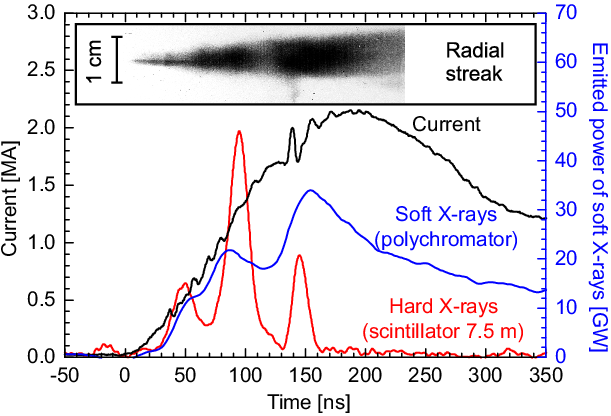}} \caption[Current waveform, emitted power of soft X--rays,
hard X--ray signal, and streak photograph, shot no.~030606-1]{Current waveform, emitted power of soft X--rays,
hard X--ray signal, and streak photograph, shot no.~030606-1.} \label{S300scope}
\end{figure}

\begin{figure}[!h]
\centerline{\includegraphics{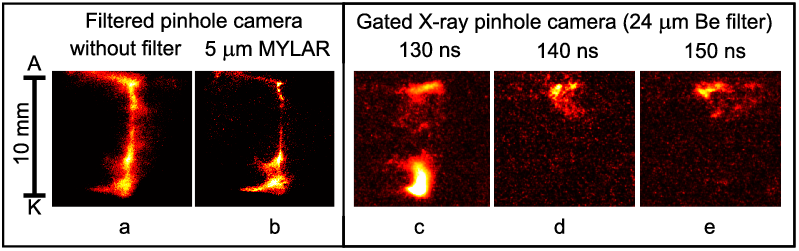}} \caption[Time--integrated X--ray pinhole images and the
sequence of gated X--ray images, shot no.~030606-1]{Time--integrated X--ray pinhole images and the sequence of
gated X--ray images, shot no.~030606-1. The used filter and/or the time of exposure is shown on the top of each
image.} \label{S300pinhole}
\end{figure}

\begin{figure}[!h]
\centerline{\includegraphics{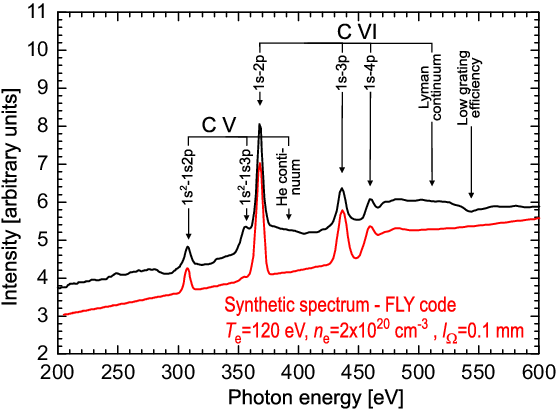}} \caption[XUV spectrum in the $200\div600$ eV spectral
range, shot no.~030606-1]{XUV spectrum in the $200\div600$ eV spectral range, shot no.~030606-1.}
\label{S300spectrum}
\end{figure}

\noindent The results of a number of diagnostic tools from the discharge no.~030606-1 are
 shown in Fig.~\ref{S300scope}--\ref{S300neutrons}.

In Fig.~\ref{S300scope} we present the current waveform, optical streak photograph, and some X--ray
characteristics. Time $t=0$ corresponds to the start of the current at the fibre. The streak photograph shows
that the optical emission began early in the discharge. The optical emission was continuing for several hundreds
of nanoseconds and its intensity was modulated by several peaks. Similarly, the soft X--ray power (measured by
the 11-channel polychromator) was peaking more than once. The peak power of sub-keV radiation reached 30 GW near
the maximum current. The total emitted energy exceeded 5 kJ.

Also, it could be observed that the pinch consisted of several distinct layers. We could distinguish a low
density coronal plasma and higher density interior layers. The coronal plasma was expanding with the radial
velocity of about $2\times10^6$ cm/s (see Fig.~\ref{S300scope}). Interior layers were displayed by the X--ray
pinhole cameras (see Fig.~\ref{S300pinhole}).

The time--integrated soft X--ray images (Fig.~\ref{S300pinhole}.a and \ref{S300pinhole}.b) show a radiating
sub-millimetre core. Several bright spots were observed near the cathode and anode. Some of them were smaller
than 100 $\muup$m in diameter, i.e.~smaller than the spatial resolution of the pinhole camera. At the peak of
the soft X--ray pulse (Fig.~\ref{S300pinhole}.d and \ref{S300pinhole}.e), the ``harder'' radiation came from
several bright spots near the anode. [Take note of the fact that the detection efficiency varied between
sections of MCP detector. For that reason the decreasing emission recorded by the pinhole camera
(Fig.~\ref{S300pinhole}.c-\ref{S300pinhole}.e) did not correspond to the increasing power of soft X--rays
(Fig.~\ref{S300scope}).] Another noticeable feature in Fig.~\ref{S300pinhole} is the radiation from the anode
which indicates runaway electrons in outer regions. On the basis of other experiments~\citep{Lebedev98}, we
assume that the strong radiating plasma near the cathode (Fig.~\ref{S300pinhole}.c) was the expanding electrode
material.

Fig.~\ref{S300spectrum} presents the temporally and spatially integrated spectrum in the $200\div600$~eV
spectral range where K--shell lines of carbon ions occurred. The plasma parameters determination is somewhat
problematic since the opacity effects with space-- and time--dependence have to be taken into account. To
simulate the spectrum with one temperature and one density was therefore impossible. Nevertheless, most features
of the obtained spectrum were simulated with the ``integrated'' electron temperature $T_\mathrm{e}$ of 120 eV
(cf.~the synthetic spectrum in Fig.~\ref{S300spectrum}). The estimation of the electron density $n_\mathrm{e}$
was ambiguous owing to its dependence on the choice of the optical path length $l_\Omega $. The only feature
that was not included in the synthetic spectrum was the strong continuum of He--like ions. This continuum could
originated from the part of plasma volume with higher density ($n_\mathrm{e}>10^{21}$ cm$^{-3}$) and lower
temperature ($T_\mathrm{e}<120$ eV). As regards the emitted energy of the carbon Ly--$\alpha$ line, the 365 eV
channel of the polychromator measured the total energy of about 30 J.

\subsection{Neutron measurements}
\noindent The results of neutron measurements are demonstrated on shot no.~030606-1 that was described above. In
this shot, an indium activation counter detected the modest neutron yield of $2\times10^7$. The yield was
calculated assuming the isotropic emission over a sphere.

\begin{figure}[!h]
\centerline{\includegraphics{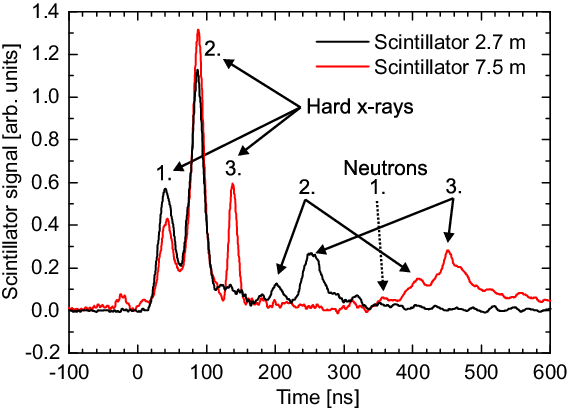}} \caption[Scintillator signals in shot no.~030606-1,
neutron yield $2\times10^7$]{Scintillator signals in shot no.~030606-1, neutron yield $2\times10^7$.}
\label{S300neutrons}
\end{figure}

Furthermore, we determined the production time and mean energy of neutrons from two time--of--flight
scintillators situated at distances of 2.70 and 7.45 m away from the pinch. The output from these two
scintillators is shown in Fig.~\ref{S300neutrons}. To make the graph in Fig.~\ref{S300neutrons} clear, X--ray
pulses from both scintillators were placed over each other. The time--of--flight of X--rays and the delay of
photomultiplier were included. Time $t=0$ corresponds to the start of the current and hence the time axis is the
same as in Fig.~\ref{S300scope}.

In Fig.~\ref{S300neutrons}, three hard X--ray pulses emitted in shot no.~030606-1 can be seen. The
time--of--flight analysis proved that each of X--ray pulses was accompanied with the neutron production. The
X--ray pulse no.~1, 2, and 3 corresponded to the neutron pulses no.~1,2, and 3, respectively. Most of the
neutrons were emitted at the third hard X--ray pulse. As Fig.~\ref{S300scope} shows, the third X--ray pulse
corresponded both to the peak of soft X--rays and to the noticeable change in the current waveform. The mean
neutron energy (measured in the axial, downstream direction) of all three pulses was $2.45\pm0.05$ MeV which
corresponds to the D--D reaction.

\section{Discussion}
\noindent The experimental results presented in this paper are similar to other fibre $Z$--pinch experiments.
There was no indication of globally collapsing pinch although the Pease--Braginskii current (600 kA for
polyethylene fibres) was exceeded. This rapid expansion is usually explained by the anomalous resistivity in a
low density corona~\citep{Chittenden95} and/or turbulent heating arising from MHD instability
growth~\citep{Riley96}. As for MHD instabilities in our experiment, the evidence of an $m=0$ mode was given by
the X--ray pinhole camera which recorded bright spots.

The key issue of our experiment was the modest neutron yield of about $10^7$ per one shot. Contrary to that, the
neutron yields between $10^9$ and $10^{10}$ were typical in other fibre $Z$--pinches even with a lower current
(cf.~\citealp{Stephanakis72, Sethian87, Kies91, Mitchell98, Lebedev98}). One of the possible explanations for
that could be a diameter of a~fibre used. \citet{Stephanakis72} reported that neutron yield was decreasing with
the increasing diameter of a~fibre. For instance, 100 $\muup$m diameter fibres produced ten times fewer neutrons
than fibres of 20 $\muup$m diameter. Another fact that could play an important role is the current generator. At
this point we could mention the disappointing results and the modest neutron yield reported by \citet{Decker89}
on a ``low'' impedance SPEED 2 generator.

Another significant result we obtained was the mean energy of neutrons, which was near the value of 2.45 MeV. In
addition to that, the time of neutron production corresponded to the soft X--ray emission
(cf.~Fig.~\ref{S300scope} and Fig.~\ref{S300neutrons}) and hence the detected neutrons could be of thermonuclear
origin. This result would be consistent with the $Z$--pinch experiments initiated from thicker ($>40$ $\muup$m)
fibres on the Gamble II \citep{Young77} and Kalif \citep{Kies91} generators. Such conclusion seems to be more
optimistic than the beam--target mechanism reported on the Poseidon~\citep{Sethian87} and
MAGPIE~\citep{Mitchell98} generators. However, the shift from 2.45 MeV could be too small to be identified.
Therefore, it is still feasible that the neutron yield was caused by beam--target interactions at a relatively
slow motion of dense plasma regions. The beam--target mechanism is also supported by the ``integrated'' electron
temperature of 120 eV that is too low even for the modest neutron yield produced by thermonuclear way. This
discrepancy was also observed by the HDZP group in Los Alamos National Laboratory. The explanation they proposed
was the model of instability heating in which the ion temperature $T_\mathrm{i}$ was substantially higher than
the temperature of electrons $T_\mathrm{e}$~\citep{Riley96}. Ion temperature $T_\mathrm{i}$ higher than the
electron temperature $T_\mathrm{e}$ was also observed during the stagnation in wire--array $Z$--pinches and was
ascribed to ion viscous heating within fine--scale interchange instabilities~\citep{Haines04}. Another
explanation how to reconcile the observed neutron yield with a low electron temperature is that only a small
part of the plasma volume was heated to a sufficiently high temperature. Then the bulk of the plasma could
remain cold. But even if the fusion mechanism were thermonuclear, it would still hold that the neutron yield was
modest in our experiment and that the yield higher than $10^{10}$ has not been reported by other research
groups.

Finally, we compare our fibre $Z$--pinch with the implosion of a wire--array onto a~deut\-erated fibre. Both
experiments were carried out on the same current generator S--300 with similar deuterated fibres and
currents~\citep{Klir04a, Klir05}. We found out that the neutron yield was one order of magnitude higher in the
case of the implosion of a wire--array. Another difference we observed was a slight shift of the neutron mean
energy from 2.45 MeV towards higher energies~\citep{Klir05}.

\section{Conclusion}
\noindent The dense $Z$--pinch formed from a deuterated polyethylene fibre was studied on the ``low'' impedance
S--300 pulsed power generator at the Kurchatov Institute. The majority of observed phenomena were in agreement
with other fibre $Z$--pinch experiments carried out on ``higher'' impedance generators. The important result was
obtained by the time--of--flight analysis which determined the mean neutron energy of about 2.45 MeV. However,
the modest neutron yield of $2\times10^7$ per shot was one order of magnitude lower than in our experiments with
the implosion of a wire--array onto a~fibre.